\documentclass[showpacs,amsmath,amssymb,aps,twocolumn,longbibliography]{revtex4-2}
\usepackage{graphicx}
\usepackage[colorlinks=true,citecolor=blue,linkcolor=magenta]{hyperref}
\usepackage[usenames]{color}
\usepackage{relsize}
\usepackage[english]{babel}
\usepackage{enumerate}

\newcommand{\ket}[1]{\left| #1 \right>} 
\newcommand{\bra}[1]{\left< #1 \right|} 

\newcommand {\grsim} {\ {\raise-.5ex\hbox{$\buildrel>\over\sim$}}\ }
\newcommand {\lessim} {\ {\raise-.5ex\hbox{$\buildrel<\over\sim$}}\ }

\newcommand{\up}{\uparrow}
\newcommand{\dn}{\downarrow}
\usepackage{xcolor}
\usepackage{siunitx}

\newcommand{\RN}[1]{%
  \textup{\uppercase\expandafter{\romannumeral#1}}%
}

\usepackage{xr}
\begin{document}

\title{Benchmarking a novel efficient numerical method for localized 1D Fermi-Hubbard systems on a quantum simulator}

\author{Bharath~Hebbe~Madhusudhana$^{1,2,3}$, Sebastian~Scherg$^{*1,2,3}$, Thomas~Kohlert$^{*1,2,3}$, Immanuel~Bloch$^{1,2,3}$,  Monika~Aidelsburger$^{1,2}$}

\affiliation{$^{1}$\,Fakult\"at f\"ur Physik, Ludwig-Maximilians-Universit\"at M\"unchen, Schellingstra{\ss}e 4, 80799 M\"unchen, Germany}
\affiliation{$^{2}$\,Munich Center for Quantum Science and Technology (MCQST), Schellingstr. 4, 80799 M\"unchen, Germany}
\affiliation{$^{3}$\,Max-Planck-Institut f\"ur Quantenoptik, Hans-Kopfermann-Stra{\ss}e 1, 85748 Garching, Germany}


\begin{abstract} 
Quantum simulators have made a remarkable progress towards exploring the dynamics of many-body systems, many of which offer a formidable challenge to both theoretical and numerical methods. While state-of-the-art quantum simulators are in principle able to simulate quantum dynamics well outside the domain of classical computers, they are noisy and limited in the variability of the initial state of the dynamics and the observables that can be measured. Despite these limitations, here we show that such a quantum simulator can be used to in-effect solve for the dynamics of a many-body system. We develop an efficient numerical technique that facilitates classical simulations in regimes not accessible to exact calculations or other established numerical techniques. The method is based on approximations that are well suited to describe localized one-dimensional Fermi-Hubbard systems. Since this new method does not have an error estimate and the approximations do not hold in general, we use a neutral-atom Fermi-Hubbard quantum simulator with $L_{\text{exp}}\simeq290$ lattice sites to benchmark its performance in terms of accuracy and convergence for evolution times up to $700$ tunnelling times. We then use these approximations in order to derive a simple prediction of the behaviour of interacting Bloch oscillations for spin-imbalanced Fermi-Hubbard systems, which we show to be in quantitative agreement with experimental results.  Finally, we demonstrate that the convergence of our method is the slowest when the entanglement depth developed in the many-body system we consider is neither too small nor too large. This represents a promising regime for near-term applications of quantum simulators. 
\end{abstract}

\maketitle

\textbf{}

\section{Introduction}
Quantum devices are on the periphery of establishing an advantage over their classical counterparts~\cite{Preskill2018quantumcomputingin}. Recently, a quantum computational advantage was demonstrated in sampling problems~\cite{2002quant.ph..5133T, aaronson2010computational} using superconducting qubits~\cite{Sycamore_2019} and a photonic quantum device~\cite{Zhong1460}. Moreover,  these and similar platforms based on neutral atoms in optical lattices~\cite{Gross995,Trotzky_2012}, trapped ions~\cite{Zhang_2017} and Rydberg atoms in optical tweezers~\cite{ebadi2020quantum, scholl2020programmable} have demonstrated high-fidelity simulations using specific model Hamiltonians in regimes that significantly challenge existing state-of-the-art classical numerical simulations. Harnessing the unique capabilities of these platforms and pushing their boundaries to even larger system sizes and evolution times paves the way towards practical applications of quantum devices in the area of quantum simulation~\cite{Preskill2018quantumcomputingin}.


The dynamics of quantum many-body systems out of equilibrium constitute fundamental questions that are both physically pertinent and computationally challenging. 
Contemporary explorations of this regime have uncovered a number of intriguing phenomena~\cite{gogolin_equilibration_2016} including many-body localization~\cite{altman_universal_2015, nandkishore_many-body_2015, abanin_colloquium_2019}, where an interacting system with quasiperiodic or random disorder defies thermalization~\cite{Schreiber_2015,smith_many-body_2016, Choi_2016, roushan_spectroscopic_2017}. Interestingly, an apparent breaking of ergodicity was also found in disorder-free models~\cite{PhysRevLett.117.240601}, e.g., in the presence of a linear potential~\cite{scherg_observing_2020, PhysRevX.10.011042, guo2020stark,morong_observation_2021}, which was attributed to a novel mechanism, known as Hilbert space fragmentation~\cite{moudgalya_thermalization_2019, khemani_localization_2020, doggen2020stark}. It constitutes one example of a rich variety of weak ergodicity-breaking models~\cite{serbyn_quantum_2020}, where the many-body Hilbert space shatters into (approximately) disconnected subspaces~\cite{khemani_local_2020, sala_ergodicity_2020}. A special example are quantum scars, where exceptional, low-entropy states in the many-body spectrum give rise to long-lived periodic orbits~\cite{bernien_probing_2017,bluvstein_controlling_2021} that resemble classical scarring.
Moreover, within the paradigm of slow thermalization, some driven systems have been shown to feature pre-thermal dynamics~\cite{de_roeck_very_2019, abanin_rigorous_2017, gromov_fracton_2020}, where two distinct thermalization timescales are found~\cite{PhysRevX.10.021044}. Adding periodic driving to the system further enables the realization of genuine out-of-equilibrium phases, a paradigmatic example being quantum time crystals~\cite{yao_discrete_2017,zhang_observation_2017,choi_observation_2017,else_discrete_2020}. Accordingly, there have been extensive experimental, theoretical and numerical efforts to study these phenomena and push the limits of current theoretical methods.

Any computation of the dynamics of quantum many-body systems is met with challenges arising from the dimension of the Hilbert space, which grows exponentially in the system size.  In other words, the quantum state may carry a large volume of information, which is impractical to store and process classically. A natural countermeasure is to relax the tolerance and seek approximate solutions which in many cases can be found efficiently~\cite{Barahona_1982}. Approximations rely on the expectation that all of the information in the many-body quantum state may not be equally important for the specific dynamics of the specific observable we are interested in.  By means of an ansatz, an approximate method identifies a part of the information in the quantum state that is most ``important" for the dynamics which can then be used to efficiently compute an approximation of the dynamics. With the advent of quantum simulators and quantum computers, it has become imperative to explore classical approximation methods that could potentially simulate quantum devices~\cite{kalai2014gaussian, PhysRevX.10.041038, pan2021simulating}.

If $\hat{H}$ is the Hamiltonian of a many-body system,  some of the most physically relevant problems include computation of thermal states $e^{-\beta \hat{H}}$ (here, $\beta = 1/kT$) or of the time evolution $e^{-i \hat{H}t }\ket{\psi}$ of a given state $\ket{\psi}$. One of the earliest numerical approximation methods developed was the cluster expansion for $2$D and $3$D lattice systems~\cite{PT_CM13, RRP_singh_high_order, 2021arXiv210112721P}. It is based on the observation that the exponential of the Hamiltonian can be written as a sum of terms representing various paths in the lattice.   Another class of approximate methods stem from a matrix product state (MPS) ansatz~\cite{Schollw_ck_2011, Verstraete_2008}.  Most commonly used MPS-based time evolution methods are the time-evolving block decimation (TEBD) and time dependent variational principle (TDVP)~\cite{PAECKEL2019167998}.  MPS based techniques have been very successful in studying both the ground state and time evolution of localized interacting many-body systems. Some bosonic many-body systems can be studied using Monte-Carlo methods. Recently, a new method has been proposed, making use of local thermalization of many-body systems~\cite{white_quantum_2018, ye_emergent_2020}.  

\begin{figure}
\includegraphics[scale=1]{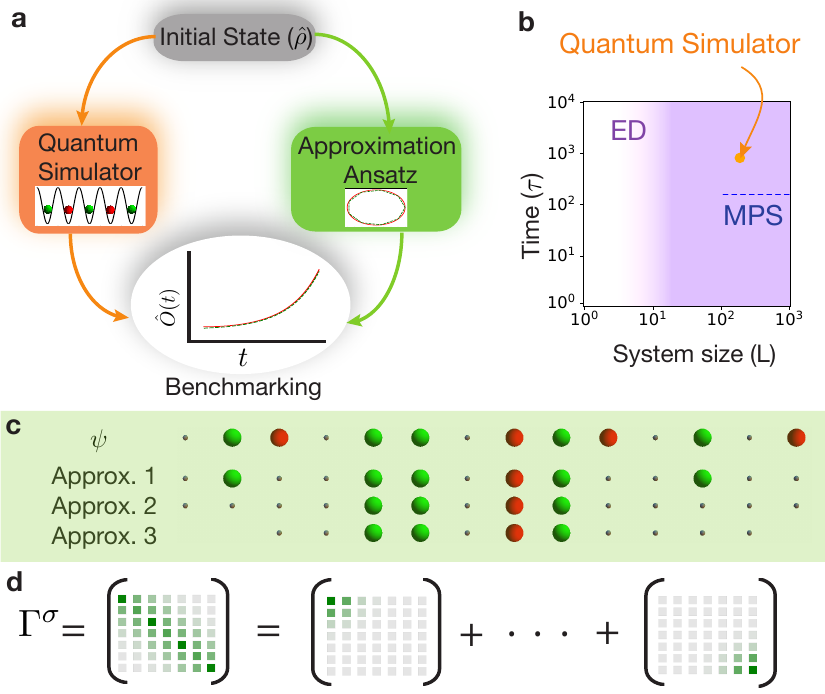}
\caption{\textbf{Quantum simulation and approximate descriptions. }\textbf{a} Schematic showing how a quantum simulator can be used to validate an efficient numerical approximate method to simulate the time evolution of a quantum many-body system,  in a regime inaccessible to current numerical methods. \textbf{b} Two-dimensional parameter space spanned by system size $L$ and evolution time in units of the tunneling time $\tau$. The violet shading represents the dimension of the Hilbert space, which is a measure of the complexity of ED. The dashed line indicates the threshold, where the local fidelity ($f_{\text{est}}^{2/L}$, where $f_{\text{est}}$ is a lower bound on the fidelity), which is an appropriate measure of reliability of TEBD (see appendix~\ref{other_numerical_methods} for details) falls below $95\%$ for the Stark Hamiltonian with $\Delta = 3J$ and $U=5J$, for practically accessible parameters. The orange point represents the experimental parameters, with $L_{\text{exp}}\simeq290$ and $t\approx 700 \tau$. \textbf{c} Schematic of the approximations used in this work. The red and the green spheres represent the two spin components (see text, Sec.~\ref{appx}). \textbf{d} Approximating the occupancy-matrix $\Gamma^{\sigma}$ [see text, Eq.~(\ref{gamma_eqn})] for spin-$\sigma$ atoms. }\label{FIG1}
\end{figure}

A key feature of approximate methods is the error estimate, which allows us to determine when it is reliable.  However, not every approximation ansatz has a well established error estimate. The bottleneck in such theories is to benchmark them, which obligates us to be able to compute the exact solution for a few instances of the many-body problem. In this work we demonstrate that a neutral atom quantum simulator can be used for this purpose [Fig.~\ref{FIG1}(a)]. 

\begin{figure*}[ht!]
\includegraphics[width=\textwidth]{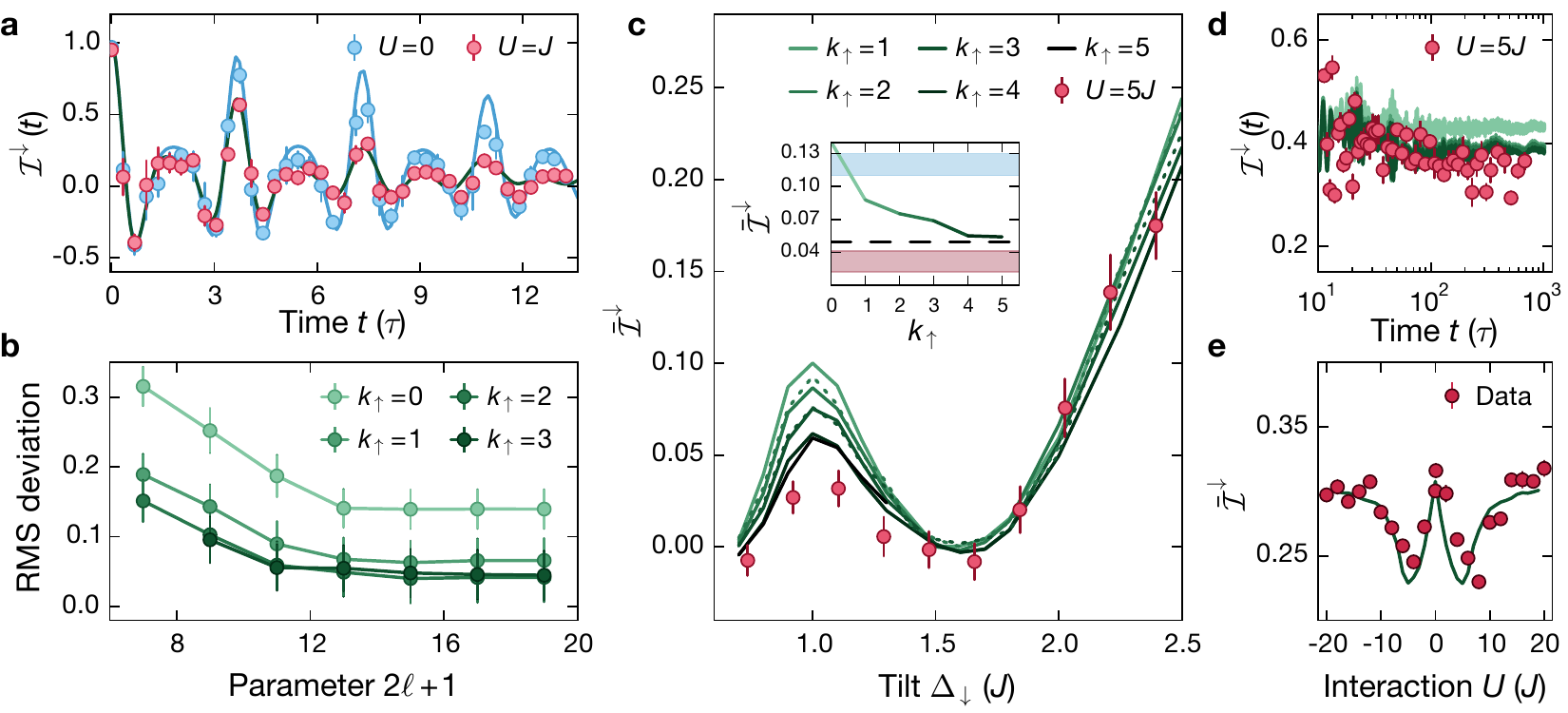}
\caption{\textbf{Benchmarking the convergence of the numerical method.} \textbf{a} A comparison of the imbalance time trace as predicted with the approximate method,  Eq.~(\ref{gamma_eqn}) (green solid curve) with a system size $L_{\text{apx}}=280$, $k_{\dn}=0, \ell = 7$ and $k_{\up}=3$ with the experimental data (red markers) for $\Delta_{\dn} = J$ and $U=J$. The precise values of $J, \Delta_{\sigma}$ and $\alpha$ are calibrated by fitting the corresponding non-interacting data (cyan markers) to the single particle theory (cyan solid curve).  In this and the following datasets, $\Delta_{\up}=0.9\Delta_{\dn}$. \textbf{b} Benchmarking the approximate method for short times (up to $25 \tau$) using the RMS deviation between the theoretical and the experimental time traces (see text) for $U=3J$ and $\Delta_{\dn} = J$, for $L_{\text{apx}}=280, k_{\dn}=0$ and various values of $\ell$ and $k_{\up}$. \textbf{c} Benchmarking the numerical method for long times ($300 \tau$) for $U=5J$ and various $\Delta_{\dn}$ measured experimentally (red markers) with the prediction of the approximate method with $L_{\text{apx}}=100$,  $\ell = 7$, $k_{\dn}=0$ and various $k_{\up}$ (green curves).  The dashed curves show the same calculations with $k_{\dn}=1$. The inset shows the convergence of the predicted steady-state imbalance $\bar{\mathcal I}^{\downarrow}$ (green curve) at $\Delta_{\dn}=1.1J$ towards the experimental value as we increase $k_{\up}$. The upper (lower) shaded band is the experimental value with errorbars for the non-interacting (interacting) case with $\Delta_{\dn}=1.1J$. The dashed line is the extrapolated convergence value of the green curve.  \textbf{d} Long time traces computed for a system size $L_{\text{apx}}=280$ using $\ell = 7$, $k_{\dn}=0$ and  $k_{\up}=2, 3, 4$ and $5$ shown as different shades for $\Delta_{\dn} = 3.3J$ and $U=5J$. The red markers represent the experimental data (data taken from Ref.~\cite{scherg_observing_2020}).  \textbf{e} The long time ($100\tau$) steady-state value of the imbalance for the Aubry-Andr\'e Hamiltonian with a detuning strength $\Delta = 3J$ for various interaction strengths,  with $L_{\text{apx}}=280$,  $\ell =7, k_{\dn}=0$ and $k_{\up}=2$ (see appendix~\ref{expt_details} for details of the data and the computation).       }\label{FIG2}
\end{figure*}

\section{Model and experimental implementation}

We consider a spinful one-dimensional (1D) Fermi-Hubbard model with $L$ sites and a spatially-dependent on-site potential $V_{ i, \sigma}$, where $\sigma = \{ \dn, \up \}$ represents the spin and $i$ represents the site index. The Hamiltonian of the system is
\begin{equation} \label{Hamiltonian}
\begin{split}
\hat{H} &= -J \sum_{i=1, \sigma =  \dn, \up}^{L} \hat{c}_{i, \sigma}^{\dagger}\hat{c}^{\phantom\dagger}_{i+1, \sigma} +\text{h.c} + \\
&\sum_{i=1, \sigma =  \dn, \up}^{L,} V_{i, \sigma} \hat{c}_{i, \sigma}^{\dagger}\hat{c}^{\phantom\dagger}_{i\sigma} + U\sum_{i=1}^{L} \hat{c}_{i, \up}^{\dagger}\hat{c}^{\phantom\dagger}_{i, \up}\hat{c}_{i, \dn}^{\dagger}\hat{c}^{\phantom\dagger}_{i,\dn}.\\
\end{split}
\end{equation}
Here $\hat{c}_{i, \sigma}^{\dagger}$ ($\hat{c}_{i, \sigma}$) denotes the fermionic creation (annihilation) operator for spin $\sigma$ on site $i$, $J$ is the tunneling matrix element and $U$ is the Hubbard interaction strength. In this work, we consider either quasiperiodic or linear on-site potentials to realize two paradigmatic models studied in the context of localization --- the Aubry-Andr\'e model and the Stark model.  In the Aubry-Andr\'e model, $V_{i, \sigma}= \Delta_{AA}\cos(2\pi \beta i + \phi) + \alpha (i-L/2)^2$, where $\Delta_{AA}$ is the strength of the detuning lattice, $\beta$ is the ratio of its wavelength to that of the primary lattice, $\phi$ a phase factor and and $\alpha$ is the harmonic trap confinement strength. For a Stark model, with a harmonic confinement, $V_{i, \sigma}= \Delta_{\sigma}i + \alpha (i-L/2)^2$, where $\Delta_{\sigma}$ is the spin-dependent tilt of the lattice. 

The neutral-atom quantum simulator we use in this work consists of a degenerate Fermi gas of $50(5) \times 10^3$ $^{40}\mathrm{K}$ atoms at temperature $T/T_F=0.15(1)$, where $T_F$ is the Fermi temperature. The gas is prepared in an equal mixture of two spin components in the $F=9/2$ manifold, with $\ket {\uparrow} = \ket{m_F = -7/2}$ and $\ket{\downarrow} = \ket{m_F = -9/2}$. The Fermi gas is loaded into a 3D optical lattice with lattice constant $d_s = \SI{266}{nm}$ along the $x$ direction and deep transverse lattices, with constant $d_\perp =  \SI{369}{nm}$, to isolate the 1D chains along $x$, with a residual coupling $<3\times10^{-4}J$. Hence, the system can be considered as a set of approximately 250 independent 1D chains, realizing Hamiltonian~(\ref{Hamiltonian}). Using an additional lattice with constant $2d_s$ we prepare an initial charge density wave (CDW), where only even sites are occupied, with an average density $\langle\hat{c}_{i,\sigma}^{\dagger}\hat{c}^{\phantom\dagger}_{i,\sigma}\rangle \lessim 0.25$. The fraction of doubly-occupied sites is suppressed to $<0.03$, by loading the gas at a repulsive scattering length of $100\,a_0$ and using an additional short off-resonant light pulse~\cite{scherg_nonequilibrium_2018}; $a_0$ is the Bohr radius. There is no spin order, therefore our initial state can be modelled as an incoherent distribution of site-localized particles with random spin configuration.

The central chain in our experiment has a length of  $L_{\text{exp}}=290 \pm 10$ sites and we can reliably simulate (experimentally) the time-evolution with the Aubry-Andr\'e or the Stark Hamiltonian up to $T\approx700 \tau$~\cite{scherg_observing_2020}, where $\tau = \hbar /J$ denotes the tunneling time and $\hbar$ is the reduced Planck's constant. Although the dynamics remains fairly localized, it is numerically inaccessible due to the quantitative value of the localization length. Exact diagonalization (ED) is impractical for $L> 21$  and methods based on MPS are impractical for $T> 200 \tau$, for our system and the parameters we use in the experiment (Fig.~\ref{FIG1}(b) and appendix~\ref{other_numerical_methods}). The error incurred in the TEBD method using MPS has been studied extensively~\cite{PhysRevB.101.035148, PhysRevX.10.041038}. In particular, the local Uhlmann fidelity~\cite{Uhlmann_1976} was found to be useful to estimate the error in local observables~\cite{PhysRevA.98.042316}.  For our system and parameters, we estimate that the TEBD error in local observables is $\sim 5\%$ after $200\tau$ for realistic bond dimensions (Appendix~\ref{other_numerical_methods}).  Based on previous experiments done on the same experimental set up with the same models~\cite{Schreiber_2015, scherg_observing_2020}, including studies of various imperfections~\cite{PhysRevX.7.011034,PhysRevLett.116.140401, PhysRevX.7.041047},  we expect that the systematic deviation in the imbalance due to experimental imperfections is comparable to the errorbars obtained in the measurements.  Therefore we use our quantum simulator to benchmark a new approximate numerical method which we develop.   For the purpose of comparison with the experiment, we restrict the on-site potential to the above mentioned models models, although our theoretical method is expected to be applicable more generally, for all localized models.  

\section{Theoretical approximations}\label{appx}
Our approximate method is built upon antisymmetrized product states and is suitable for fermionic systems. 
A basis state $|\psi\rangle$ of the Fock space of two component fermions on a 1D lattice of size $L$ with $N_{\sigma}$ atoms in spin-$\sigma$ can be represented in second-quantized notation as $|\psi\rangle=\hat{c}_{j_1, \up}^{\dagger}\cdots \hat{c}_{j_{N_{\up}}, \up}^{\dagger}\hat{c}_{i_1, \dn}^{\dagger}\cdots \hat{c}_{i_{N_{\dn}}, \dn}^{\dagger}|\text{vac}\rangle$, where $|\text{vac}\rangle$ denotes the state of the empty lattice. Here $1\leq i_1 < i_2 \cdots < i_{N_{\dn}}\leq L$ and $1\leq j_1 < j_2 \cdots < j_{N_{\up}}\leq L$. States of multiple non-interacting particles, (e.g., $N_{\dn}>1, N_{\up}=0$) can always be written as antisymmetrized products. For instance, if the spin-$\dn$ atoms start at sites $i_1, \cdots, i_{N_{\dn}}$, we can compute the full time-evolved state $|\psi(t)\rangle$ under the Hamiltonian~(\ref{Hamiltonian}) as the antisymmetrized product of $|\psi_1(t)\rangle, \cdots, |\psi_{N_{\dn}}(t)\rangle$, where $|\psi_r(t)\rangle$ is the time-evolved single-particle state  of an atom starting at the site $i_r$,  $r \in \{1, 2, \cdots, N_\downarrow\}$.  If the system is localized, we can compute each $|\psi_r(t)\rangle$ by restricting the lattice to a finite size,  including only sites $i \in \{ i_r-\ell, \cdots, i_r+\ell \}$ for some positive integer $\ell$. As we increase $\ell$, $|\psi_r(t)\rangle$ converges quickly to the exact value, if the system is localized. This is an efficient approximation for many-particle non-interacting dynamics.

We now construct an efficient approximation for interacting many-body systems (i.e., $N_{\dn}, N_{\up} >0$), where the dynamics is localized. Although the many-body state $|\psi\rangle$ consists of a large number of variables, the experimentally relevant information can be summarized in two $L\times L$  ``occupancy matrices" $\Gamma^{\dn}$ and $\Gamma^{\up}$. The $ij$-th elements of these two matrices are defined as~\cite{PhysRevLett.115.046603}, 
\begin{equation}
\Gamma^{\sigma}_{ij}=\langle \psi | \hat{c}_{i, \sigma}^{\dagger}\hat{c}^{\phantom\dagger}_{j, \sigma}|\psi\rangle. 
\end{equation}
The diagonal entries of $\Gamma^{\sigma}$ represent the on-site occupation density of spin $\sigma$ atoms and the off-diagonal terms represent correlators. For instance, in the previous example of multiple non-interacting particles, $\Gamma^{\dn}(t)$ can be written as (see appendix~\ref{occupancy_matrix} for a derivation).
\begin{equation}\label{gamma_eqn_ni}
\Gamma^{\dn}(t)= |\psi_1(t)\rangle \langle \psi_1(t)|+\cdots + |\psi_{N_{\dn}}(t)\rangle \langle \psi_{N_{\dn}}(t)|.
\end{equation}
If the system is localized, each of the operators $|\psi_r(t)\rangle \langle \psi_r(t)|$ are dominated by elements around a particular diagonal entry~\cite{PhysRev.109.1492}. Thus, their sum $\Gamma^{\dn}(t)$ will be dominated by elements around the diagonal, although the diagonal itself may be uniform~\cite{PhysRevA.93.031601}~[Fig.~\ref{FIG1}(d)].  


In the many-body case if the system remains localized, the occupancy matrices $\Gamma^{\dn}(t)$ and $\Gamma^{\up}(t)$  are again dominated by elements in and around the diagonal. Therefore, we seek an approximate representation for these matrices, of the form:
\begin{equation}\label{gamma_eqn}
\Gamma^{\sigma}(t)=\Gamma^{\sigma, 1}(t)+\cdots +\Gamma^{\sigma, N_{\sigma}}(t)
\end{equation}

Where, $\Gamma^{\sigma ,r}(t)$ are density matrices to be defined, loosely representing the state of the $r$-th atom. Unlike in the non-interacting case, the matrices $\Gamma^{\sigma, r}(t)$ need not be pure or orthonormal  [see appendix~\ref{reconstruction} for a detailed discussion of Eq.~(\ref{gamma_eqn})]. We use this representation to approximate the time evolution of a separable initial state $|\psi\rangle=\hat{c}_{j_1, \up}^{\dagger}\cdots \hat{c}_{j_{N_{\up}}, \up}^{\dagger}\hat{c}_{i_1, \dn}^{\dagger}\cdots \hat{c}_{i_{N_{\dn}}, \dn}^{\dagger}|\text{vac}\rangle$ under the Hamiltonian Eq.~(\ref{Hamiltonian}). For such a state, the matrices $\Gamma^{\sigma, r}(t=0)$ are well defined.  Below, we summarize the approximations used to define $\Gamma^{\sigma, r}(t)$~[Fig.~\ref{FIG1}(c)]. Hereafter, we use $\sigma$ to represent the spin component for which we are computing the occupancy matrix and $\bar{\sigma}$ to represent the other spin component. 

\begin{itemize}
\item[1.]\textbf{Approximation-$\mathbf{1}$:} Choose an integer $\kappa_{\sigma}\geq 0$ and construct a $\kappa_{\sigma}$\textit{-shell} around the site $i_r$., i.e.,  a set of $\kappa_{\sigma}$ sites nearest to $i_r$~\cite{k_sigma_shell}.  Replace all spin $\sigma$ atoms in the initial state $|\psi\rangle$, outside this $\kappa_{\sigma}$-shell with holes. This would result in a new state $|\psi'\rangle$ with at most $\kappa_{\sigma}+1$ spin $\sigma$ atoms.
\item[2.]\textbf{Approximation-$\mathbf{2}$:} Choose an integer $\kappa_{\bar{\sigma}}\geq 0$ and replace all spin $\bar{\sigma}$ atoms in $|\psi'\rangle$ outside a $\kappa_{\bar{\sigma}}$-shell with holes. This would result in a new state $|\psi''\rangle$ with at-most $\kappa_{\sigma}+1$ spin $\sigma$ atoms and at-most $\kappa_{\bar{\sigma}}+1$ spin $\bar{\sigma}$ atoms. 
\item[3.]\textbf{Approximation-$\mathbf{3}$:} Choose an integer $\ell\geq \kappa_{\sigma}/2,  \kappa_{\bar{\sigma}}/2$ and truncate the lattice to $\{i_r-\ell, \cdots,  i_r+\ell\}$.
\end{itemize}

The final state $|\psi^{\prime \prime}\rangle$ after applying the above three approximations on $|\psi\rangle$ is a few-body state on a lattice of size $2\ell +1$, $|\psi^{\prime\prime}\rangle= \hat{c}_{\alpha_1, \sigma}^{\dagger}\cdots \hat{c}_{i_r, \sigma}^{\dagger} \cdots  \hat{c}_{{\alpha_{q_{\sigma}}}, \sigma}^{\dagger}\hat{c}_{\beta_1, \bar{\sigma}}^{\dagger}\cdots \hat{c}_{\beta_{q_{\bar{\sigma}}}, \bar{\sigma}}^{\dagger}|\text{vac}\rangle$ where $q_{\sigma}+1$ (the $+1$ accounts for the spin-$\sigma$ atom that starts at site $i_r$) and $q_{\bar{\sigma}}$ are the number of atoms of spin $\sigma$ and spin $\bar{\sigma}$ that remain after approximations $1$ and $2$ respectively and $\alpha_i, \beta_i$ are their positions. During a time evolution under the Hamiltonian Eq.~(\ref{Hamiltonian}), $|\psi^{\prime \prime}\rangle$ will remain in a smaller Hilbert space $\mathcal H^{\prime\prime}$,  the dimension of which is independent of $L$, polynomial in $\ell$ and  exponential in $q_{\sigma}, q_{\bar{\sigma}}$.   We evolve $|\psi^{\prime \prime }\rangle$ in time under the projection of the  Hamiltonian, Eq.~(\ref{Hamiltonian}),  to the space $\mathcal H^{\prime\prime}$, using ED. We then use the time-evolved state $|\psi^{\prime\prime}(t)\rangle$ to compute $\Gamma^{\sigma, r}$:
\begin{equation}\label{single_atom_ms}
\Gamma_{ij}^{\sigma, r}(t)= \frac{1}{1+q_{\sigma}}\langle \psi^{\prime\prime}(t) | \hat{c}_{i, \sigma}^{\dagger}\hat{c}^{\phantom \dagger}_{j, \sigma}|\psi^{\prime\prime}(t)\rangle .
\end{equation}

\noindent This matrix is in fact the occupancy matrix of spin $\sigma$ atoms corresponding to $|\psi^{\prime\prime}(t)\rangle$.  The factor $\frac{1}{1+q_{\sigma}}$  stems from the  $q_{\sigma}+1$ spin $\sigma$ atoms in this state (see appendix~\ref{reconstruction} for a further discussion).  We follow this procedure for $r=1, 2, \cdots, N_{\sigma}$ to obtain $\Gamma^{\sigma, 1}(t), \cdots, \Gamma^{\sigma, N_{\sigma}}(t)$, from which we compute $\Gamma^{\sigma}(t)$ following Eq.~(\ref{gamma_eqn}).  

The possible choices for the free parameters $\ell, \kappa_{\sigma}$ and $\kappa_{\bar{\sigma}}$ can be represented by integer points in a $3$D region defined by $\kappa_{\sigma}, \kappa_{\bar{\sigma}}\leq \text{min}\{2\ell, L\}$ (see Fig.~\ref{FIGS2} in the appendix).  At the extreme point $\ell, \kappa_{\sigma}, \kappa_{\bar{\sigma}}  =L$, the approximate method is exact.  If we use periodic boundary conditions on the full lattice,  the approximate method is exact when $2\ell, \kappa_\sigma, \kappa_{\bar{\sigma}}=L$. In another limiting case when $\kappa_{\sigma}=\kappa_{\bar{\sigma}}=2\ell$,  represented by a line in the $3$D space, the approximate method reduces to a possible adaptation of the standard cluster expansion method to $1$D Fermi-Hubbard systems~\cite{PhysRevLett.97.187202, PhysRevE.95.033302, PhysRevLett.113.195302}, also discussed in appendix~\ref{cluster_comparison}. This extreme case can also be related to TEBD with a spatially varying bond dimension optimized for a local observable (see appendix~\ref{other_numerical_methods} and~\ref{MPS_reformulation} for more details.  Also see Ref.~\cite{white_quantum_2018} for a related idea). Depending on the nature of the dynamics being studied, the convergence rate of the approximation may be the fastest along a non-trivial path towards the extreme point $2\ell, \kappa_{\sigma}, \kappa_{\bar{\sigma}}  =L$ in this $3$D space.  The practical utility of the approximate method depends on the rate of convergence. We show that the convergence can occur for relatively small values of $\ell, \kappa_{\sigma}$ and $\kappa_{\bar{\sigma}}$ in the localized regime by benchmarking the numerical results with a neutral-atom quantum simulator.


\begin{figure*}
\includegraphics[width=\textwidth]{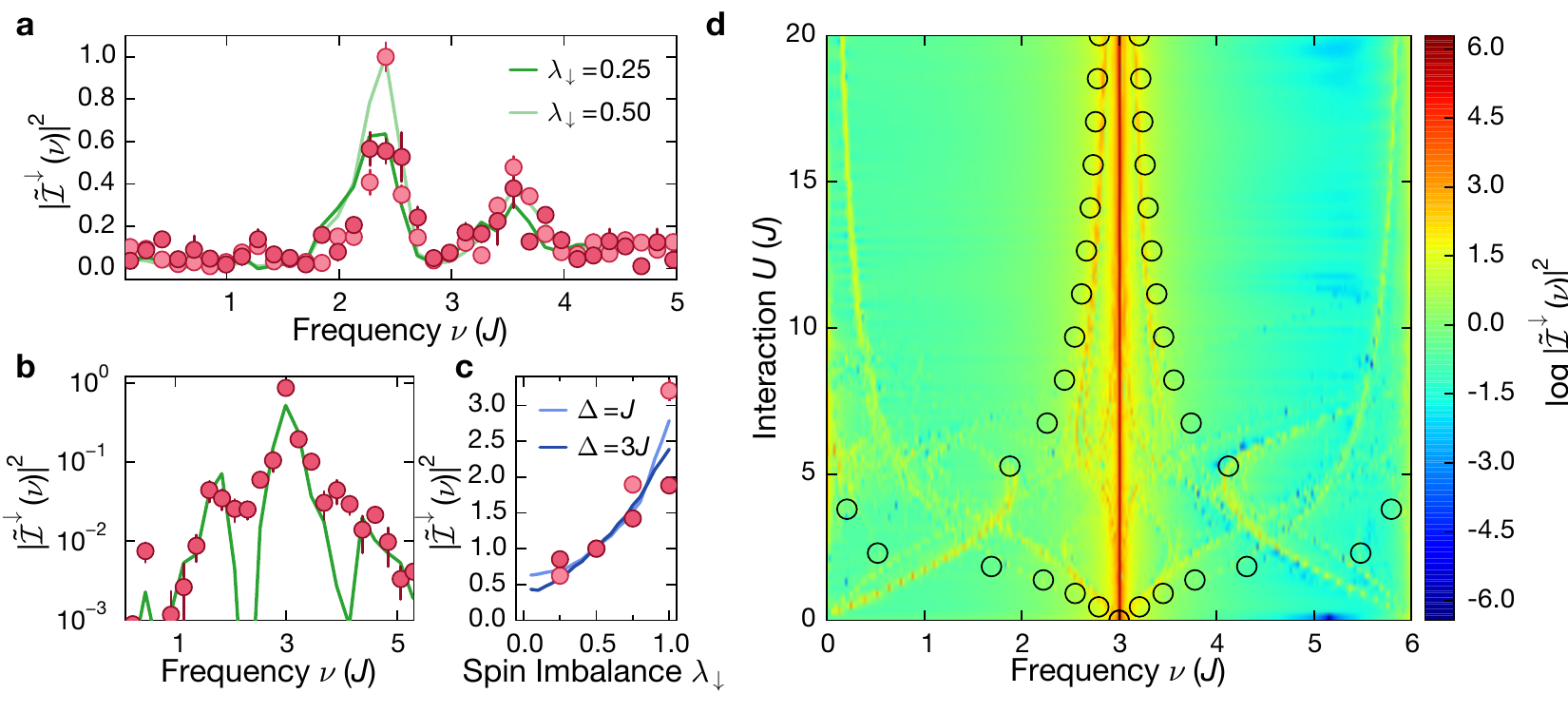}
\caption{\textbf{Interacting short-time dynamics versus spin imbalance. } \textbf{a} Fourier spectrum $\tilde{\mathcal I}^{\dn}(\nu)$ of the imbalance time trace of spin-$\dn$ atoms for a spin-imbalanced CDW, with $\lambda_\downarrow = 0.5$ (light shade) and $\lambda_\downarrow = 0.25$ (dark shade).  The spectrum was obtained from a time trace up to $t=\SI{8}{ms} = 25 \tau$. Throughout, the solid lines are predictions from the approximate method and the circular markers are the experimental data. The reduction in the peak value of the Fourier spectrum with $25\%$ spin-$\dn$ shows the enhancement in the interaction effect due to the spin imbalance, as predicted by Eq.~(\ref{spin_imb}). \textbf{b} Fourier spectrum for $\Delta_{\dn}= 3J$, $U=3J$ and $\lambda_\downarrow = 0.25$. The enhancement in the interaction effect results in the prominent side peak at $\nu = 1.5 J$. \textbf{c} The primary peak in the Fourier spectrum of the imbalance of spin $\up$ for $U=3J$, $\Delta_{\dn} = 1.1 J$ (light shade) and $\Delta_{\dn} = 3J$ (dark shade)  as a function of the fraction of the atoms in spin $\up$, showing a clear interaction effect enhancement as a result of the spin imbalance.  In the experimental data shown in \textbf{a}-\textbf{c}, $\Delta_\up=0.9\Delta_\dn$. The computations in \textbf{a-c} were performed with a system size $L_{\text{apx}}=100$,  $\ell = 4, $ $k_{\dn}=0$ and $k_{\up}=2$.  \textbf{d} Fourier spectra of the imbalance time trace with a $10^{-2}J$ step (i.e.,  the time evolution was up to $700 \tau$) for various interaction strengths with $\Delta =3J$.  The computations were performed with a system size $L_{\text{apx}}=280$,  $\ell = 7, $ $ k_{\dn}=0$ and $k_{\up}=4$. The red line at the center corresponds to the main peak of the imbalance oscillation.  The black circles represent the perturbative estimate,  valid in the limit $J\ll \Delta$, of the side peak, $3J \pm 4J^2U/(\Delta^2-U^2)$ (see appendix~\ref{expt_details} for more details) . }\label{FIG3}
\end{figure*}

\section{Benchmarking}

In the experiment, we prepare the system in a CDW initial state, i.e., a uniform incoherent sum of states where each even site is occupied by a spin-$\up$ or a spin-$\dn$ atom as described above (with the total numbers fixed to $N_{\up}$ and $N_{\dn}$ respectively) and each odd site is empty. Starting from a spin-polarized sample in the $\ket{\dn}$-state we generate a spin imbalanced gas where the spin imbalance is adjusted by varying the RF power during a sweep that couples the two spin states for constant atom number, $N=N_{\up}+ N_{\dn}$.  After a variable time evolution, we measure the spin-resolved imbalance $\mathcal I^{\sigma}(t)$ between the even- and odd-site occupancies, defined as
\begin{equation}
\mathcal I^{\sigma}(t) = \frac{N_{\sigma}^{\text{even}}(t)-N_{\sigma}^{\text{odd}}(t)}{N_{\sigma}}\ .
\end{equation}
Here, $N_{\sigma}^{\text{even}}(t)$ $ [N_{\sigma}^{\text{odd}}(t)]$ is the number of spin $\sigma$ atoms on even [odd] sites. We use the approximate method to compute the spin-resolved imbalance for both, the Stark and Aubry-Andr\'{e} model, and benchmark it with experimental results. The parameters in the two models are calibrated by realizing the non-interacting limit.



For simplicity, we use $\sigma= \dn$, i.e., we will compute the imbalance of $\dn$ atoms using the proposed method and compare it with the experimental data.  Note that in a CDW, since only even sites can be occupied, the number of even sites within the $\kappa_\sigma$-shell is more relevant than $\kappa_\sigma$ itself.  Therefore, we label the shells by $k_{\sigma}$, the number of even sites in the $\kappa_\sigma$-shell and $k_{\bar{\sigma}}$, the number of even sites in a $\kappa_{\bar{\sigma}}$ -shell.  Note that $k_\sigma=0$ when $\kappa_\sigma = 0, 1$ or $2$.  And $k_{\sigma}=1$ when $\kappa_\sigma = 3$.  In general, when $\kappa_\sigma = 4n+3$, $k_{\sigma}=2n+1$ and when $\kappa_\sigma = 4n, 4n+1$ or $4n+2$, $k_{\sigma}=2n$.  Hereafter, we will label a shell by $k_\sigma$.  See table~\ref{T1} in the appendix for a list of $k_\sigma$ values and the corresponding $\kappa_\sigma$ values.  For the Stark Hamiltonian, we consider two time scales --- a short time scale where one can observe coherent Bloch oscillations in the imbalance~\cite{scherg_observing_2020} and a long time scale where the oscillations are dephased and a steady-state value is reached. To quantify the disagreement between the experiment and the numerical method, we use the root-mean-square (RMS) deviation, defined as $\sqrt{\int |\mathcal I_{\text{expt}}-\mathcal I_{\text{apx}}|^2 \text{d}t}$.  Here, $\mathcal I_{\text{apx}}$ is computed using Eq.~(\ref{gamma_eqn}).  In Fig.~\ref{FIG2}(b) we show the RMS deviation as a function of $\ell$ for various $k_{\up}$, with $k_{\dn}=0$, upto $t=\SI{8}{ms}=25 \tau$. The data indicates that the variation of the approximate method with respect to the experiment is negligible for $\ell >7$, and therefore, we use $\ell =7$ for further computations. While there is a significant difference between $k_{\up}=0$ and $k_{\up}=2$, increasing it further has no significant impact on the RMS deviation, suggesting that much of the interaction effects on short-time Bloch oscillations stem from three-atom processes.  For a CDW initial state, increasing $k_{\dn}$ does not have a significant effect.  This can be attributed to the averaging in the CDW. In particular, the dynamics of a spin-$\dn$ island is unaffected by increasing $k_{\dn}$.  The effect of $k_{\dn}$ is most significant for a N\'eel type initial state (see appendix~\ref{role_of_kdn} for more details).

For long evolution times, we directly compare the steady-state imbalance values $\bar{\mathcal I}^{\dn}$, defined as the imbalance averaged over a time window (see appendix~\ref{expt_details} for details).  Fig.~\ref{FIG2}(c) shows the benchmarking results for long-time dynamics, averaged over $10$ points between $300\tau$ and $330\tau$. We find that for $\Delta >2J$, numerical results with $k_{\up}=4$ are already sufficient to predict the long-time dynamics within errorbars (Fig~\ref{FIG2}d). However, the convergence is slower at smaller $\Delta$, owing to a larger single-particle localization length (Fig~\ref{FIG2}c) of more than two sites for $\Delta < 2J$, where we expect our method to fail to converge within the accessible range of the parameters $\ell, k_{\up}$ and $k_{\dn}$.  Similarly, we find that for the Aubry-Andr\'{e} Hamiltonian at $4J$ a value of $k_{\up}=2$ is already sufficient to reproduce the well-known interaction dependence within errorbars~\cite{Schreiber_2015}. This observation is in agreement with the analytical localization length~\cite{Aubry80} of 1.4 lattices sites obtained at a detuning strength of $\Delta = 3J$.

In all the datasets, $\Delta_\up =0.9 \Delta_\dn$ and the tilt factor $0.9$ is determined by the differential tilts experienced by the $\ket{F=9/2: m_F=-9/2}$ and $\ket{F=9/2; m_F=-7/2}$ states.  For a study of the effect of varying this tilt difference and the performance of our approximate method in those cases see Ref.~\cite{HS_fragmentation}.


\section{Computations using the approximate method}
We now use the approximate method to obtain further insights into the interacting dynamics of the Stark Hamiltonian. For a CDW initial state, we can use the approximations described above to derive the following expression for the imbalance (see appendix~\ref{imbalace_computation} for a derivation.  Also see Ref.~\cite{PhysRevA.93.031601} for a similar expression for equal spin populations).
\begin{equation}\label{cluster}
\begin{split}
\mathcal I^{\dn}(t)= &   \mathcal{I}^{\dn}(t; U=0) + \sum_{q_{\up}=k_{\dn}-q_{\dn}}^{k_{\up}} \sum_{q_{\dn}=0}^{k_{\dn}} \binom{k_{\dn}}{q_{\dn}}  \binom{k_{\up}-k_{\dn}}{q_{\up}+q_{\dn}-k_{\dn}} \\
& \times  \lambda_{\up}^{q_{\up}} (1-\lambda_{\up})^{k_{\up}-q_{\up}} \left( \mathcal{I}^{\dn}_{q_{\up}, q_{\dn}}(t; U)- \mathcal{I}^{\dn}(t; U=0)\right),\\
\end{split}
\end{equation}

\noindent where, $\lambda_{\up} = 2N_{\up}/L$ is the filling of spin-$\up$ atoms and $\mathcal{I}^{\dn}_{q_{\up}, q_{\dn}}(t; U)$ is the imbalance of spin-$\dn$ atoms in a few-body interacting state with $q_{\up}$ spin-$\up$ atoms and $q_{\dn}+1$ spin-$\dn$ atoms.  In the case $k_{\up}=0$, this expression reduces to
\begin{equation}\label{spin_imb}
\begin{split}
\mathcal{I}^{\dn}(t; U) =& \mathcal{I}^{\dn}(t; U=0)
+\sum_{q_{\up}}  \binom{k_{\up}}{q_{\up}}\lambda_{\up}^{q_{\up}}(1-\lambda_{\up})^{k_{\up}-q_{\up}}\\
&\times \left( \mathcal{I}^{\dn}_{q_{\up}}(t; U)- \mathcal{I}^{\dn}(t; U=0)\right).\\
\end{split}
\end{equation}

\noindent The first term in the above expression is the non-interacting imbalance, while the second one corresponds to the interaction effect. Intuitively, for any spin-$\dn$ atom, the probability that it interacts with a spin-$\up$ atom is proportional to $N_{\up}$ and therefore, the first term in the sum is linear in $\lambda_{\up}$.  A CDW initial state with unequal spin populations must therefore show an enhanced (reduced) interaction effect on the minority (majority) spin component. This simple prediction can be directly tested using our cold-atom quantum simulator.

\begin{figure}
\includegraphics[width=3.3in]{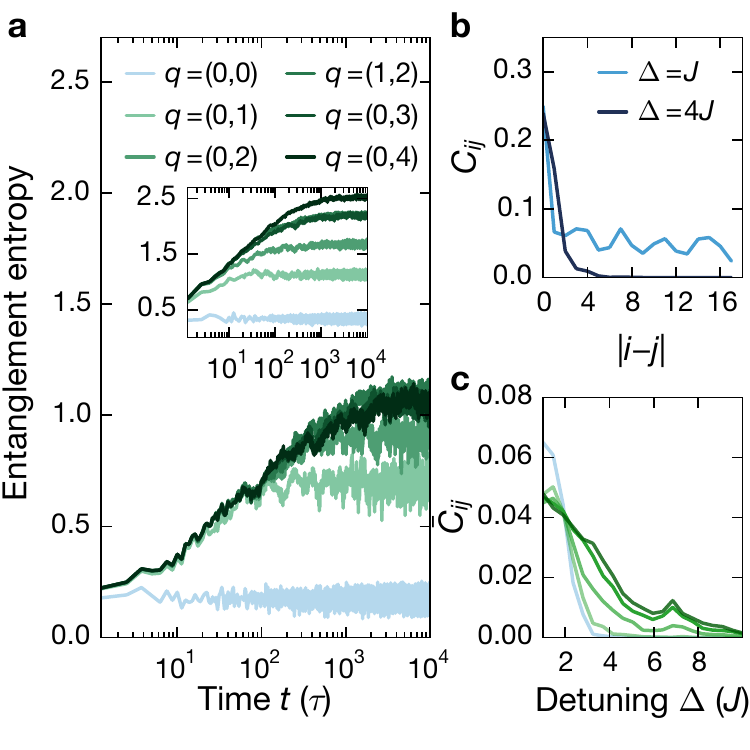}
\caption{\textbf{Limitations of the approximate description.} \textbf{a} The growth of the bipartite entanglement entropy  computed for a \textit{single few-body state } in the approximate method using $\ell = 7$ and various values of $q=(q_{\up}, q_{\dn})$ for the Aubry-Andr\'e Hamiltonian with  $\Delta = 8J$. Similar plot for $\Delta = 4J$ is shown in the inset.   See Fig.~\ref{FigS4} in the appendix for a comparison of the entropy with TEBD computations.  \textbf{b} The correlation $C_{ij}=\langle \hat{n}_i \hat{n}_j\rangle - \langle \hat{n}_i \rangle \langle \hat{n}_j \rangle $ in the single particle case for the Aubry-Andr\'e model. \textbf{c} The steady-state value of the correlation $\bar{C}_{ij}$ for various $\Delta$ and $q$ with $\ell=7$ for an interacting system,  showing that the convergence is the slowest near the critical point, $\Delta \approx 3.5 J$. The same color code as in \textbf{a} is used.  The interaction strength used in all the computations in this figure is $U=5J$ (see appendix~\ref{expt_details} for more details).}\label{FIG5}
\end{figure}

To quantify the effect of interactions,  we consider the height of the primary peak in the Fourier spectra of the measured imbalance time traces~\cite{scherg_observing_2020}. As shown in Fig.~\ref{FIG2}(a), finite interactions induce additional frequency components in the dynamics, thereby reducing the strength of the primary peak in Fourier space. We study the interaction effect for initial CDWs with variable spin imbalance $\lambda_\downarrow$. Fig.~\ref{FIG3}(a) and Fig.~\ref{FIG3}(c) show that a spin imbalance in the CDW indeed produces an enhanced interaction effect on the minority spin component and the results agree quantitatively with the predictions of the method. The interaction-induced side-peak in the Fourier spectrum is clearly visible, both in the data and the numerical simulations [Fig.~\ref{FIG3}(b)]. These side peaks correspond to energy shifts in the many-body eigenspectra that appear due to interactions~\cite{scherg_observing_2020}.  

Having an efficient numerical method at hand, we can now systematically explore the frequency-resolved features that appear in the interacting dynamics, which would otherwise be an arduous task. Following Fig.~\ref{FIG2}, the time dynamics of the imbalance is well approximated by the approximate method with $\ell = 7$, $k_{\dn}=0$ and $k_{\up}=4$ for $\Delta =3J$, up to $700 \tau$.  This corresponds to a resolution of $10^{-2}J$ in the Fourier spectrum.  In Fig.~\ref{FIG3}(d), we show the Fourier spectrum of the imbalance for $\Delta = 3J$ for $100$ values of the interaction strength ranging between $0$ and $20J$. The main peak at $3J$ corresponds to the primary frequency of the Bloch oscillations. The side peaks are placed fairly symmetrical around this main peak. In the limit of weak interactions ($U<2J$) the dominant side peaks are $\propto U$ away from the main peak and in the limit of strong interactions, they are $\propto 1/U$ away from the main peak. In these two limits, the prediction of the approximate method agrees with the perturbative estimate, $4J^2U/(\Delta^2-U^2)$ of the side peaks.  This is the first non-zero correction to the energy obtained by treating $J$ as a perturbation~\cite{scherg_observing_2020, HS_fragmentation}.  Note that at $U= \Delta$, when this perturbative estimate breaks down, the approximate method reveals a rich structure in the spectrum.

So far, we have considered charge imbalance as the main observable to study the many-body dynamics, since it is easily accessible in experiments~\cite{Schreiber_2015}. We now turn to the bipartite entanglement entropy (EE), which is widely employed to study localization dynamics in many-body systems. It is known that for thermal systems, the EE rapidly increases and saturates to a thermal value, proportional to the volume of the smaller of the two parts of the lattice. In many-body localized systems, the EE grows logarithmically in time before saturating to a value lower than the thermal value~\cite{PhysRevLett.109.017202,Bauer_2013, PhysRevLett.124.243601, luitz2020slow}. 


Here we show that the approximations behind our efficient simulations allow us to gain additional insight into the microscopic processes underlying the dynamics and growth of the EE.  Indeed, we find that the logarithmic growth at initial times stems mostly from few-body physics. As a reminder, the many-body quantum state in our description is constructed by patching together several few-body states reduced into a single-atom mixed state [Eq.~(\ref{single_atom_ms})].  Each few-body state is associated with a short sublattice of size $2\ell + 1$ centered at the initial position of the corresponding atom.  Computing the EE would require us to compute the full many-body state, which can only be defined by a canonical choice (see appendix~\ref{reconstruction} for details). However, constructing the full many-body state and computing its EE is hindered by significant mathematical and computational challenges.  Therefore, here we use a simple estimate of the EE. If this short lattice does not contain the center of the full lattice, then we expect that this particular few-body state does not affect the total half-chain EE.  Therefore in order to understand the qualitative properties of the EE, we consider only one few-body state, whose associated short lattice is placed at the center of the full lattice.  This state would have the largest EE among all the few body states, whose associated short lattice intersects with the center of the full lattice.  

We show that the characteristic logarithmic growth of the bipartite EE (computed from a single few-body state with $q_{\dn}+1$ spin-$\dn$ atoms and $q_{\up}$ spin-$\up$ atoms) in the localized case arises from few-body processes [Fig.~\ref{FIG5}(a)] and quantitatively agrees with TEBD calculations, shown in Fig.~\ref{FigS4} in the appendix. While the logarithmic growth is visible already for $q_{\up}=1$, i.e. two particles, the steady-state value is near convergence by $q_{\up}=4$. 

\section{Limitations of the approximate description}
We next consider the conditions under which our approximation ansatz is ineffective, i.e., fails to converge within the accessible range of the parameters $\ell, k_{\dn}$ and $k_{\up}$, suggesting future experimental work to explore many-body dynamics under these conditions.  The parameter $\ell$ represents the dynamical range,  i.e., the spatial extent explored by each atom. $k_{\sigma}+k_{\bar{\sigma}}+1$ can be interpreted as the maximum entanglement depth of the state~\cite{PhysRevLett.107.180502}. Therefore,  the dynamics that break the approximate method would necessarily involve a large dynamical range for each atom and produce a large entanglement depth~\cite{PhysRevLett.100.030504, PhysRevA.89.062110}.  Therefore, one of the technological challenges in developing quantum simulators is the high fidelity creation and control of states with a large entanglement~\cite{Omran_2019}.


The above arguments appear to suggest that the many-body dynamics considered in this paper is the hardest to simulate using our method when the system is fully delocalized.  However, on the contrary, we demonstrate  that the approximate method is effective in simulating local or few body observables when the localization length is small or large,  but ineffective when it is intermediate.  It has been identified  that the vicinity of the phase transition offers many physically interesting effects such as slow dynamics~\cite{PhysRevLett.119.260401,  PhysRevB.92.014208,  PhysRevB.93.060201,PhysRevB.93.134206} and anomalous diffusion~\cite{PhysRevLett.114.160401, PhysRevLett.117.040601}. 

We consider the Aubry-Andr\'{e} model and study the convergence of the density-density correlation $C_{ij}=\langle \hat{n}_i \hat{n}_j\rangle - \langle \hat{n}_i \rangle \langle \hat{n}_j \rangle$ between site $i$ and site $j$. The correlation decays down to zero for large $|i-j|$ when the system is localized and to a non-zero value when the system is delocalized as illustrated in Fig.~\ref{FIG5}(b) for a simple non-interacting Aubry-Andr\'e model.  Based on this observation,  we consider an interacting Aubry-Andr\'e model and study the convergence of the plateau value, $\bar{C}_{ij}$ of the correlation $C_{ij}$ for various disorder strengths.  Although we can compute the correlation $C_{ij}$ for the full many-body state using our approximate method, for the purpose of the study of the convergence it suffices to consider a single few body state.
 Note that if the absolute change in the correlator due to increasing $k_{\sigma}, k_{\bar{\sigma}}$ is small, then the corresponding change in the quantum state will also be small.  As shown in Fig.~\ref{FIG5}(c), the convergence of the quantum state is fast for small and large disorders, and slow in the intermediate regime, i.e., when $3J<\Delta < 4J$.  

In a nutshell, many-body dynamics with small and large number of modes is likely to be approximable either by an effective model or by a thermal ensemble, when we consider local or few body observables. Dynamics with intermediate number of modes, however, constitute the quintessential hard to simulate regime using the approximate method.  While the correlator $C_{ij}$ is not a local observable (i.e., it cannot be extracted from the on-site reduced density matrix), it can still be extracted using the reduced density matrix corresponding to two sites --- it is a few-body observable. If the many-body system is thermal, the reduced density matrix corresponding to two sites $i$ and $j$ lies in the vicinity of the corresponding Gibbs state (as long as the localization length is larger than the distance $i-j$). Therefore, approximation methods can be quite effective in studying local observables and two point correlations~\cite{PhysRevB.91.045138, white_quantum_2018, Wu_2019}.

\section{Conclusions}

To conclude, we have shown how a near-term quantum device can be used to quantitatively solve for the time dynamics of a quantum many-body system, which is otherwise inaccessible.  In particular, we have developed an efficient approximate numerical method for localized Fermi-Hubbard systems and benchmarked it using a quantum simulator. Although our method is built to study time dynamics, based on a Wick rotation~\cite{srednicki_2007} $e^{-i\hat{H}t}\rightarrow e^{-\beta \hat{H}}$ we can adapt it to study thermodynamic properties in the intermediate to low temperature regime. Moreover we can use our method to study the effect of open system dynamics in a localized Fermi-Hubbard system~\cite{HS_fragmentation}, which involves solving Lindblad equations.

One of the tools that has not been used in this method so far is acceleration of convergence, which is effective in extrapolating well-behaved sequences~\cite{PT_CM13, brezinski2013extrapolation}. In the future, it would be interesting to apply this technique to systems where the dynamics does not converge within the accessible range of $k_{\up}$ and $k_{\dn}$, for instance, $2$D Fermi-Hubbard and Bose-Hubbard models, where the number of participating atoms is not strictly bounded by the dynamical range. Moreover, one can combine MPS methods with our approximate method to enhance the efficiency for such systems. Eq.~(\ref{spin_imb}) indicates that the imbalance of $\dn$-atoms is polynomial in $\lambda_{\up}$ and its degree represents the entanglement depth that develops in the dynamics. Therefore, exploring the effect of interactions in relation to the spin-imbalance would reveal the number of atoms participating in the dynamics.  

One of the central technological challenges in neutral atom quantum simulators is to benchmark and optimize the experimentally accessible coherence length of the dynamics. Our approximate method provides a quantitative estimate of the range  (i.e., the value of $\ell$ at convergence) and the number of atoms (i.e., the value of $k_{\up}$ and $k_{\dn}$ at convergence) contributing to the observed dynamics.  We can therefore use our method to set a lower bound on the coherence length accessible in the experimental system. Moreover, using our method we can identify dynamical features that require a certain coherence range to be observed, which can then be used to optimize the experimental system.



\paragraph*{\textbf{Acknowledgments}}
We thank D.~Abanin,  E.~Mueller,  P.~Sala and N. ~Y. ~Yao for illuminating discussions.  We thank F.~Pollmann  for insightful discussions regarding the relation to MPS. This work was supported by Deutsche Forschungsgemeinschaft (DFG, German Research Foundation) under Germany's Excellence Strategy -- EXC- 2111 -- 39081486. The work at LMU was additionally supported by DIP and B.~H.~M acknowledges support from the European Union (Marie Curie, Pasquans).

\paragraph*{\textbf{Data availability}}
The data used in this work is  available from the corresponding author upon a reasonable request. 

\paragraph*{\textbf{Code availability}}
The code corresponding to the approximate method presented in this work is available at \url{https://gitlab.physik.uni-muenchen.de/LDAP_ag-bec-fermi1/approximate-method-for-1d-fermi-hubbard-model}

\paragraph*{\textbf{Competing interests}} The authors declare no competing interests.

\appendix

\section{Analysing the limitations of ED and TEBD}\label{other_numerical_methods}

In this section, we discuss the limitations of two of the most common numerical methods, --- exact diagonalization (ED) and time-evolved block decimation (TEBD), when applied to the problem of computing the time evolution of our system.  

The limitations of ED fundamentally come from the exponentially growing size of the Hilbert space.  On our desktop computer with a RAM of $32$ GB, we are at-best able to simulate a system with $L=21$ sites and $N_{\up} = N_{\dn} =5$ atoms, without using a Lanczos algorithm (the code is available at \url{https://gitlab.physik.uni-muenchen.de/LDAP_ag-bec-fermi1/exact-diagonalization-for-fermi-hubbard-model}). The corresponding  Hilbert space dimension is $\sim 4 \times 10^8$. See appendix~\ref{code_description} for the algorithm used.  Given the exponential growth of the Hilbert space dimension 
with $N_{\up}$ and $N_{\dn}$, hardware improvements and methods such as Lanczos algorithm~\cite{golub2013matrix} will only marginally enhance the accessible values of $L$. 

\begin{figure}[h!]
\includegraphics[scale=1]{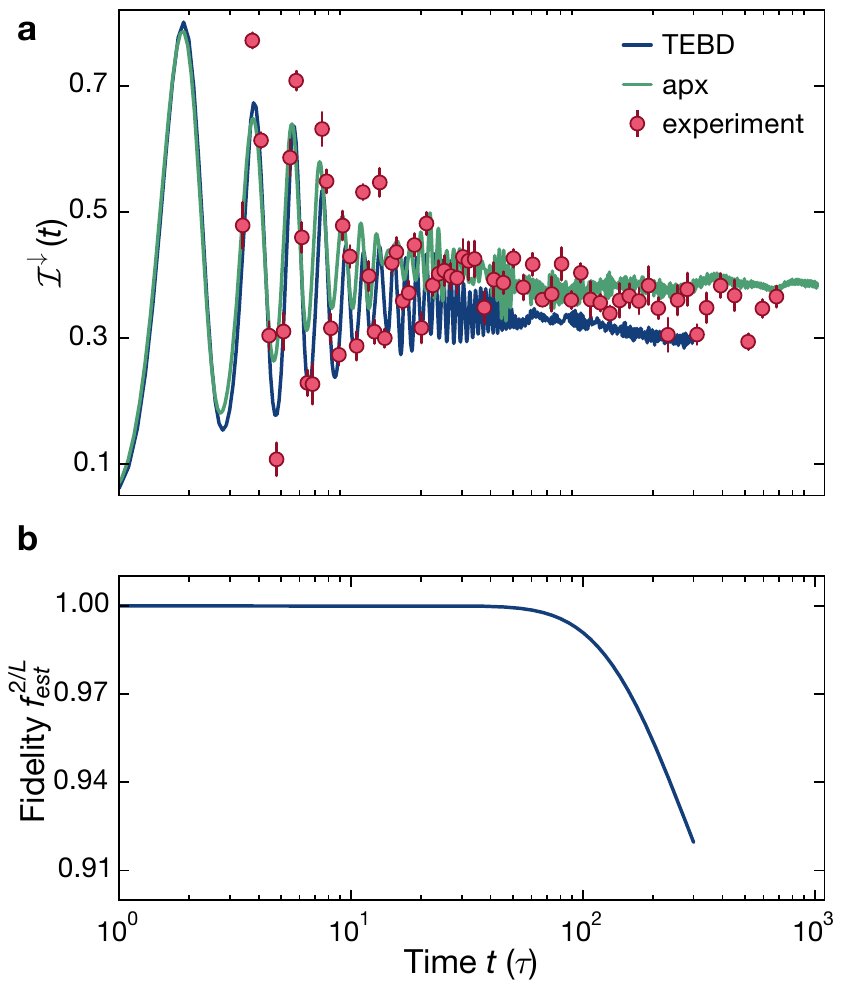}
\caption{\textbf{TEBD computation of the imbalance time trace:}\textbf{ a }the imbalance time trace for $\Delta_{\dn} = 3J$ and $U=5J$. The red circles represent the experimental data and the errorbars represent the standard deviation extracted from $12$ samples. This data is the same as the one presented in Fig.~$2$(d) of the main text.  The green solid curve represents a computation using our approximate method with $L_{\text{apx}}=280, \ell =6, k_{\up}=5$ and $k_{\dn}=0$.  This computation took about $7$ hours on our desktop computer. The blue solid curve represents a TEBD calculation of the same time trace with $L_{\text{tebd}}=100$ and $\chi = 500$, done on TeNPy~\cite{Hauschild_2018}.  \textbf{b} Estimate for the local fidelity in the above TEBD calculation, $f_{\text{est}}^{\frac{2}{L_{\text{tebd}}}}$ (see text), where $f_{\text{est}}$ is a lower bound for the global fidelity. }\label{FIGS1}
\end{figure}

Unlike ED, it is not straightforward to estimate the limitations of TEBD for a specific problem. In TEBD, the precision of computation is traded out for system size. One can in principle compute the time dynamics for a larger system, while incurring a higher error.  Moreover the corresponding error cannot be estimated accurately. One can obtain a lower bound for the fidelity of the full quantum state using the discarded weights in each step in TEBD.  However, the global fidelity can be low for several reasons and therefore it does not reliably capture the error in local observables like the imbalance, that we are actually interested in.  The error in a local observable is better captured by the \textit{local fidelity}. If $|\psi_1\rangle$ and $|\psi_2\rangle$ are two many-body states, their local fidelity corresponding to a given site is defined using the Uhlmann fidelity~\cite{Uhlmann_1976,  Uhlmann_2011, PhysRevA.98.042316} between the reduced density matrices $\rho_1, \rho_2$ corresponding to that site,
\begin{equation}
F(\rho_1, \rho_2) = \left|\text{Tr}\left(\sqrt{\sqrt{\rho_1}\rho_2 \sqrt{\rho_1}}\right)\right|^2.
\end{equation} 
Here, the square-root of a positive-semidefinite hermitian matrix $\rho$ is defined as the unique hermitian matrix whose eigenvalues are the positive square-roots of the eigenvalues of $\rho$ and whose eigenvectors are the same as that of $\rho$.  While the local fidelity is bounded from below by the global fidelity,  nothing more can be said of the relation between the two. For instance, while the many-body states $|\psi_1\rangle = \frac{1}{\sqrt{2}}(\ket{\up}^{\otimes N}+ \ket{\dn}^{\otimes N})$ and $|\psi_2\rangle = \frac{1}{\sqrt{2}}(\ket{\up}^{\otimes N}- \ket{\dn}^{\otimes N})$ have zero global fidelity, they agree on all local observables -- their local fidelity is $1$. As a second example, consider  $|\psi_1\rangle =\ket{\up}^{\otimes N}$ and $|\psi_2\rangle =(\sqrt{1-\epsilon}\ket{\up}+ \sqrt{\epsilon}\ket{\dn})^{\otimes N}$. The global fidelity is $f=(1-\epsilon)^N$ but the \textit{local} fidelity is $f_{loc}=(1-\epsilon)$.  As a third example, consider $|\psi_1\rangle = \ket{\up}^{\otimes N}$ and $|\psi_2\rangle =(\sqrt{1-\epsilon}\ket{\up}^{\otimes N}+ \sqrt{\epsilon} \ket{\dn}^{\otimes N})$. Both the global fidelity and the local fidelity are $(1-\epsilon)$. 

An interesting problem, therefore, is to find an estimate for the local fidelity in a time evolution under TEBD. While we leave a comprehensive investigation of this problem for a future work, here we make a heuristic argument to suggest that for a localized system with a localization length of $\xi$,  $f^{\frac{2\xi}{L}}$, where $f$ is the global fidelity, is a likely estimate for the local fidelity in a TEBD computation. We assume that, if the system remains localized with a localization length of $\xi$, a truncation at the bond between sites $i$ and $i+1$ will affect the local fidelity at site $j$ only when $|i-j|\leq \xi$. Based on this assumption we use $f^{\frac{2\xi}{L}}$ as an estimate for the local fidelity. In Fig.~\ref{FIGS1} we show that the imbalance time trace computed using TEBD deviates from the experimental data roughly around the same time when the estimate $f_{\text{est}}^{\frac{2}{L}}$ deviates from $1$ (we set $\xi=1$ because the single particle localization length for $\Delta_{\dn} = 3J$ is about one site. $\xi$ is expected to be slightly higher due to interactions, but this will only lower $f_{\text{est}}^{\frac{2}{L}}$ ).  Here $f_{\text{est}}$ is the lower bound on the global fidelity.  We therefore use this quantity in Fig.~$1$ of the main text.  See refs.~\cite{PhysRevLett.100.080601, zhou2007renormalization} for some related ideas. 

The time trace in Fig.~\ref{FIGS1}, going up to $300 \tau$ was computed using a TEBD with a bond dimension $\chi = 500, L_{\text{tebd}}=100$ and it took $269$ hours on our desktop computer. We note that while it maybe possible to compute a time trace for $L_{\text{exp}}=290$ (the experimental system size) and improve the precision to longer time by choosing a larger bond dimension, it will take an inconvenient amount of computational time.  The number of pure CDW states is $\binom{L/2}{L/4}$, exponential in the system size.  Averaging over all of these states, therefore, would not be possible in a TEBD computation.  However, this average can be approximated using the following technique.  We map the four possible states of a single site to four states of chain of $4-$level systems: $\ket{\text{vac}}\rightarrow \ket{0}$, $\ket{\up}\rightarrow \ket{1}$, $\ket{\dn}\rightarrow \ket{2}$ and a doublon, $\ket{\up\dn}\rightarrow \ket{3}$. We then construct the product state
\begin{equation}
\begin{split}
\ket{\psi} =& \frac{1}{2^{L_{\text{tebd}}/4}}\left(\ket{1}+e^{i\phi_1}\ket{2}\right)\otimes \ket{0}\otimes\left(\ket{1}+e^{i\phi_2}\ket{2}\right) \\
 & \otimes \ket{0}\otimes\cdots \otimes \left(\ket{1}+e^{i\phi_{L_{\text{tebd}}/2}}\ket{2}\right)\otimes \ket{0}
\end{split}
\end{equation}
Here, the phases $\phi_1,  \phi_2, \cdots$ are randomly chosen. For large system sizes,  a few avarages of such states quickly converges to a mixed CDW. 

We describe how this issue of averaging over all pure CDW states is handled  in our approximate method in appendix~\ref{imbalace_computation}.

\section{Reformulating the approximate method using MPS}\label{MPS_reformulation}
In this section, we provide an intuitive explanation of how our approximate method can be linked to TEBD. The imbalance can be expressed as a linear sum of local observables. In fact, if $\hat{n}_{i, \dn}$ is the on-site number operator for spin-$\dn$,  the imbalance can be written as:
\begin{equation}
\mathcal I^{\dn} = \frac{1}{N_{\dn}}\sum_i (-1)^i \hat{n}_{i, \dn}.
\end{equation} 
Therefore, the error in the computation of the imbalance using TEBD will depend only on the error in the local observable $\hat{n}_{i, \dn}$.  The error in $\langle \hat{n}_{i, \dn}(t)\rangle $ would depend only on the local fidelity at site $i$.  The standard TEBD is optimized on the global fidelity rather than a local fidelity.  That is, the bond dimension is chosen so as to minimize the loss in the global fidelity, while paying no special attention to any particular local fidelity.  We can modify the standard TEBD method so that it can be optimized on specific local fidelities.  For reasons that will be clear soon, let us assume that we use independent TEBD computations to evaluate $\langle \hat{n}_{i, \dn}(t)\rangle $ for each $i$. Following the arguments in the previous section, truncations of bonds far away from site $i$ will not contribute to the error in $\langle \hat{n}_{i, \dn}(t)\rangle $. Therefore, to optimize the computational performance, we may choose a spatially varying bond dimension, that takes a high value around site $i$ and gets lower as we move away from this site~\cite{white_quantum_2018}.  In order to implement such a TEBD scheme, we will have to compute each $\langle \hat{n}_{i, \dn}(t)\rangle $ independently. This will add  $N_{\dn}=O(L)$ overhead to the computation which can be easily offset by the significant advantage that we obtain by choosing  the bond dimension $\chi$  to be spatially varying.  In particular, if we choose $\chi = 4^{\ell}$ between sites $i-\ell$ and $i+\ell$ and $\chi=1$ for the rest of the bonds,  it reduces to our approximate method with $\kappa_{\dn}=\kappa_{\up}=2\ell$.

The key feature of our approximate method, however is that $\kappa_{\up}$ and $\kappa_{\dn}$ can be chosen to be different from $\ell$. This feature does not appear naturally in the above described TEBD scheme.  These considerations suggest an alternate formulation of the approximate method.  By replacing approximation-$3$ with an MPS ansatz, we trade the parameter $\ell$ for the bond dimension.  In this alternative formulation, we compute the time evolved few-body state $\ket{\psi^{\prime\prime}(t)}$ on the full lattice of size $L$ using TEBD, with an appropriately chosen spatially varying bond dimension. 

\begin{figure}
\includegraphics[scale=1]{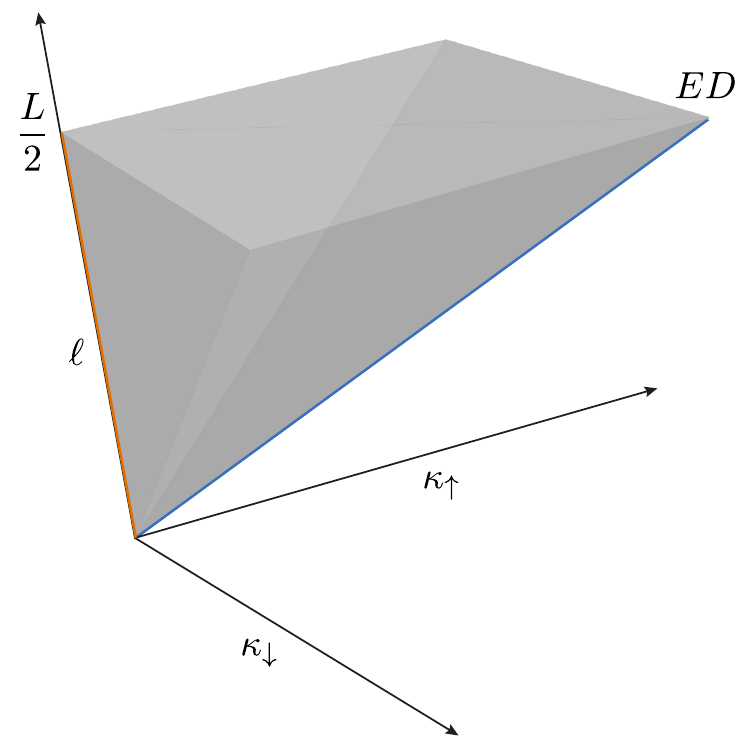}
\caption{\textbf{Our approximate method and cluster expansion:} The relevant $3$D space spanning the parameters $\ell, k_{\up}, k_{\dn}$.  The blue solid line corresponds to $2\ell = \kappa_{\up}=\kappa_{\dn}$, where the approximate method reduces to the standard cluster expansion.  At $2\ell = \kappa_{\up}=\kappa_{\dn}=L$, it reduces to exact diagonalization (ED), for periodic boundary conditions on the full lattice.  In the non-interacting case, the approximate method reduces to the exact calculation when $\ell =L/2, \kappa_\up = \kappa_\dn =0$. The orange line indicates the points with $\kappa_\up = \kappa_\dn=0$,  that can be used to approximate the non-interacting dynamics (see Fig.~\ref{FigR1} for a convergence analysis for this case).  }\label{FIGS2}
\end{figure}

\begin{figure*}
\includegraphics[scale=1]{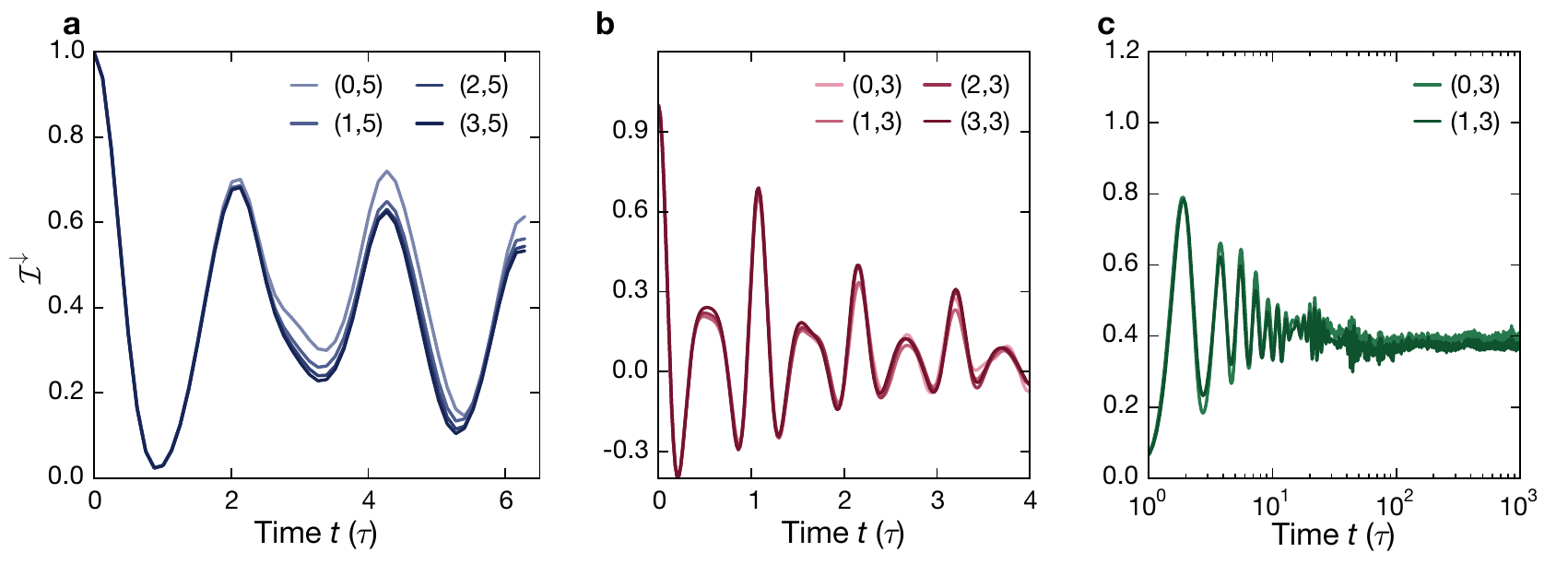}
\caption{\textbf{The effect of $k_{\dn}$.} \textbf{a} Imbalance time trace with a N\'eel type initial state computed using the approximate method for various values of $(k_{\dn}, k_{\up})$.   We can see that the effect of increasing $k_{\up}$ on the time trace decreases progressively.  The parameters used are $\Delta =3J$ and $U=5J$. \textbf{b}  Similar computation for an incoherent CDW initial state with the parameters used in Fig.~$2$(a) of the main text.  Note that the effect of $k_{\dn}$ is weaker.  \textbf{c} similar computation of a long time trace, for the parameters used in Fig.~$2$(d) of the main text.  }\label{FIGS5}
\end{figure*}

\section{Contrasting our work with cluster expansion}\label{cluster_comparison}
In this section, we discuss how our approximate method is related to the well-known cluster expansion method~\cite{PT_CM13, RRP_singh_high_order} for higher dimensional systems.  While to our best knowledge there are no non-trivial cluster expansions for a $1$D system with only nearest neighbour hopping~\cite{PhysRevLett.97.187202, PhysRevE.95.033302, PhysRevLett.113.195302},  it is possible to construct non-trivial clusters in the case of spinful fermions. We show below that in comparison to this approach, our approximate method allows for a better optimization of the computational performance.

The cluster expansion method is suitable for $2$D and $3$D lattice systems. It comes from the observation that in the expansion for $e^{-\beta H} = 1 - \beta H  + \frac{1}{2}\beta^2 H^2 + \cdots$, the $r$-th term (i.e., $(-1)^r\beta^r H^r/r!$) corresponds to paths in the lattice with $r$ steps.  In $2$D and $3$D the paths with $r$ steps form a non-trivial collection of subsets of the lattice.  One can use it to construct a nested collection of subsets of the lattice and  build a cluster expansion.

Although our system is a $1$D lattice we can make use of the internal states of the atoms to map it to a ladder-like system. This ladder, to some extent, allows for a non-trivial cluster expansion. While this idea has not been explored so far,  a related idea has been studied for the case of a $1$D system with next-nearest neighbour hopping~\cite{PhysRevE.95.033302}.   

We may represent the parameters $\ell, \kappa_{\up}, \kappa_{\dn}$ of our approximate method in a $3$D space (Fig.~\ref{FIGS2}). The direct (trivial) application of a cluster expansion to our system  is represented by the line $2\ell = \kappa_{\up}=\kappa_{\dn}$.  The line $\kappa_\up=\kappa_\dn=0$ represents parameters that can be used to approximate non-interacting dynamics. See Fig.~\ref{FigR1} for a convergence analysis on this line.  As we mentioned before, the key feature of our approximate method is that the atom numbers ($\kappa_{\up}, \kappa_{\dn}$) can be varied independent of $\ell$. This not only allows for a better optimization of computational performance, but also reveals a physically meaningful information, i.e., the entanglement depth in the system.  

\begin{table}
\begin{tabular}{c|c}
Number of even sites ($k_{\sigma}$)& Shell size ($\kappa_{\sigma}$)\\
\hline
$1$ & $3$\\
$2$ & $4$\\
$3$ & $7$\\
$4$ & $8$\\
$5$ & $11$\\
\end{tabular}\caption{A few values of $k_{\sigma}$ with the corresponding values of $\kappa_\sigma$. 
}\label{T1}
\end{table}

\section{The role of $k_{\dn}$}\label{role_of_kdn}
One of the parameters whose effect we have not explored in detail in the main text is $k_{\dn}$. In this section we briefly discuss this parameter.  As we mentioned before, in the $\kappa_{\up}-\kappa_{\dn}$ parameter space for a fixed $\ell$, the point $(\kappa_{\up}, \kappa_{\dn})=(2\ell, 2\ell)$ is where the error in the computation is minimized.  Depending on the nature of the dynamics and the observable of interest, there may be an optimal trajectory towards this point in the  $\kappa_{\up}-\kappa_{\dn}$ that maximizes the convergence rate. We note that in general, finding this optimal trajectory can be very complicated. We therefore restrict our discussion to the data shown in Fig.~$2$ of the main text.  As before, we work with $k_{\dn}$, i.e., the number of even sites inside the $\kappa_\dn$-shell, as this parameter is more convenient and physically meaningful.  See Table~\ref{T1} for a list of $k_\sigma$ values and the corresponding $\kappa_\sigma$ values. In Fig.~\ref{FIGS5}(b,c), we demonstrate that the impact of having a higher $k_{\dn}$ is not significant on the scale of the experimental errorbars for the parameters from Fig.~$2$ of the main text.  We attribute this to the spin polarized islands that may be present in an incoherent CDW initial state. Note that in a polarized spin-$\dn$ system the other spin-$\dn$ atoms are irrelevant for the dynamics of a given spin-$\dn$ atom.  Indeed, if we replace the initial incoherent CDW with a N\'eel type CDW initial state (i.e., $\circ\up\circ\dn\circ\up\circ\dn\circ \cdots $), which prevents spin-polarized islands, the effect of $k_{\dn}$ is more significant (Fig.~\ref{FIGS5}(a)).  

\begin{figure}
\includegraphics[scale=1]{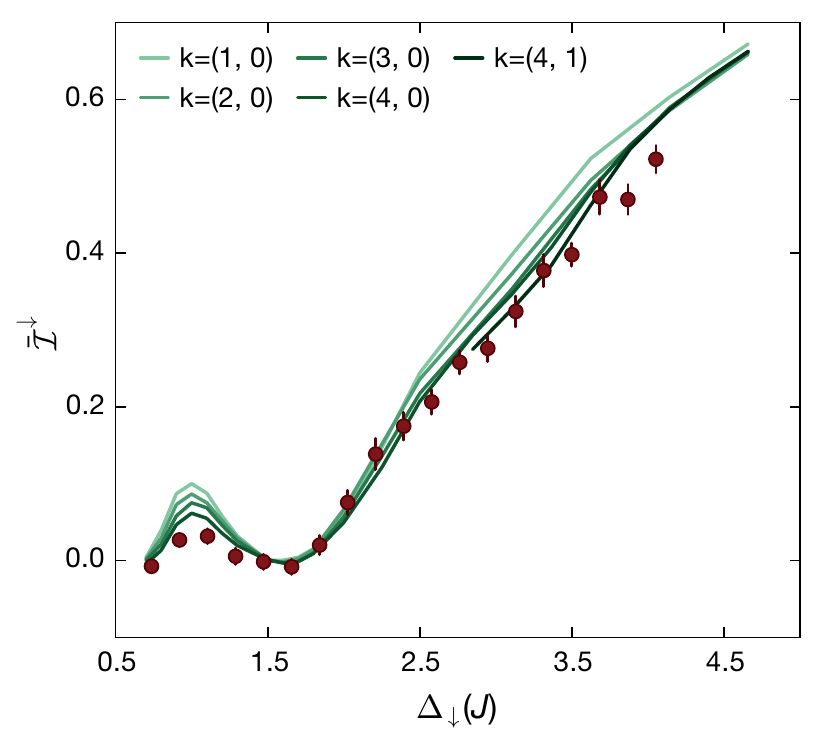}
\caption{\textbf{Higher tilt values:} We show the data and our calculations extending those shown in Fig. $2$c of the main text to higher values of $\Delta_\dn$.  The legend indicates the values of $k=(k_{\up}, k_{\dn})$. $\ell=7$ for all the curves. The system size used is $L_{apx}=100$. }\label{FigR2}
\end{figure}

\begin{figure}
\includegraphics[scale=1]{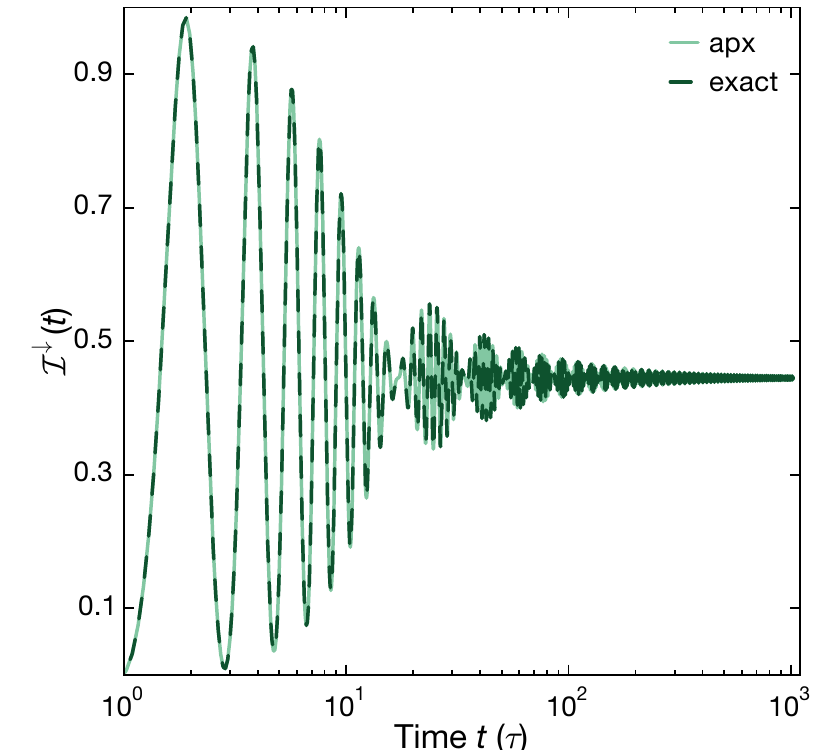}
\caption{\textbf{Non-interacting time traces: }A long time trace, up to $1000 \tau$ with $\Delta =3.3J$ computed using $L_{apx}=280, \ell=7$, $k_\up =k_\dn =0$, and compared with the exact calculation.  The exact calculation is equivalent to using $\ell = 140$.  }\label{FigR1}
\end{figure}

In Fig.~\ref{FigR2}, we show the imbalance steady state value, averaged between $300\tau$ and $330\tau$ as in Fig.~$2$c of the main text, but for a larger range of $\Delta_\dn$.  Betwen $\Delta_\dn=2.5 J$ and $\Delta_\dn = 3.5J$, one can see a weak effect of $k_\dn$, as the experiment agrees better with the computation corresponding to $\ell =7, k_\up =4$ and $k_\dn=1$.  This also shows a  feature around $\Delta =3J$,  corresponding to a $U=2\Delta$ resonance, explored in detail in Ref.~\cite{HS_fragmentation}.

\section{Experimental details}
The quantum simulator, i.e., the experimental system consists of a $3$D lattice formed by three laser beams, one of which along the $x$-axis, has a wavelength of $\SI{532}{nm}$ (this is the primary lattice) and the other two along the $y$ and $z$-axes have a wavelength of $\SI{738}{nm}$ each.  The latter two lattices are deep, preventing any hopping in the $y$ or the $z$-directions within the timescale of the experiment.  If $J$ is the hopping rate along the primary axis the hopping along the orthogonal axes is about $\sim 3\times  10^{-4} J$. Therefore, our system can be considered as a set of independent $1$D tubes with a lattice spacing of $\SI{266}{nm}$ and length $L\approx 290 \pm 10$ sites~\cite{scherg_observing_2020}. We have an additional lattice in the $x$-axis created by a $\SI{1064}{nm}$ laser to form a bichromatic superlattice, which we use for the preparation of a charge density wave and measurement of the imbalance (see appendix~\ref{initial_state} and ~\ref{measurement}). We load the lattice with about $\sim 50(5)\times 10^3$ $^{40}$K atoms at a temperature of $=  0.15(1) T_F$, where $T_F$ is the Fermi temperature. The internal states $|F=9/2; m_F = -9/2\rangle $ and $|F=9/2; m_F=-7/2\rangle$ of $^{40}$K are used as the spin-$\dn$ and spin-$\up$ states respectively. We use the magnetic Feshbach resonance at $\SI{202.1}{G}$ between these two states to control the Hubbard  interaction in the lattice. 

We apply two classes of on-site potentials; a quasi-random potential, $V_{i, \sigma}= \Delta \cos (2\pi \beta i+ \phi) +\alpha(i-L/2)^2$, i.e., the Aubry-Andr\'e model and a linear potential $V_{i, \sigma}= i \Delta_{\sigma} +\alpha(i-L/2)^2$ i.e., the Stark model. The Aubry-Andr\'e model is realized by an incommensurate lattice along the $x$-axis, with a wavelength of $\SI{738}{nm}$, which introduces a quasi periodic on-site potential.   The Stark model is realized by a magnetic field gradient along the $x$-axis, produced by a single current carrying coil. The weak quadratic term, $\alpha(i-L/2)^2$ stems from an additional harmonic confinement induced by the optical dipole potentials.  Typical values are $\alpha \approx h\times \SI{216}{\milli \hertz}$. In the tight-binding limit, the Hamiltonian of this system is given by Eq.~$(1)$ of the main text. The nearest neighbour hopping $J$ is controlled by the depth of the primary lattice. It is typically between $h\times\SI{200}{\hertz}$ and $h\times\SI{ 500}{\hertz}$.

\subsection{Initial state preparation}\label{initial_state}

To prepare the initial state, we load the atoms repulsively, at a scattering length of $a=100 a_{0}$ into a three-dimensional (3D) optical lattice~\cite{scherg_observing_2020}.  Loading the atoms repulsively suppresses the formation of doublons to a large extent and we are left with $\lessim  15\%$ doubly occupied sites. Moreover, we eliminate any residual doublons by applying a $\SI{100}{\micro \second}$ near-resonant light pulse right after loading the deep lattice~\cite{scherg_nonequilibrium_2018}, which causes light assisted collisions, removing the  doublons without affecting the singlons. With deep orthogonal lattices, we can consider the system as a collection of $1$D tubes.  We characterize the $4 \sigma$ width of the central tubes to $L_\text{exp}=290$ sites, using a Gaussian fit to an in-situ image of the atoms.  Along the two orthogonal axes, we estimate about $150$ sites and $22$ sites respectively.

In order to obtain a CDW pattern in our initial state,  we make use of an adiabatic ramp of the phase between the lasers forming the short ($\lambda_s = \SI{532}{\nano \meter}$) and the long ($\lambda_l = \SI{1064}{\nano \meter}$) lattices along the $x$-axis,  also known as the superlattice phase. The atoms are loaded into the long lattice. We then ramp up the power of the short lattice with a superlattice phase of $\phi=0.44\pi$ in about $\SI{200}{\micro\second}$.  A symmetric double-well potential is realized for $\phi=k\cdot\pi$, for integer $k$. The chosen phase creates strongly tilted double wells with the atom that was previously loaded into the long lattice located on the low-energy site of each double well (even site), while the high energy site (odd site) is empty.  We then ramp down the power of the long lattice, which remains switched off during the time evolution.

\begin{figure}
\includegraphics[scale=1]{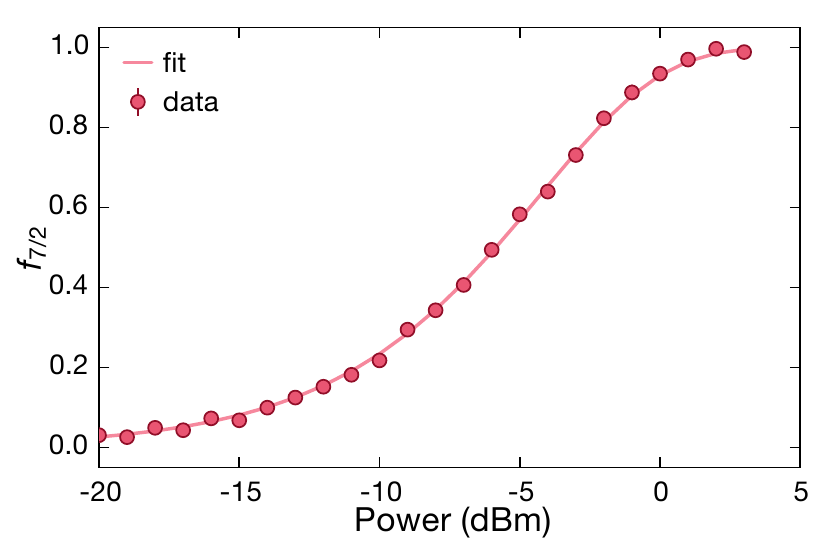}
\caption{\textbf{Preparing a spin-imbalanced CDW:} The proportion of the atoms in the state $|F=9/2; m_F=-7/2\rangle$ after the RF sweep (see text) as a function of the power of the RF signal.  The markers represent the experimental data and the solid line represents a fit with the Landau-Zener equation (see text). }\label{FIGS3}
\end{figure}

In order to prepare a spin-imbalanced charge-density wave, we make use of an adiabatic sweep of an RF signal coupling the states $|F=9/2; m_F = -9/2\rangle $ and $|F=9/2; m_F=-7/2\rangle$. The atoms are initially cooled in the $|F=9/2; m_F = -9/2\rangle $ state and before loading into the lattice,  we induce an RF sweep that produces the desired incoherent mixture of the two internal states. We control the weights in the mixture using the power of the RF signal.  Such a sweep of the RF frequency transports the system  
across an avoided level crossing, where the minimal gap is controlled by the power of the RF signal. Therefore, the probability of finding the atoms in the $|F=9/2; m_F= -7/2\rangle$ after the sweep is given by the Landau-Zener equation,  $f_{7/2} = e^{-x/a}$, where $x$ is the power of the RF signal and $f_{7/2}$ is the fraction of the atoms in the $|F=9/2; m_F=-7/2\rangle$ state (Fig.~\ref{FIGS3}).  We obtain a fit value of $a=\SI{0.44}{mW}$.

The adiabatic RF sweep is applied before loading the atoms into the lattice.  After loading, the state of a single $1$D tube can be described as

\begin{equation}
\begin{split}
|\psi \rangle =& |0\rangle \otimes \left(\sqrt{f_{7/2}}\ket{\up} +\sqrt{1-f_{7/2}} e^{i\phi_1}\ket{\dn}\right)\otimes |0\rangle \\
&\otimes \left(\sqrt{f_{7/2}}\ket{\up} +\sqrt{1-f_{7/2}} e^{i\phi_2}\ket{\dn}\right) \otimes \cdots \\
& \otimes  \left(\sqrt{f_{7/2}}\ket{\up} +\sqrt{1-f_{7/2}} e^{i\phi_{L_{exp}/2}}\ket{\dn}\right)
\end{split}
\end{equation}

Note that a global phase factor has been dropped and $|0\rangle $ represents an empty site.  The phases $ \phi_i$ are acquired during the sweep, depending on the local magnetic field. Therefore they phases vary along the lattice, across the $250$ tubes and between the experimental realizations and can be considered as independent random values.  Accounting for this averaging, the initial state is modelled as the following mixed state:

\begin{equation}
\begin{split}
\rho = &\ket{0}\bra{0}\otimes \left(f_{7/2} \ket{\up} \bra{\up}+ (1-f_{7/2})\ket{\dn}\bra{\dn} \right)\\
&\otimes  \ket{0}\bra{0} \otimes \cdots \otimes \left(f_{7/2} \ket{\up} \bra{\up}+ (1-f_{7/2})\ket{\dn}\bra{\dn} \right)
\end{split}
\end{equation}

This is a mixture of all incoherent CDWs with various spin imbalance, weighted by a binomial distribution, i.e.,  a CDW with $N_{\up}$ spin $\up$ atoms appears with a probability $\binom{L_{\text{exp}}/2}{N_{\up}}p^{N_{\up}}(1-f_{7/2})^{N_{\dn}}$ where $N_{\dn}=L_{\text{exp}}/2-N_{\up}$ .  With $L_{\text{exp}}/2 \approx 145 $, this distribution has a very sharp peak at $N_{\up}=f_{7/2}L_{\text{exp}}/2$. Thus, the mixed state is very well modelled by an incoherent mixed state of all spin configurations with  $N_{\up}=f_{7/2}L_{\text{exp}}/2$ and $N_{\dn}=(1-f_{7/2})L_{\text{exp}}/2$.

\subsection{Calibration}
We use the non-interacting dynamics to calibrate the parameters $J, \Delta_{\sigma}$ and $\alpha$ in Eq.~$(1)$ of the main text. In the Stark model the imbalance $\mathcal{I}^{\sigma}$ oscillates at frequency $\Delta_{\sigma}$ (see Ref. ~\cite{scherg_observing_2020} for details). In the non-interacting case (i.e., when $U=0$), this dynamics can be described analytically~\cite{hartmann_dynamics_2004} using Bessel functions. For an atom starting on the $i$-th site at $t=0$, the occupation probability at site $j$ is $\left|\mathcal{J}_{i-j}\left(\frac{4J}{\Delta_{\sigma}}\sin(\pi \Delta_{\sigma}t/h) \right)\right|^2$where $\mathcal J_{\nu}$ is the $\nu$-th order Bessel function of the first kind. The confinement $\alpha$ introduces a damping of the Bloch oscillations. In fact, the Bloch oscillations develop a beat-note envelope with a frequency $L\alpha$. This is the result of the effective tilt varying from $\Delta_{\sigma}-L\alpha/2$ to $\Delta_{\sigma}+L\alpha/2$, through the lattice. In our experiment, $L\alpha \approx h\times \SI{60}{\hertz}$. We choose the tilt $\Delta_{\sigma}$ between $\SI{0.5}{\kilo \hertz}$ and $\SI{2.0}{\kilo \hertz}$. We calibrate these parameters using the imbalance time trace in the non-interacting case (Fig.~$2$a of the main text).

\subsection{Measurement}\label{measurement}
In order to measure the spin resolved imbalance,  we apply a band transfer technique in the superlattice~\cite{sebby-strabley_lattice_2006, foelling_direct_2007} along with a Stern-Gerlach gradient.  In the band transfer technique, we map the atoms on odd sites into the third band of the long lattice,  and atoms on even sites remain in the first band.  In order to accomplish this, we use a superlattice phase of  $\phi=0.15\pi$,  while ramping up the long lattice and ramping down the short lattice.  Finally, we perform bandmapping and Stern-Gerlach resolved absorption imaging to evaluate the spin-resolved imbalance, at a finite time of flight. We use the same coil to produce the Stern-Gerlach gradient and the magnetic field gradient during time evolution.

The spatial separation between $|F=9/2; m_F = -9/2\rangle $ and $|F=9/2; m_F= -7/2\rangle$ achieved during the Stern-Gerlach separation is not large enough to obtain absorption images where we can distinguish between the two spin states. Therefore, prior to the band transfer, we apply a Landau-Zener sweep to convert atoms from $|F=9/2; m_F = -7/2\rangle $ to $|F=9/2; m_F=-5/2\rangle$. The ambient magnetic field during this sweep is $231.6\,\mathrm{G}$, corresponding to the zero crossing of the Feshbach resonance between the two states  $|F=9/2; m_F = -9/2\rangle $ and $|F=9/2; m_F=-5/2\rangle$, centered around $224.2\,\mathrm{G}$. We perform a linear frequency ramp with a duration of $\SI{10}{\milli \second}$ centered at $\SI{51.87}{\mega \hertz}$ with a deviation of $\SI{1}{\mega\hertz}$. Switching off interactions between these two states ensures the absence of interband oscillations after the transfer to the third band. 

To calibrate out the systematic imperfections in the detection sequence,  we take two different sets of images. The first set is a measurement with no evolution time.  Ideally this measurement should give an imbalance of $1$. Due to systematics, we obtain normalized populations $n_{e, 1}^{\sigma}$ and $n_{o, 1}^{\sigma}$ in the even and odd sites and the raw imbalance  is around $0.92(2)$. The second set measures the imbalance after $\SI{25}{\milli \second}$ evolution time with zero tilt, which should ideally correspond to a zero imbalance.  We obtain populations  $n_{e, 2}^{\sigma}$ and $n_{o, 2}^{\sigma}$ in even and odd sites in this measurement. We then construct a matrix that maps the measured populations for these two sets to the ideal value.  That is, we determine the  $2\times 2$-matrix $A^\sigma$, for each state $\sigma=\uparrow, \downarrow$, such that 

\begin{equation}
\begin{split}
A^{\sigma} \left(\begin{array}{c}
n^{\sigma}_{e, 1}\\
n^{\sigma}_{o, 1}\\
\end{array}\right) &=  \left(\begin{array}{c}
1\\
0\\
\end{array}\right) \text{ and }\\
A^{\sigma} \left(\begin{array}{c}
n^{\sigma}_{e, 2}\\
n^{\sigma}_{o, 2}\\
\end{array}\right) &=  \left(\begin{array}{c}
0.5\\
0.5\\
\end{array}\right) 
\end{split}
\end{equation}
This matrix is used to rescale the measured populations for each spin component.

\section{Details of computations and measurements}\label{expt_details}
In this section, we provide the details of the measurements and computations shown in the main text. The following sub sections refer to figures in the main text. 
\subsection{Main text Figure 1}
In Fig.~$1$(b) of the main text, the dashed line represents the value of $t$ at which the estimate of the local fidelity,  i.e.,  $f_{\text{est}}^{\frac{2\xi}{L}}$ with $\xi=1$ dips below $0.9$ for typical bond dimensions. Here, $f_{\text{est}}$ is the fidelity estimated based on the truncations in TEBD.  See appendix~\ref{other_numerical_methods} for details of why this expression is used.  The typical value of $\Delta_{\dn}$ for the data used in the paper is $3J$ and $\xi \sim 1$ for this value.  We used a N\'eel-type initial state and a Stark Hamiltonian with $\Delta = 3J, U=5J$ for this calculation, since this represents the typical parameter regime studied in the rest of the paper. In this plot we use the local fidelity estimated for $\chi = 500$ and $L=100$.  The violet shades represent the dimension of the Hilbert space for $N_{\up} = N_{\dn} = L/4$. 
\subsection{Main text Figure 2}
In Fig.~$2$(a) of the main text, the blue curve represents the least square fit for data corresponding to $U =0$. The fit values were $\Delta_{\dn} =\SI{0.928(2)}{k\hertz}$, $J=\SI{0.77(1)}{k\hertz}$ and the revival time $T_r=\SI{14.510(6)}{ms}$ (the confinement $\alpha$ is related to the revival time as $\alpha=\frac{1}{2 L T_r}$).  The dataset included $101$ points in time evenly distributed between $t=0$ and $t=\SI{10}{ms}$. Each data point was averaged $4$ times and the errorbars represent the standard deviation. These values of the parameters were used for the computation of a time trace with $U=J$, also shown in Fig.~$2$(a). The initial state was an incoherent CDW (see appendix~\ref{imbalace_computation} for details on how the mixed CDW state is constructed).  The system size used in the computation was $L_{\text{apx}}=280$.  The parameters used were $\ell = 7, k_{\up}=3$ and $k_{\dn}=0$.  

In Fig.~$2$(b), the fit parameters obtained using a non-interacting dataset were $J=\SI{1.502(2)}{k\hertz}$, $\Delta_{\dn}=\SI{1.260(5)}{k\hertz}$ and $T_r =\SI{8.41(2)}{ms}$.  Both the interacting and the non-interacting datasets consists of $n=81$ points in time sampled uniformly between $t=0$ and $t=\SI{4}{ms}$, averaged $4$ times at each point.  The interaction strength was $U=3J$.  In the computation using our approximate method, we use $L_{\text{apx}}=280$ and $k_{\dn}=0$ for various values of $\ell$ and $k_{\up}$.   The RMS deviation is computed in its discrete form:
\begin{equation}
\text{RMS}=\frac{1}{\sqrt{n}}\sqrt{\sum_{i=1}^n |\mathcal I_{\text{exp}}(t_i) - \mathcal I_{\text{apx}}(t_i)|^2}
\end{equation}
Here, $t_i$ are the points in time at which the data is taken and $n=81$.  The errorbars on the RMS is defined as 
\begin{equation}
\Delta(\text{RMS}) = \frac{1}{n\times \text{RMS}}\sum_{i=1}^n |\mathcal I_{\text{exp}}(t_i) - \mathcal I_{\text{apx}}(t_i)|\Delta \mathcal I_{\text{exp}}(t_i)
\end{equation}
Here, $\Delta \mathcal I_{\text{exp}}(t_i)$ is the standard deviation at $t_i$, extracted from $4$ data points. 
We show this RMS for $\ell =3, \cdots, 9$, $k_{\dn}=0$ and $k_{\up}=0, 1, 2, 3$ in Fig.~$2$(b) of the main text.

In Fig.~$2$(c) the fit parameter is $ J=\SI{0.54(1)}{k\hertz}$.  Here, the experimental data was averaged $20$ times and the errorbars represent the standard error of the mean. We use $L_{\text{apx}}=100$ and $T_r =\SI{7.5}{ms}$ for all computations in Fig. $2$(c). We also account for an averaging over $J$, caused due to a Gaussian beam profile of the primary lattice laser~\cite{PhysRevLett.122.170403}.  Different $1$D tubes have different value of $J$ in the $x$-axis due to variation of the power of the lattice laser in the $y-z$ plane. To account for this effect, we computed a weighted average over $4$ values of $J$ ranging up to $0.8$ times the value at the center.  In the inset, the black dashed line is obtained by a polynomial extrapolation~\cite{brezinski2013extrapolation} of the sequence of imbalance computed for different $k_{\up}$.  In this extrapolation, we assume that the convergence of the approximate method is described by a power series expansion in $1/k_{\up}$. Accordingly, we fit the imbalance to $a + b/k_{\up}$ an use $a$ as the extrapolated convergence value.

In Fig.~$2$(d) the fit parameters are $J=\SI{0.54(1)}{k\hertz}, $ $\Delta_{\dn}=\SI{3.30(3)}{}J = \SI{1.80(2)}{k\hertz}, $ $\Delta_{\up}=\SI{1.62(2)}{k\hertz}, $  and $U=5J$. The experimental data was sampled at $67$ points between $t=\SI{1}{ms}$ and $t=\SI{200}{ms} = \SI{682}{\tau}$. The experimental data was averaged $ 12 $ times at each point and the errorbars represent the standard deviation.  In the numerical calculation we use $T_r =\SI{10}{ms}$,  $L_{\text{apx}}=280$, $\ell=7$, $k_{\up}=0$ and $k_{\dn}=2, 3, 4, 5$.  We compute the imbalance at $15000$ uniformly spaced points between $t=0$ and $t=\SI{300}{ms}\approx \SI{1000}{\tau}$.  

In Fig.~$2$(e), the parameters are $J=\SI{0.560(6)}{k\hertz}$.  This value was calibrated using Kapitza-Dirac scattering with a Bose-Einstein condensate of $^{87}$Rb atoms loaded into the same lattice.  The detuning strength was $\Delta = \SI{3.00(6)}{J}$ also calibrated using the same method.  Each data point was averaged $10$ times and the errorbars represent the standard error of the mean.  The numerical calculation used $L_{\text{apx}}=280$ and was averaged over $48$ evenly spaced values of the detuning phase between $0$ and $2\pi$. The range of the interaction strength $U$ was $[-20J, 20J]$ with a step size of $J$.  In this dataset, we use a calibration factor of $1.1$ on the imbalance to account for the underestimation due to imperfections.  In all other datasets, we use a more sophisticated calibration method as described in appendix~\ref{measurement}.

\subsection{Main text Figure 3}In Fig.~$3$(a),and (b), the data shown is a Fourier transform of  the imbalance with $80$ data points in time ranging between $0$ and $t=\SI{8}{ms}  = 25\tau$.  The data was averaged four times. The errorbars represent the standard deviation, propagated appropriately.  The system parameters, obtained by fitting a non-interacting dataset are, $J=\SI{0.54(1)}{k\hertz}$, $\Delta_{\dn}=\SI{1.60(1)}{k\hertz},  \Delta_{\up}=\SI{1.44(1)}{k\hertz}$ and $T_r= \SI{9.00(3)}{ms}$. The interaction strength was $U=3J$.  In the calculation (i.e., solid lines), we use a system size of $L_{\text{apx}}=100$ and parameters  $\ell = 4, $ $k_{\dn}=0$ and $k_{\up}=2$. The numerical sampling was the same (i.e., $80$ data points) as the experiment so that we can make a comparison.

In Fig.~$3$(c) all the parameters were the same as in Fig.~$3$(a) for the case of $\Delta =3J$. For the case of $\Delta =1.1J$, the system parameters were $J=\SI{0.90(2)}{k\hertz}$, $\Delta_{\dn}=\SI{1.043(8)}{k\hertz} \Delta_{\up}=\SI{0.943(7)}{k\hertz}$ and $T_r= \SI{5.61(4)}{ms}$. The interaction strength was $U=3J$. 

In Fig.~$3$(d), we compute the Fourier spectra of the imbalance for various interaction strengths. We use a system size of $L_{\text{apx}}=280$ and  parameters $\ell = 7, $ $ k_{\dn}=0$ and $k_{\up}=4$. We begin with a CDW (see appendix~\ref{imbalace_computation} for more details) with $50$\% population in each spin and time evolve it under the interacting Stark Hamiltonian with $\Delta = 3J$. and $\alpha =0$. We use $J=\SI{0.5}{k\hertz}$ and go up to $\sim 700 \tau$, sampled at $2000$ points. Therefore, in the Fourier space it corresponds to a step of $\SI{5}{Hz}$ or $10^{-2}J$.  This computation is done for $100$ values of the interaction, placed uniformly between $U=0$ and $U=20J$.

\subsection{Main text Figure 4}

In Fig.~$4$(a), we compute the entanglement entropy for one few-body state. We begin with a lattice of size $L=15$ (this corresponds to $\ell =7$) with a spin-$\dn$ atom on site $i_0 = 8$ and $q_{\up}$  spin-$\up$ atoms filled in  $i_0+2, i_0-2, i_0+4, i_0-4, \cdots $, in that order.  In the case $q_{\dn}=1$, the other spin-$\dn$ atom is placed on site $i_0+4$. We then compute the time evolution of this state under the interacting Aubry-Andr\'e model with $U=5J$ and two values of $\Delta$ ($4J$ and $8J$). We then compute the time-dependent entanglement entropy sampled approximately  after each tunnelling time. We then average this entanglement entropy over $12$ values of the detuning phase, placed uniformly between $0$ and $2\pi$. 

\begin{figure}
\includegraphics[scale=1]{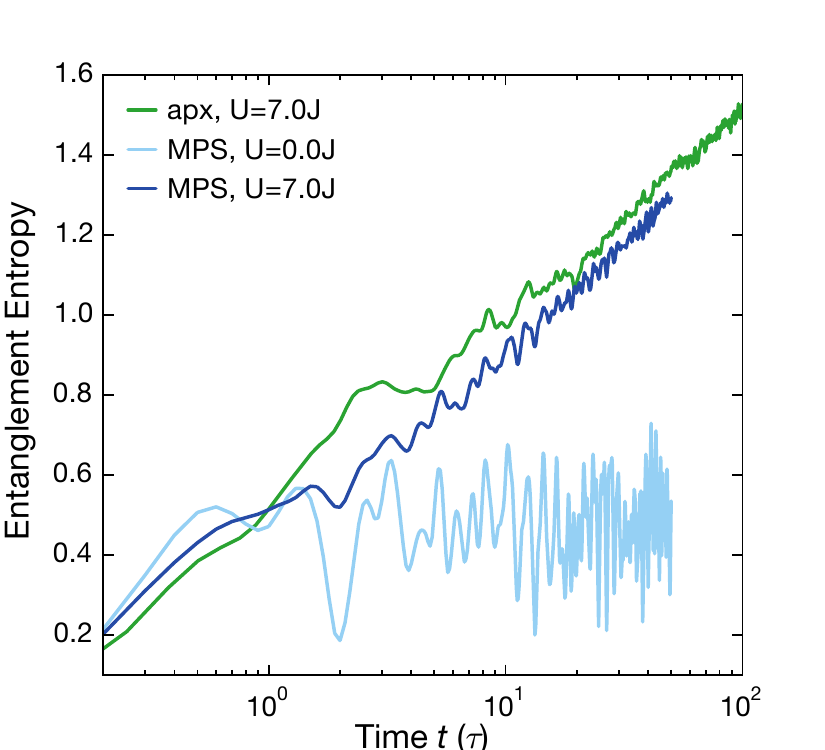}
\caption{\textbf{Entanglement entropy. } The entanglement entropy computed using one few-body state with $\ell=7, q_{\sigma}=0$ and $q_{\bar{\sigma}}=4$. The parameters were $\Delta = 5J$ and $U=7J$.  The dark blue curve shows the TEBD calculation for a similar parameter regime reproduced from  Ref. ~\cite{Schreiber_2015}. }\label{FigS4}
\end{figure}

For a comparison,  we show the entanglement entropy computed as described above with the entanglement entropy computed using TEBD, taken from Ref. ~\cite{Schreiber_2015} in Fig.~\ref{FigS4}. Here we use $\Delta = 5J$ and $ U=7J$ and average it over $24$ detuning phases. 

In Fig.~$4$(b), we compute the non-interacting, time-averaged correlation for two detuning strengths of the Aubry-Andr\'e model. We begin with one atom positioned on site $i_0=2$ of a lattice with $L=19$ sites. We compute the time evolution of this state under the Aubry-Andr\'e Hamiltonian up to $700 \tau$, sampled at $500$ points using exact diagonalization. We then compute the time dependent correlator $C_{i_0, j}(t) = \langle \hat{n}_{i_0} \hat{n}_j \rangle  - \langle \hat{n}_{i_0}\rangle \langle \hat{n}_j \rangle$. Here, $\hat{n}_j$ is the total occupancy (i.e., both the spins included) on site $j$.  We then compute the time average followed by the disorder phase average  over $20$  detuning phases placed uniformly between $0$ and $2\pi$, of $|C_{i_0 j}|$. We then plot  this average against $j-i_0$ in the figure. 

\begin{figure}
\includegraphics[scale=1]{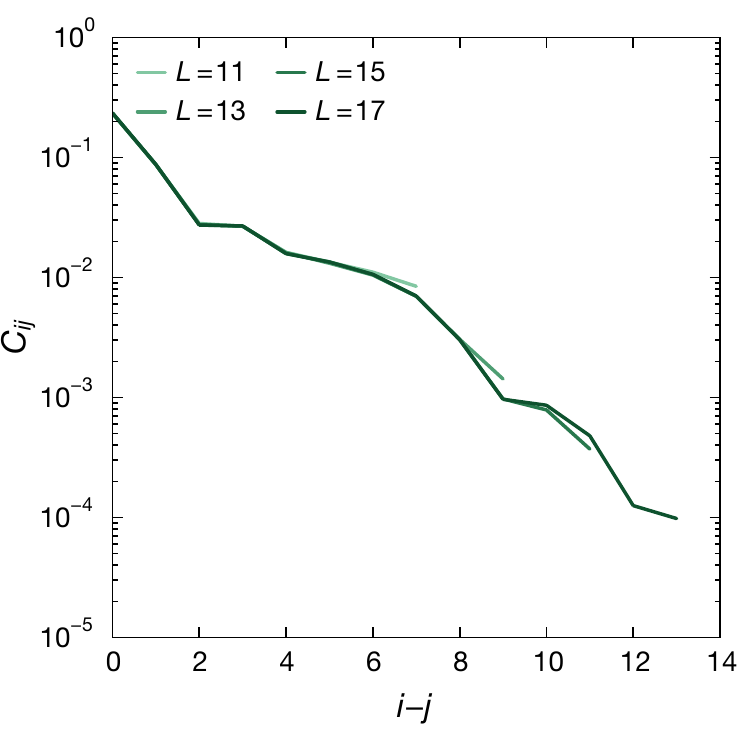}
\caption{\textbf{System size scaling of $C_{ij}$:}  The correlations $C_{ij}$ plotted for the Aubry-Andr\'e model with $\Delta =4J, U=5J$ and $q=(0, 3)$ for various sizes of the short lattice.  }\label{FigR4}
\end{figure}

In Fig.~$4$(c), we compute the plateau value of the correlation for an interacting system for various values of detuning strength.  Similar to the previous figure, we begin with an initial state with a spin-$\dn$ atoms at position $i_0=1$ on a lattice with size $L=13$ sites (see Fig.~\ref{FigR4} for a convergence analysis).  We place $q_{\up}$ number of spin-$\up$ atoms in the odd sites after $i_0$, that is, on $i_0+2, i_0+4, \cdots$.  We use $q_{\up}=0, 1, 2, 3$ and $4$. We evolve this initial state under the interacting Aubry-Andr\'e Hamiltonian for $t=700 \tau$, sampled at $500$ points with an interaction strength $U=5J$. Similar to the previous figure, we compute the time and detuning phase average of the correlator $C_{i_0, j}$ over $20$ values of the latter. Additionally, for this figure, we extract the plateau value of the averaged $C_{i_0, j}$ by further averaging it over the second half of the range of $j$. That is, over $j=7, \cdots, 13$. We compute this plateau value for $25$ values of the detuning strength, $\Delta$, placed uniformly between $\Delta = J$ and $\Delta = 10J$.  We show this plateau value against $\Delta$ in the figure. 

\section{The occupancy matrix}\label{occupancy_matrix}
In this section, we derive Eq.~$(3)$ in the main text. Let us assume $N_{\dn}=0$ and the $N_{\up}$ spin-$\up$ atoms at $t=0$ are on sites $i_1, \cdots, i_{N_{\up}}$ of the lattice. Let  $\phi_{i_1}(t), \phi_{i_2}(t), \cdots, \phi_{i_{N_{\up}}}(t)$ be the time evolved states of the atoms. Clearly, $\langle \phi_i(t), \phi_j(t)\rangle =\delta_{ij}$. The many-body state $\psi$ is given by the antisymmetrized product 
of $\phi_{i_1}(t), \phi_{i_2}(t), \cdots, \phi_{i_{N_{\up}}}(t)$.  That is
\begin{equation}\label{anti_symm}
\psi(t) = \sum_{\mu \in S_{N_{\up}}} \text{sgn}(\mu) \phi_{\mu(1)}(t)\cdots \phi_{\mu(N_{\up})}(t)
\end{equation}
Here $S_{N_{\up}}$ is the symmetric group and sgn($\mu$) is the sign of a permutation $\mu \in S_{N_{\up}}$. For convenience, we define vectors $\psi_i = \hat{c}_{i, \up}\psi$. The occupancy matrix $\Gamma^{\up}$ is given by 
\begin{equation}
\begin{split}
\Gamma^{\up}_{ij}&= \langle \psi | \hat{c}_{i, \up}^{\dagger} \hat{c}_{j, \up}| \psi \rangle =  \langle \psi_i, \psi_j\rangle \\
= & \sum_{\mu, \mu'}\langle \hat{c}_{i, \up}\phi_{\mu(1)}(t), \hat{c}_{j, \up}\phi_{\mu'(1)}(t)\rangle  \langle \phi_{\mu(2)}(t), \phi_{\mu'(2)}(t)\rangle  \times\\
& \cdots \times \langle \phi_{\mu(N_{\up})}(t), \phi_{\mu'(N_{\up})}(t)\rangle  \text{sgn}(\mu)  \text{sgn}(\mu')\\
\end{split}
\end{equation}
Following $\langle \phi_i, \phi_j\rangle = \delta_{ij}$, we may write, 
\begin{equation}
\Gamma^{\up}_{ij}= \sum_{\mu}  \langle \hat{c}_{i, \up}\phi_{\mu(1)}(t), \hat{c}_{j, \up}\phi_{\mu(1)}(t)\rangle = \sum_k |\phi_k (t)\rangle \langle \phi_k(t)|_{ij}
\end{equation}

\section{Canonical construction of the many-body quantum state}\label{reconstruction}
The approximate method produces $L\times L$ single-particle density matrices $\Gamma^{\up, 1}, \cdots, \Gamma^{\up, N_{\up}}$ and $\Gamma^{\dn, 1}, \cdots, \Gamma^{\dn, N_{\dn}}$ corresponding to the individual atoms at any given time. These matrices represent a very small proportion of the information in the quantum system. As a result, while it is desirable to construct a many-body quantum state starting from these single-particle density matrices, it cannot be done in a unique way. There will be multiple many-body quantum states that all map to the same set of single-particle density matrices. Nevertheless, it is worthwhile to construct a canonical many-body quantum state starting from the set of single--particle density matrices. Moreover, we are helped by the fact that the many-body quantum state is pure --- this reduces the ambiguity. 

Here, we discuss one possible method of  constructing such a many-body quantum state.  We also discuss Eq. ~$(4)$ in the main text and possible corrections to it.  For simplicity, we restrict to the case $k_{\dn}=0$ in the computation of $\Gamma^{\dn, j}$ and $k_{\up}=0$ in the computation of $\Gamma^{\up, j}$. The general case is more involved. 

\begin{figure}
\includegraphics[scale=1]{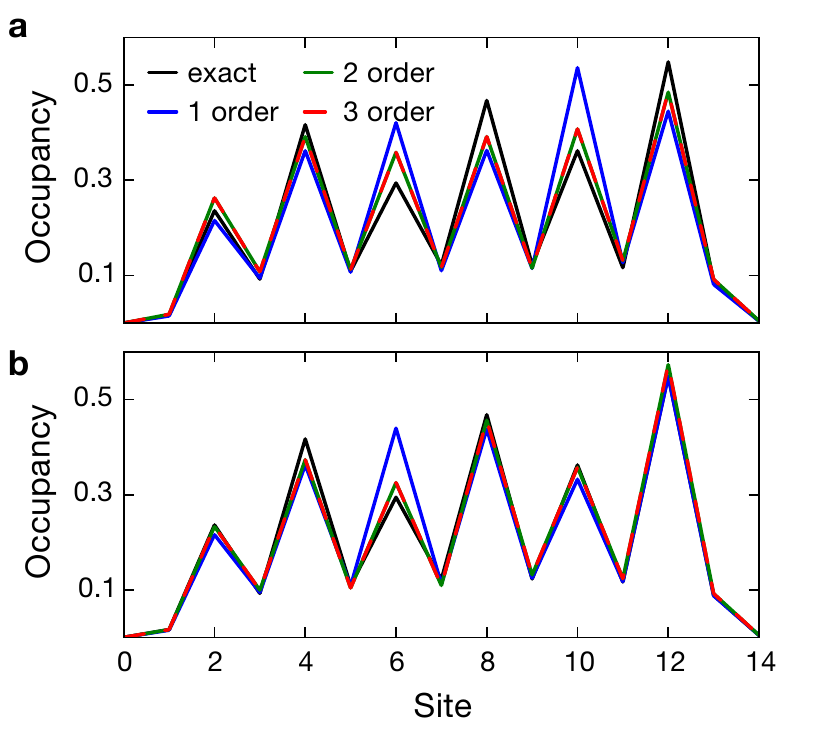}
\caption{\textbf{Higher order contributions to the occupancy matrix.} The on-site occupancy of spin-$\up$ atoms, averaged over time up to $300$ tunnelings, starting from a Neel-type CDW under the Stark model, i.e., with $3$ spin-$\up$ and $3$ spin-$\dn$ atoms on alternating even sites. \textbf{a} The density for $k_{\up}=0, k_{\dn}=2$ and using various  orders in Eq.~(\ref{Gamma}) along with the exact calculation. Note that the difference between the second order and third order is insignificant.  \textbf{b} The density for $k_{\up}=1, k_{\dn}=2$. The parameters used are $\Delta_{\sigma} = 4J$ and $U=2J$. }\label{Gamma_orders}
\end{figure}

We begin by constructing two single-spin-many-particle density matrices $\rho^{\up}$ and $\rho^{\dn}$. These are $\binom{L}{N_{\up}}\times \binom{L}{N_{\up}}$ and $\binom{L}{N_{\dn}}\times \binom{L}{N_{\dn}}$ matrices respectively.  In the lexicographically ordered basis of $N_{\sigma}$ identical atoms in $L$ sites (see main text), let $\alpha = \{i_1, i_2, \cdots, i_{N_{\sigma}}\}$ and $\beta = \{j_1, j_2, \cdots, j_{N_{\sigma}}\}$ represent two basis elements. Following Eq.~(\ref{anti_symm}), we define
\begin{equation}\label{anti_symm_mixed}
\rho^{\sigma}_{\alpha \beta} =\eta  \sum_{\mu, \mu' \in S_{N_{\sigma}}} \text{sgn}(\mu)\text{sgn}(\mu') \Gamma^{\sigma, 1}_{i_{\mu'(1)}, j_{\mu(1)}}\cdots \Gamma^{\sigma, N_{\sigma}}_{i_{\mu'(N_{\sigma})}, j_{\mu(N_{\sigma})}}
\end{equation} 

Here, $\eta$ is the normalization factor to ensure that $\text{Tr}(\rho^{\sigma})=1$. Note that this construction is justified only when $k_{\sigma}=0$ and $k_{\bar{\sigma}}\geq 0$.  We may now compute the occupancy matrix $\Gamma^{\sigma}_{ij}=\text{Tr}(\rho^{\sigma} \hat{c}_{i, \sigma}^{\dagger}\hat{c}_{j, \sigma}) $. 
\begin{equation}\label{Gamma}
\begin{split}
\Gamma^{\sigma}=&\eta  \sum_{j=1}^{N_{\sigma}}\Gamma^{\sigma, j} -\eta  \sum_{j\neq j'}^{N_{\sigma}}\Gamma^{\sigma, j}\Gamma^{\sigma, j'} \\
&+ \eta \sum_{j \neq j' \neq j'' }\Gamma^{\sigma, j}\Gamma^{\sigma, j'} \Gamma^{\sigma, j''} + \cdots
\end{split}
\end{equation}
Note that, insofar as $\Gamma^{\sigma, i}$ represents localized states, the higher order terms are insignificant (see Fig.~\ref{Gamma_orders}).

We now address the question of how to construct a many--body state corresponding to $N_{\up}$ spin-$\up$ atoms and $N_{\dn}$ spin-$\dn$ atoms in the lattice with $L$ sites, starting from the density matrices $\rho^{\up}$ and $\rho^{\dn}$. Ideally, we would seek a pure ``parent" quantum state $\psi$ satisfying the equations
\begin{equation}\label{trace_eqn}
\text{Tr}_{\up}|\psi \rangle \langle \psi |  = \rho^{\dn}\text{ and }
\text{Tr}_{\dn}|\psi \rangle \langle \psi |  = \rho^{\up}.
\end{equation}
Here, $\text{Tr}_{\sigma}$ represents a partial trace with respect to the subspace of spin-$\sigma$ atoms. The above equations have solutions if and only if $\rho^{\up}$ and $\rho^{\dn}$ have the same set of non-zero eigenvalues. To see this, let us consider the Schmidt decomposition $\psi= \sum_i \lambda_i |\psi_i^{\up} \rangle \otimes |\psi_i^{\dn} \rangle$, where $\langle \psi_i^{\sigma}|\psi_j^{\sigma}\rangle = \delta_{ij}$ for $\sigma= \up, \dn$. The reduced density matrices, $\text{Tr}_{\dn}|\psi \rangle \langle \psi | = \sum_i \lambda_i^2 |\psi_i^{\up} \rangle \langle \psi_i^{\up}|$ and  $\text{Tr}_{\up}|\psi \rangle \langle \psi | = \sum_i \lambda_i^2 |\psi_i^{\dn} \rangle \langle \psi_i^{\dn}|$ have the same set of non-zero eigenvalues.  Therefore, in general, if $\rho^{\up}$ and $ \rho^{\dn}$ have different sets of non-zero eigenvalues, Eq.~(\ref{trace_eqn}) has no solution. Accordingly, we seek the \textit{purest possible} mixed parent state $\rho$ that satisfies
\begin{equation}\label{trace_eqn_mixed}
\text{Tr}_{\up}\rho  = \rho^{\dn}\text{ and }
\text{Tr}_{\dn}\rho  = \rho^{\up}.
\end{equation}
We will define ``purest possible" shortly. Eq.~(\ref{trace_eqn_mixed}), unlike Eq.~(\ref{trace_eqn}),  has infinitely many solutions in general. Let $W_{\rho^{\up}, \rho^{\dn}}=\{\rho: \ \text{Tr}_{\up}\rho  = \rho^{\dn}\text{ and } \text{Tr}_{\dn}\rho  = \rho^{\up}\}$ be the set of all possible parent states of $\rho^{\up}$ and $\rho^{\dn}$, i.e., the set of all solutions of Eq.~(\ref{trace_eqn_mixed}). Trivially, $\rho^{\up}\otimes \rho^{\dn} \in W_{\rho^{\up}, \rho^{\dn}}$, i.e., there is always at least one solution. Moreover, $W_{\rho^{\up}, \rho^{\dn}}$ is convex, i.e., convex sum of two solutions is also a solution of Eq.~(\ref{trace_eqn_mixed}). If $\rho, \rho^{\prime} \in W_{\rho^{\up}, \rho^{\dn}}$, then $\mu \rho + (1-\mu)\rho^{\prime} \in W_{\rho^{\up}, \rho^{\dn}}$ for $0\leq \mu \leq 1$. 

In the special case when at least one of $\rho^{\up}$ and $\rho^{\dn}$ is pure, the parent state is unique, i.e., $W_{\rho^{\up}, \rho^{\dn}}= \{\rho^{\up}\otimes \rho^{\dn}\}$.  To see this let $\rho^{\up}= |\phi^{\up}\rangle \langle \phi^{\up}|$ for some pure state $\phi^{\up}$. For any $\rho \in W_{\rho^{\up}, \rho^{\dn}}$, we may consider the spectral decomposition $\rho =\sum_i  \mu_i |\psi_i \rangle \langle \psi_i|$. It follows from $\text{Tr}_{\dn}\rho = \sum_i  \mu_i \text{Tr}_{\dn}|\psi_i \rangle \langle \psi_i| = |\phi^{\up}\rangle \langle \phi^{\up}|$ that $ \text{Tr}_{\dn}|\psi_i \rangle \langle \psi_i| = |\phi^{\up}\rangle \langle \phi^{\up}|$ for each $i$. Thus, each $\psi_i$ is separable into $\psi_i = \phi^{\up}\otimes \psi^{\dn}_i$. Therefore, $\rho = |\phi^{\up}\rangle \langle \phi^{\up}| \otimes \rho^{\dn}$. 

In all other cases, however, $ W_{\rho^{\up}, \rho^{\dn}}$ is an infinite set. One canonical choice is to pick a $\rho^{*}\in W_{\rho^{\up}, \rho^{\dn}}$ with the lowest Von-Neumann entropy. This is justified because Von-Neumann entropy is a measure of purity;  pure states have zero Von-Neumann entropy. However, states in $W_{\rho^{\up}, \rho^{\dn}}$ with the lowest Von-Neumann entropy are still not unique. For instance, consider the special case where $\rho^{\up}$ and $\rho^{\dn}$ have the same set of non-zero eigenvalues.  Let $\rho^{\up}= \sum_i \lambda_i |\psi^{\up}_i\rangle \langle \psi_i^{\up}|$ and $\rho^{\dn}= \sum_i \lambda_i |\psi^{\dn}_i\rangle \langle \psi_i^{\dn}|$
with $\lambda_i $ arranged in decreasing order. It follows that $\rho=|\psi \rangle \langle \psi |$ with
\begin{equation}\label{pure_parent}
\psi = \sum_i \sqrt{\lambda_i} e^{i\theta_i}|\psi_i^{\up}\rangle \otimes |\psi^{\dn}_i\rangle
\end{equation}
satisfies Eq.~(\ref{trace_eqn_mixed}) for arbitrary $\theta_i$. Thus, there is an infinitude of pure parent states, all having a zero Von-Neumann entropy. Therefore, even in the general case we may expect there to be multiple parent states with a minimal Von-Neumann entropy. However, constructing a parent state with minimal Von-Neumann entropy is in general fairly non-trivial. 

Below, we present a natural generalization of  Eq.~(\ref{pure_parent}) to the case where $\rho^{\up}$ and $\rho^{\dn}$ don't necessarily have the same set of non-zero eigenvalues, resulting in an impure state $\rho^* \in W_{\rho^{\up}, \rho^{\dn}}$, but nevertheless having a relatively low Von-Neumann entropy. 

Let us assume that 
\begin{equation}
\rho^{\up}= \sum_{j=1}^n \lambda_j^{\up} |\psi^{\up}_j\rangle \langle \psi_j^{\up}| \text{ and } \rho^{\dn}= \sum_{j=1}^m \lambda_j^{\dn} |\psi^{\dn}_j\rangle \langle \psi_j^{\dn}|
\end{equation}
where $\lambda_1^{\up}\geq \lambda_2^{\up} \geq \cdots \geq  \lambda_n^{\up} > 0 $ and  $\lambda_1^{\dn}\geq \lambda_2^{\dn} \geq \cdots \geq \lambda_m^{\dn} > 0 $. $n$ and $m$ are the ranks of $\rho^{\up}$ and $\rho^{\dn}$ respectively. 

While there is no pure parent many-body state ,  we show that $\rho^{\up}$ and $\rho^{\dn}$ can be written as a sum of matrices that have pure parents:
\begin{equation}
\rho^{\sigma} = \rho^{\sigma, 1}+ \cdots + \rho^{\sigma, r}
\end{equation}
Such that each $\rho^{\sigma, i}$ is positive semi-definite (not necessarily normalized) matrices and $\rho^{\up, i}$ and $\rho^{\dn, i}$ have the same set of non-zero eigenvalues, i.e., they have a pure parent $\psi_i$ (again, not necessarily normalized). In other words,
\begin{equation}
\rho^{\up, i} = \text{Tr}_{\dn} |\psi_i \rangle \langle \psi_i| \text{ and } \rho^{\dn, i} = \text{Tr}_{\up} |\psi_i \rangle \langle \psi_i|
\end{equation}   
We define
\begin{equation}\label{canonical_state}
\rho^* = |\psi_1 \rangle \langle \psi_1|+ \cdots + |\psi_r \rangle \langle \psi_r|
\end{equation}
as the many-body quantum state. We will argue that this is likely to be one of the parent states with minimal Von-Neumann entropy. We begin by defining $\rho^{\sigma, \nu}$ recursively. For convenience of notation, we introduce remainder matrices, $REM_{\sigma}(\nu)= \rho^{\sigma, \nu+1}+ \cdots + \rho^{\sigma, r}$.
\begin{itemize}
\item[1.] For $\nu =1$, 
\begin{equation}\label{seed}
\begin{split}
&\rho^{\sigma, 1}=\sum_{i}\mathrm{min}(\lambda_i^{\up}, \lambda_i^{\dn}) |\psi_i^{\sigma}\rangle \langle \psi_i^{\sigma}| \\
&REM_{\sigma}(1) = \rho^{\sigma}-\rho^{\sigma, 1}
\end{split}
\end{equation} 
\item[2.] For $\nu >1$, we diagonalize $REM_{\sigma}(\nu -1)$, 
\begin{equation}
REM_{\sigma}(\nu-1) =  \sum_{i=1}^{n_{\sigma}} \lambda_i^{\sigma, \nu -1} |\psi^{\sigma, \nu -1}_i\rangle \langle \psi_i^{\sigma, \nu -1}|
\end{equation}
We assume an ordering $ \lambda_1^{\sigma, \nu-1}\geq \cdots \geq  \lambda_{n_{\sigma}}^{\sigma, \nu -1}$And then we define
\begin{equation}\label{induction}
\begin{split}
&\rho^{\sigma, \nu}=\sum_{i}\mathrm{min}(\lambda_i^{\up, \nu-1}, \lambda_i^{\dn, \nu -1}) |\psi_i^{\sigma, \nu -1}\rangle \langle \psi_i^{\sigma, \nu -1}| \\
&REM_{\sigma}(\nu) = REM_{\sigma}(\nu-1)-\rho^{\sigma, \nu}
\end{split}
\end{equation} 

\end{itemize}
It follows from Eq.~(\ref{seed}) and Eq.~(\ref{induction}) that $REM_{\sigma}(\nu)$ is positive semi-definite. Moreover, it also follows that $\text{Tr}(REM_{\up}(k)) =  \text{Tr}(REM_{\dn}(k))$. Therefore, if $REM_{\up}(\nu)=0$, then $REM_{\dn}(\nu)=0$, i.e., this iteration terminates when $REM_{\sigma}(\nu)$ is zero. 

Furthermore, it follows from Eq.~(\ref{induction}) that,
\begin{equation}
\begin{split}
\text{rank}&(REM_{\up}(\nu))+\text{rank}(REM_{\dn}(\nu))=\\
&\text{max}(\text{rank}(REM_{\up}(\nu -1)), \text{rank}(REM_{\dn}(\nu-1)))\\
\end{split}
\end{equation}
Thus, the sum of the ranks of $REM_{\up}(\nu)$ and $REM_{\dn}(\nu)$ is a strictly decreasing function of $\nu$. Therefore, the iteration terminates after a finite number of steps.

The Von-Neumann entropy of $\rho^*$ [Eq.~(\ref{canonical_state})]is 
\begin{equation}
S[\rho^*] = -\sum_{i=1}^r  \bra{\psi_i}\psi_i\rangle \log (  \bra{\psi_i}\psi_i\rangle)
\end{equation}
This follows from the observation that $\langle \psi_i |\psi_j\rangle = \bra{\psi_i}\psi_i\rangle\delta_{ij}$. Note that if $\rho^{\up}$ and $\rho^{\dn}$ have the same set of non-zero eigenvalues, the above construction terminates at $r=1$, and $\rho^{*}$ will be pure. 

We have discussed one possible algebraic way of constructing a many-body state starting from the single particle occupancy matrices. Another possible way would be to exploit the continuity of the many-body state $\rho(t)$ in time, given that we know $\rho(0)$, to reduce the ambiguity introduced by Eq.~(\ref{trace_eqn_mixed}). 

Next, we make a few comments on the general case, $k_{\sigma}>0$ when computing $\Gamma^{\sigma, j}$. The single--particle occupancy matrix $\Gamma^{\sigma, j}(t)$ cannot be unambiguously defined in this case. Consider for instance,  a few--body state $\ket{\psi^{\prime \prime}(t)}$ (see main text) corresponding to $q_{\sigma}+1$ spin-$\sigma$ atoms and $q_{\bar{\sigma}}$ spin-$\bar{\sigma}$ atoms. The spin-$\sigma$ occupancy matrix computed using this state describes $q_{\sigma}+1$ atoms. The simple prescription used in the main text, i.e., to divide the occupancy matrix by $q_{\sigma}+1$ does not affect the first term in Eq.~(\ref{Gamma}). However, this could lead to repetitions in $\Gamma^{\sigma, 1}, \cdots, \Gamma^{\sigma, N_{\sigma}}$ making Eq.~(\ref{anti_symm_mixed}) unjustified. Moreover, the higher-order terms in Eq.~(\ref{Gamma}) would include squares. 

In Fig.~\ref{Gamma_orders}, we show a computation for $k_{\sigma}=1$, where we resolve the above issue by simply dropping all terms that contain squares in Eq.~(\ref{Gamma}). While the result appears to agree with the exact calculation, this procedure is not on firm mathematical grounds. An alternative method would be to resolve the occupancy matrix coming from $\ket{\psi^{\prime\prime}(t)}$ into a sum of $q_{\sigma}+1$ maximally distinct occupancy matrices, each with unit trace.

\section{Computing the imbalance time trace}\label{imbalace_computation}

In this section, we describe how the imbalance of spin-$\sigma$ atoms at time $t$, $\mathcal I^{\sigma}(t)$, is computed using the approximate method for a mixed singlon charge density wave (CDW) initial state that we start with in the experiment. In particular, we show how Eq.~$(7)$ of the main text is derived. We retrict to a CDW with $N_{\up}+N_{\dn}=L/2$ and no doubly-occupied sites. The more general derivation for a CDW with doublons and holes follows the same procedure. 

The mixed singlon CDW is an incoherent sum of all pure states with $N_{\sigma}$ atoms in the spin-$\sigma$ state, all of them in even sites of a lattice of size $L$, with no double occupancies. The number of such pure states is given by $\binom{L/2}{N_{\sigma}}=\binom{L/2}{N_{\bar{\sigma}}}$, where $\sigma$ and $\bar{\sigma}$ are the two spins. The time evolved imbalance of this mixed state is the average of the time evolved imbalances of these pure states. Indeed, if $\psi$ is a pure state, one can compute its time evolved imbalance, denoted by $\mathcal I^{\sigma}_{\psi}(t)$ using the time-evolved occupancy matrix, denoted by $\Gamma_{\psi}^{\sigma}(t)$ (see main text for a definition of the occupancy matrix). If $\hat{\mathcal I}^{\sigma} = \sum_{i=1}^L(-1)^{i} \hat{c}_{i, \sigma}^{\dagger}\hat{c}_{i, \sigma}$ is the single-particle imbalance operator,  then by restricting to the first term in Eq.~(\ref{Gamma}) we obtain $\mathcal I_{\psi}^{\sigma}(t) =\frac{1}{N_{\sigma}}\text{Tr}(\Gamma_{\psi}^{\sigma}(t) \hat{\mathcal I}^{\sigma} )=  \frac{1}{N_{\sigma}}\sum_{r=1}^{N_{\sigma}} \text{Tr}(\Gamma_{\psi}^{\sigma, r}(t) \hat{\mathcal I}^{\sigma} )$.

Each of the $\text{Tr}(\Gamma_{\psi}^{\sigma, r}(t) \hat{\mathcal I}^{\sigma} )$ are independently computed and therefore it is most useful to express the total imbalance as a sum of them. Loosely speaking, $\text{Tr}(\Gamma_{\psi}^{\sigma, r}(t) \hat{\mathcal I}^{\sigma} )$ is  the imbalance of the $r$-th spin-$\sigma$ atom in the initial state $\psi$. We can now express the imbalance of the CDW in terms of these quantities. 
\begin{equation}\label{imb_apx}
\mathcal I^{\sigma}(t) = \frac{1}{N_{\sigma}\binom{L/2}{N_{\sigma}}} \sum_{\psi} \sum_{r=1}^{N_{\sigma}} \text{Tr}(\Gamma_{\psi}^{\sigma, r}(t) \hat{\mathcal I}^{\sigma} )
\end{equation}

Here, the outer sum is over all pure singlon CDW states $\psi$. This expression is a sum of $N_{\sigma}\binom{L/2}{N_{\sigma}}$ terms. Despite its appearance, this does not require an exponentially (in $L$) large number of independent computations, for not all of   $\text{Tr}(\Gamma_{\psi}^{\sigma, r}(t) \hat{\mathcal I}^{\sigma})$ are distinct. This quantity depends on the configuration of atoms in the neighborhood of the $r$-th atom in the state $\psi$. The same configuration could appear multiple times for different values of $r$ and $\psi$. Indeed, making use of this, we show that the number of distinct terms in the sum Eq.~(\ref{imb_apx}) is at-most linear in $L$. 

Consider, for instance, the case of $k_{\sigma}=0$ and $k_{\bar{\sigma}}>0$. The time trace $\text{Tr}(\Gamma_{\psi}^{\sigma, r}(t) \hat{\mathcal I}^{\sigma} )$ depends solely on the configuration of spin- $\bar{\sigma}$ atoms in the neighborhood of the $r$-th spin-$\sigma$ atom. There are $2^{k_{\bar{\sigma}}}L/2$ distinct such traces. To see this, let the position of the $r$-th atom be $2j$ for some $j \in \{1, 2, \cdots, L/2\}$ (the initial position of an atom is always even in a CDW).  $j$ can take $L/2$ distinct values. Corresponding to each of them, there are $2^{k_{\bar{\sigma}}}$ configurations of spin-$\bar{\sigma}$ atoms in the $k_{\bar{\sigma}}$-neighborhood of $2j$ (note that we are ignoring the boundary effects due to finite size). Thus, the number of distinct terms in the sum Eq.~(\ref{imb_apx}) is linear in $L$. 

The multiplicity of one such term depends on the number of spin-$\bar{\sigma}$ atoms in the $k_{\bar{\sigma}}$-neighborhood of $2j$. We denote this number by $q_{\bar{\sigma}}$. The multiplicity $C_{q_{\bar{\sigma}}}$ then is,
\begin{equation}
C_{q_{\bar{\sigma}}}=\binom{L/2-k_{\bar{\sigma}}-1}{N_{\bar{\sigma}}-q_{\bar{\sigma}}}
\end{equation}
This is the multiplicity of terms in Eq.~(\ref{imb_apx}) corresponding to a configuration with $q_{\bar{\sigma}}$ spin-$\bar{\sigma}$ atoms in the $k_{\bar{\sigma}}$-neighborhood of $2j$, and there are $\binom{k_{\bar{\sigma}}}{q_{\bar{\sigma}}}$ of such configurations.  The time evolved imbalance corresponding to such a configuration is computed by performing exact diagonalization of a system with $q_{\bar{\sigma}}$ spin-$\bar{\sigma}$ atoms  and one spin-$\sigma$ atoms in a short lattice of length $2\ell +1$ (as mentiomned in the main text, $\ell \geq k_{\bar{\sigma}}$). Note that the Hilbert space and the unitary operator corresponding to the time evolution is the same for all $\binom{k_{\bar{\sigma}}}{q_{\bar{\sigma}}}$ initial configurations with $q_{\bar{\sigma}}$ spin-$\bar{\sigma}$ atoms in the $k_{\bar{\sigma}}$-neighborhood. Therefore,  for practical convenience, we compute the time evolved imbalance of all of these configurations together. Accordingly, we define $\mathcal I^{\sigma}_{2j, q_{\bar{\sigma}}}(t)$ as the imbalance averaged over all of these states. 

Thus, the total imbalance of the mixed CDW is
\begin{equation}
\mathcal I^{\sigma}(t)= \frac{1}{N_{\sigma}\binom{L/2}{N_{\sigma}}}\sum_{j=1}^{L/2} \sum_{q_{\bar{\sigma}}=0}^{k_{\bar{\sigma}}} C_{q_{\bar{\sigma}}}\binom{k_{\bar{\sigma}}}{q_{\bar{\sigma}}} \mathcal I^{\sigma}_{2j, q_{\bar{\sigma}}}(t).
\end{equation}

This contains $2^{k_{\bar{\sigma}}}L/2$ terms. In fact, we may drop the outer sum over $j$ in the case of the Stark model,  since the dynamics of the system does not depend on the position of the center of the short lattice with $2\ell + 1$ sites. The number of terms is then independent of $L$. This expression maybe compared with the cluster expansion presented in Ref.~\cite{PhysRevA.93.031601}.  Note that the most complex computation is the one corresponding to $q_{\bar{\sigma}}=k_{\bar{\sigma}}$, where the dimension of the Hilbert space of the spin-$\bar{\sigma}$ atoms is $\binom{2\ell + 1}{k_{\bar{\sigma}}}$.

The above derivation was for $k_{\sigma}=0$. In the general case, assuming that $k_{\sigma}<k_{\bar{\sigma}}$ (which is usually the chosen case, based on the convergence rates shown in the main text), we obtain the following expression upon using the same derivation procedure. For a given spin-$\sigma$ atom in position $2j$, and $q_{\sigma}$ additional spin-$\sigma$ atoms in its $k_{\sigma}$ neighborhood and $q_{\bar{\sigma}}$ spin-$\bar{\sigma}$ atoms in its $k_{\bar{\sigma}}$ neighborhood, the multiplicity of this configuration is
\begin{equation}
C_{q_{\sigma}, q_{\bar{\sigma}}}=\binom{L/2-k_{\bar{\sigma}}-1}{N_{\bar{\sigma}}-q_{\bar{\sigma}}}
\end{equation}

The number of configurations of $q_{\sigma}$ spin-$\sigma$ atoms in the $k_{\sigma}$ neighborhood of a site and $q_{\bar{\sigma}}$ spin-$\bar{\sigma}$ atoms in its $k_{\bar{\sigma}}$ neighborhood is $\binom{k_{\sigma}}{q_{\sigma}} \binom{k_{\bar{\sigma}}-k_{\sigma}}{q_{\bar{\sigma}}+q_{\sigma}-k_{\sigma}} $. And thus, the total imbalance is
\begin{equation}\label{imbalance_calc}
\begin{split}
\mathcal I^{\sigma}(t)= \frac{1}{N_{\sigma}\binom{L/2}{N_{\sigma}}}& \sum_{j=1}^{L/2} \sum_{q_{\bar{\sigma}}=k_{\sigma}-q_{\sigma}}^{k_{\bar{\sigma}}} \sum_{q_{\sigma}=0}^{k_{\sigma}} C_{q_{\sigma}, q_{\bar{\sigma}}}\binom{k_{\sigma}}{q_{\sigma}}  \\
& \times  \binom{k_{\bar{\sigma}}-k_{\sigma}}{q_{\bar{\sigma}}+q_{\sigma}-k_{\sigma}}  \mathcal I^{\sigma}_{2j, q_{\sigma},  q_{\bar{\sigma}}}(t)\\
\end{split}
\end{equation}

Here, $\mathcal I^{\sigma}_{2j, q_{\sigma},  q_{\bar{\sigma}}}(t)$ is the time evolved imbalance of spin-$\sigma$ atoms averaged over all configurations of $q_{\sigma}$ spin-$\sigma$ atoms in a $k_{\sigma}$ neighborhood and  $q_{\bar{\sigma}}$ spin-$\bar{\sigma}$ atoms in a $k_{\bar{\sigma}}$ neighborhood of the $2j$-th site with a spin-$\sigma$ atom at the center. There are  $2^{k_{\bar{\sigma}}}$ such configurations for a given $j$ \big(this is simply the sum $\sum_{q_{\sigma}, q_{\bar{\sigma}}} \binom{k_{\sigma}}{q_{\sigma}} \binom{k_{\bar{\sigma}}-k_{\sigma}}{q_{\bar{\sigma}}+q_{\sigma}-k_{\sigma}}$\big). Thus, the number of imbalance time trace computations is $2^{k_{\bar{\sigma}}}L/2$. 

When $N_{\sigma} = \Omega (L)$, it is convenient to introduce $\lambda_{\sigma}=2N_{\sigma}/L$.  We can now simplify Eq.~(\ref{imbalance_calc}) using the Stirling approximation. For $a=o(n)$, it follows that $(n+a)!/n! \approx n^a $. Thus, 
\begin{equation}
\frac{ C_{q_{\sigma}, q_{\bar{\sigma}}}}{\binom{L/2}{N_{\sigma}}} \approx \lambda_{\bar{\sigma}}^{q_{\bar{\sigma}}} \lambda_{\sigma}^{k_{\bar{\sigma}}-q_{\bar{\sigma}}+1}
\end{equation}
Using $\lambda_{\sigma} = 1-\lambda_{\bar{\sigma}}$, Eq.~(\ref{imbalance_calc}) reads
\begin{equation}\label{cluster}
\begin{split}
\mathcal I^{\sigma}(t)= &\frac{1}{N_{\sigma}} \sum_{j=1}^{L/2} \sum_{q_{\bar{\sigma}}=k_{\sigma}-q_{\sigma}}^{k_{\bar{\sigma}}} \sum_{q_{\sigma}=0}^{k_{\sigma}} \binom{k_{\sigma}}{q_{\sigma}}  \\
& \times  \binom{k_{\bar{\sigma}}-k_{\sigma}}{q_{\bar{\sigma}}+q_{\sigma}-k_{\sigma}} \lambda_{\bar{\sigma}}^{q_{\bar{\sigma}}} (1-\lambda_{\bar{\sigma}})^{k_{\bar{\sigma}}+1-q_{\bar{\sigma}}} \mathcal I^{\sigma}_{2j, q_{\sigma},  q_{\bar{\sigma}}}(t)\\
\end{split}
\end{equation}

Note that $(1-\lambda_{\bar{\sigma}})/N_{\sigma} = 2/L$. It is straightforward to see that the above expression reduces to Eq.~$(7)$ of the main text.


\section{Description of the code}\label{code_description}

The code corresponding to the approximate method is available at \url{https://gitlab.physik.uni-muenchen.de/LDAP_ag-bec-fermi1/approximate-method-for-1d-fermi-hubbard-model}. The central part of the code is the time evolution of the few-body state $\ket{\psi^{\prime \prime}}$, which we do using the version of exact diagonalization  presented in our previous work,  Ref. ~\cite{scherg_observing_2020}. We briefly discuss the technique and discuss its efficiency. 

The few-body state $\ket{\psi^{\prime \prime}}$ consists of $q_{\up}+1$ spin-$\up$ atoms and $q_{\dn}$ spin-$\dn$ atoms on a lattice with $2\ell +1$ sites. Thus, the relevant Hilbert space can be written as a tensor product $\mathcal{H}_{\up}\otimes \mathcal{H}_{\dn}$ of the Hilbert spaces corresponding to the two spin components. Their dimensions are $d_{\up}=\binom{2\ell+1}{q_{\up}+1}$ and $d_{\dn}=\binom{2\ell+1}{q_{\dn}}$ respectively. The dimension of the full Hilbert space is therefore $\binom{2\ell+1}{q_{\up}+1} \binom{2\ell+1}{q_{\dn}}$. We use the basis consisting of states of the form $\hat{c}_{i_1, \up}^{\dagger}\cdots \hat{c}_{i_{q_{\up}+1}, \up}^{\dagger}|\text{vac}\rangle $ with $1\leq i_1 < i_2 <\cdots <i_{q_{\up}+1}\leq 2\ell + 1$ for $\mathcal{H}_{\up}$, which we represent as $\mathcal{V}_{\up}=\{\{i_1, \cdots, i_{q_{\up}+1}\}:\ 1\leq i_1 < i_2 <\cdots <i_{q_{\up}+1}\leq 2\ell + 1\}$, where the states are ordered lexicographically. Similarly we define the basis $\mathcal{V}_{\dn}=\{\{j_1, \cdots, j_{q_{\dn}}\}:\ 1\leq j_1 < j_2 <\cdots <j_{q_{\dn}}\leq 2\ell + 1\}$. 

In order to study the efficiency of an ED-based time evolution in this system, we classify the relevant parameters in $3$ categories, summarized in Table~\ref{Table1}. 
\begin{table}
\begin{tabular}{ |c|c|c| } 
 \hline
 Category & Parameter(s) & Typical \\
 \hline
 Small $(\mathcal S)$& $\ell$, $q_{\up}$, $q_{\dn}$ & $10$ \\ 
 Medium $(\mathcal M)$& $d_{\up}$, $d_{\dn}$ & $10^3$ \\ 
 Large $(\mathcal L)$ & $d_{\up}d_{\dn}$ & $10^7$ \\ 
 \hline
\end{tabular}
\caption{The computationally relevant numbers that appear in the code. The dimension of the full Hilbert space is an $\mathcal L$-sized number. }\label{Table1}
\end{table}
The technique that we use involves exponentiation of only $\mathcal S$-sized matrices and multiplication of $\mathcal M$-sized matrices. We exploit the sparsity of the matrices and the structure of the fermionic Hamiltonian to make the computation more efficient than a standard sparse matrix approach by a factor of $\sim q_{\dn}$. 

The Hamiltonian can be written as 
\begin{equation}
\hat{H} = \hat{H}_{\up}\otimes\textbf{1} + \textbf{1}\otimes \hat{H}_{\dn} + \hat{H}_{\up \dn}
\end{equation}
where, $\hat{H}_{\sigma}$ is the Hamiltonian of spin-$\sigma$ atoms acting on $\mathcal{H}_{\sigma}$ and $\hat{H}_{\up\dn}$ is the interacting part. $\hat{H}_{\sigma}$ are $\mathcal M$-sized matrices (size  $d_{\sigma}\times d_{\sigma}$) and $\hat{H}_{\up\dn}$ is an $\mathcal L$-sized diagonal matrix. We define an $\mathcal M$-sized matrix $V$ (size  $d_{\up}\times d_{\dn}$) that stores the diagonal entries of $\hat{H}_{\up\dn}$. If $\alpha \equiv \{i_1, \cdots, i_{q_{\up}+1}\}$ and $\beta \equiv \{j_1, \cdots, j_{q_{\dn}}\}$ are basis elements in $\mathcal V_{\up}$ and $\mathcal V_{\dn}$ respectively, we define $V_{\alpha \beta} = U | \{i_1, \cdots, i_{q_{\up}+1}\} \cap  \{j_1, \cdots, j_{q_{\dn}}\}|$. Here, $|\cdot|$ represents the cardinality of a set; the element $V_{\alpha\beta}$ is simply the interaction energy of the many-body state $|\alpha \otimes \beta \rangle$. We reshape the quantum state $\ket{\psi^{\prime\prime}}$, which is an $\mathcal L$-sized vector into an $\mathcal M$-sized matrix $M$ (size  $d_{\up}\times d_{\dn}$) defined as $M_{\alpha\beta} = \langle \alpha \otimes \beta | \psi^{\prime\prime}\rangle$. For $M$ and $V$, the rows correspond to the basis elements of $\mathcal{H}_{\up}$ and the columns correspond to basis elements of $\mathcal{H}_{\dn}$. Thus, all the relevant objects are expressed as $\mathcal M$-sized matrices, i.e., $\hat{H}_{\up}, \hat{H}_{\dn}, V$ and $M$. In terms of these, we may write the Schr\"odinger equation as 
\begin{equation}
\dot{M} = -i \hat{H}_{\up} M - i M\hat{H}_{\dn} - i V\circ M
\end{equation}
Here $\circ$ represents the Hadamard product. See Ref. ~\cite{scherg_observing_2020}. for a derivation. Using Trotter-Suzuki approximation, a time step taking the state $M(t)$ to $M(t+\delta t)$ can be written as
\begin{equation}
M(t+\delta t) = e^{-i\delta t \circ V}\circ e^{-i\delta t \hat{H}_{\up}}M (t)e^{-i\delta t \hat{H}_{\dn}}
\end{equation}
Here, $ e^{-i\delta t \circ V}$ is the element-wise exponentiation of $V$. The above expression consists of three mutually commuting operations on $M(t)$-- Hadamard product with $ e^{-i\delta t \circ V}$, left multiplication with $e^{-i\delta t \hat{H}_{\up}}$ and right multiplication with $e^{-i\delta t \hat{H}_{\dn}}$. The latter two are multiplications of $\mathcal M$-sized matrices. The former is an element-by-element multiplication of $\mathcal M$-sized matrices.  

It remains to compute the unitaries $e^{-i\delta t \hat{H}_{\up}}$ and $e^{-i\delta t \hat{H}_{\dn}}$. Although they are both $\mathcal M$-sized matrices, computing them does not require an exponentiation of $\mathcal M$-sized matrices. We can make use of the structure of Fermionic systems to compute them.  Let $U^{\up}(\delta t)$ and $U^{\dn}(\delta t)$ be the $(2\ell +1)\times (2\ell + 1)$ unitary propagator corresponding to time step $\delta t$ of a single spin-$\up$ atom and a single spin-$\dn$ atom respectively. These two are $\mathcal S$-sized matrices  obtained by exponentiating $\mathcal S$-sized matrices. We can now express $e^{-i\delta t \hat{H}_{\up}}$ and $e^{-i\delta t \hat{H}_{\dn}}$ in terms of these matrices. For instance, if $\beta = \{j_1, \cdots, j_{q_{\dn}}\}$ and $\beta^{\prime} = \{j'_1, \cdots, j'_{q_{\dn}}\}$
\begin{equation}
(e^{-i\delta t \hat{H}_{\dn}})_{\beta\beta^{\prime}} = \sum_{\mu \in S_{q_{\dn}}} \text{sgn}(\mu)U^{\dn}(\delta t)_{j_1 j'_{\mu (1)}}\cdots U^{\dn}(\delta t)_{j_{q_{\dn}} j'_{\mu (q_{\dn})}}
\end{equation}
Thus, the time dynamics can be computed involving only multiplication of $\mathcal M$-sized matrices which is done every timestep and an exponentiation of $\mathcal S$-sized matrices which is done once for all. One can also absorb the above expression into the product $M (t)e^{-i\delta t \hat{H}_{\dn}}$ thereby never having to multiply even $\mathcal M$-sized matrices. Nevertheless, we find that the advantage gained from this additional optimization is negligible. 

In the code, starting at $r=1$, we construct the few body state $\ket{\psi^{\prime \prime}(0)}$ corresponding to the $r$-th spin-$\sigma$ atom. We then use the above described procedure to compute its time evolution. Using the time evolved few-body state $\ket{\psi^{\prime\prime}(t)}$, we compute the occupancy matrix $\Gamma^{\sigma, r}(t)$.  We repeat this process for $r=1, 2, \cdots , N_{\sigma}$ to compute the total occupancy matrix of spin-$\sigma$ atoms, $\Gamma^{\sigma}(t)$. Further, we repeat the whole process for the other spin component to compute its occupancy matrix. 

In order to compute the imbalance time trace, we use Eq.~(\ref{imbalance_calc}) and the above described ED technique.




\bibliography{References}

\begin{thebibliography}{94}%
\makeatletter
\providecommand \@ifxundefined [1]{%
 \@ifx{#1\undefined}
}%
\providecommand \@ifnum [1]{%
 \ifnum #1\expandafter \@firstoftwo
 \else \expandafter \@secondoftwo
 \fi
}%
\providecommand \@ifx [1]{%
 \ifx #1\expandafter \@firstoftwo
 \else \expandafter \@secondoftwo
 \fi
}%
\providecommand \natexlab [1]{#1}%
\providecommand \enquote  [1]{``#1''}%
\providecommand \bibnamefont  [1]{#1}%
\providecommand \bibfnamefont [1]{#1}%
\providecommand \citenamefont [1]{#1}%
\providecommand \href@noop [0]{\@secondoftwo}%
\providecommand \href [0]{\begingroup \@sanitize@url \@href}%
\providecommand \@href[1]{\@@startlink{#1}\@@href}%
\providecommand \@@href[1]{\endgroup#1\@@endlink}%
\providecommand \@sanitize@url [0]{\catcode `\\12\catcode `\$12\catcode
  `\&12\catcode `\#12\catcode `\^12\catcode `\_12\catcode `\%12\relax}%
\providecommand \@@startlink[1]{}%
\providecommand \@@endlink[0]{}%
\providecommand \url  [0]{\begingroup\@sanitize@url \@url }%
\providecommand \@url [1]{\endgroup\@href {#1}{\urlprefix }}%
\providecommand \urlprefix  [0]{URL }%
\providecommand \Eprint [0]{\href }%
\providecommand \doibase [0]{https://doi.org/}%
\providecommand \selectlanguage [0]{\@gobble}%
\providecommand \bibinfo  [0]{\@secondoftwo}%
\providecommand \bibfield  [0]{\@secondoftwo}%
\providecommand \translation [1]{[#1]}%
\providecommand \BibitemOpen [0]{}%
\providecommand \bibitemStop [0]{}%
\providecommand \bibitemNoStop [0]{.\EOS\space}%
\providecommand \EOS [0]{\spacefactor3000\relax}%
\providecommand \BibitemShut  [1]{\csname bibitem#1\endcsname}%
\let\auto@bib@innerbib\@empty
\bibitem [{\citenamefont {Preskill}(2018)}]{Preskill2018quantumcomputingin}%
  \BibitemOpen
  \bibfield  {author} {\bibinfo {author} {\bibfnamefont {J.}~\bibnamefont
  {Preskill}},\ }\bibfield  {title} {\bibinfo {title} {Quantum {C}omputing in
  the {NISQ} era and beyond},\ }\href
  {https://doi.org/10.22331/q-2018-08-06-79} {\bibfield  {journal} {\bibinfo
  {journal} {{Quantum}}\ }\textbf {\bibinfo {volume} {2}},\ \bibinfo {pages}
  {79} (\bibinfo {year} {2018})}\BibitemShut {NoStop}%
\bibitem [{\citenamefont {{Terhal}}\ and\ \citenamefont
  {{DiVincenzo}}(2004)}]{2002quant.ph..5133T}%
  \BibitemOpen
  \bibfield  {author} {\bibinfo {author} {\bibfnamefont {B.~M.}\ \bibnamefont
  {{Terhal}}}\ and\ \bibinfo {author} {\bibfnamefont {D.~P.}\ \bibnamefont
  {{DiVincenzo}}},\ }\bibfield  {title} {\bibinfo {title} {Adaptive quantum
  computation, constant depth quantum circuits and arthur-merlin games},\
  }\href {https://ui.adsabs.harvard.edu/abs/2002quant.ph..5133T} {\bibfield
  {journal} {\bibinfo  {journal} {Quant. Inf. Comp.}\ }\textbf {\bibinfo
  {volume} {4}},\ \bibinfo {pages} {134} (\bibinfo {year} {2004})}\BibitemShut
  {NoStop}%
\bibitem [{\citenamefont {Aaronson}\ and\ \citenamefont
  {Arkhipov}(2010)}]{aaronson2010computational}%
  \BibitemOpen
  \bibfield  {author} {\bibinfo {author} {\bibfnamefont {S.}~\bibnamefont
  {Aaronson}}\ and\ \bibinfo {author} {\bibfnamefont {A.}~\bibnamefont
  {Arkhipov}},\ }\bibfield  {title} {\bibinfo {title} {The computational
  complexity of linear optics},\ }\href {https://arxiv.org/pdf/1011.3245.pdf}
  {\bibfield  {journal} {\bibinfo  {journal} {arXiv:1011.3245}\ } (\bibinfo
  {year} {2010})}\BibitemShut {NoStop}%
\bibitem [{\citenamefont {{Arute, F.}}\ \emph {et~al.}(2019)\citenamefont
  {{Arute, F.}}, \citenamefont {{Arya, K.}}, \citenamefont {{Babbush, R.}},\
  and\ \citenamefont {{et al.}}}]{Sycamore_2019}%
  \BibitemOpen
  \bibfield  {author} {\bibinfo {author} {\bibnamefont {{Arute, F.}}}, \bibinfo
  {author} {\bibnamefont {{Arya, K.}}}, \bibinfo {author} {\bibnamefont
  {{Babbush, R.}}},\ and\ \bibinfo {author} {\bibnamefont {{et al.}}},\
  }\bibfield  {title} {\bibinfo {title} {Quantum supremacy using a programmable
  superconducting processor},\ }\href
  {https://doi.org/10.1038/s41586-019-1666-5} {\bibfield  {journal} {\bibinfo
  {journal} {Nature}\ }\textbf {\bibinfo {volume} {574}},\ \bibinfo {pages}
  {505} (\bibinfo {year} {2019})}\BibitemShut {NoStop}%
\bibitem [{\citenamefont {Zhong}\ \emph {et~al.}(2020)\citenamefont {Zhong},
  \citenamefont {Wang}, \citenamefont {Deng}, \citenamefont {Chen},
  \citenamefont {Peng}, \citenamefont {Luo}, \citenamefont {Qin}, \citenamefont
  {Wu}, \citenamefont {Ding}, \citenamefont {Hu}, \citenamefont {Hu},
  \citenamefont {Yang}, \citenamefont {Zhang}, \citenamefont {Li},
  \citenamefont {Li}, \citenamefont {Jiang}, \citenamefont {Gan}, \citenamefont
  {Yang}, \citenamefont {You}, \citenamefont {Wang}, \citenamefont {Li},
  \citenamefont {Liu}, \citenamefont {Lu},\ and\ \citenamefont
  {Pan}}]{Zhong1460}%
  \BibitemOpen
  \bibfield  {author} {\bibinfo {author} {\bibfnamefont {H.-S.}\ \bibnamefont
  {Zhong}}, \bibinfo {author} {\bibfnamefont {H.}~\bibnamefont {Wang}},
  \bibinfo {author} {\bibfnamefont {Y.-H.}\ \bibnamefont {Deng}}, \bibinfo
  {author} {\bibfnamefont {M.-C.}\ \bibnamefont {Chen}}, \bibinfo {author}
  {\bibfnamefont {L.-C.}\ \bibnamefont {Peng}}, \bibinfo {author}
  {\bibfnamefont {Y.-H.}\ \bibnamefont {Luo}}, \bibinfo {author} {\bibfnamefont
  {J.}~\bibnamefont {Qin}}, \bibinfo {author} {\bibfnamefont {D.}~\bibnamefont
  {Wu}}, \bibinfo {author} {\bibfnamefont {X.}~\bibnamefont {Ding}}, \bibinfo
  {author} {\bibfnamefont {Y.}~\bibnamefont {Hu}}, \bibinfo {author}
  {\bibfnamefont {P.}~\bibnamefont {Hu}}, \bibinfo {author} {\bibfnamefont
  {X.-Y.}\ \bibnamefont {Yang}}, \bibinfo {author} {\bibfnamefont {W.-J.}\
  \bibnamefont {Zhang}}, \bibinfo {author} {\bibfnamefont {H.}~\bibnamefont
  {Li}}, \bibinfo {author} {\bibfnamefont {Y.}~\bibnamefont {Li}}, \bibinfo
  {author} {\bibfnamefont {X.}~\bibnamefont {Jiang}}, \bibinfo {author}
  {\bibfnamefont {L.}~\bibnamefont {Gan}}, \bibinfo {author} {\bibfnamefont
  {G.}~\bibnamefont {Yang}}, \bibinfo {author} {\bibfnamefont {L.}~\bibnamefont
  {You}}, \bibinfo {author} {\bibfnamefont {Z.}~\bibnamefont {Wang}}, \bibinfo
  {author} {\bibfnamefont {L.}~\bibnamefont {Li}}, \bibinfo {author}
  {\bibfnamefont {N.-L.}\ \bibnamefont {Liu}}, \bibinfo {author} {\bibfnamefont
  {C.-Y.}\ \bibnamefont {Lu}},\ and\ \bibinfo {author} {\bibfnamefont {J.-W.}\
  \bibnamefont {Pan}},\ }\bibfield  {title} {\bibinfo {title} {Quantum
  computational advantage using photons},\ }\href
  {https://doi.org/10.1126/science.abe8770} {\bibfield  {journal} {\bibinfo
  {journal} {Science}\ }\textbf {\bibinfo {volume} {370}},\ \bibinfo {pages}
  {1460} (\bibinfo {year} {2020})}\BibitemShut {NoStop}%
\bibitem [{\citenamefont {Gross}\ and\ \citenamefont {Bloch}(2017)}]{Gross995}%
  \BibitemOpen
  \bibfield  {author} {\bibinfo {author} {\bibfnamefont {C.}~\bibnamefont
  {Gross}}\ and\ \bibinfo {author} {\bibfnamefont {I.}~\bibnamefont {Bloch}},\
  }\bibfield  {title} {\bibinfo {title} {Quantum simulations with ultracold
  atoms in optical lattices},\ }\href {https://doi.org/10.1126/science.aal3837}
  {\bibfield  {journal} {\bibinfo  {journal} {Science}\ }\textbf {\bibinfo
  {volume} {357}},\ \bibinfo {pages} {995} (\bibinfo {year}
  {2017})}\BibitemShut {NoStop}%
\bibitem [{\citenamefont {Trotzky}\ \emph {et~al.}(2012)\citenamefont
  {Trotzky}, \citenamefont {Chen}, \citenamefont {Flesch}, \citenamefont
  {McCulloch}, \citenamefont {Schollwöck}, \citenamefont {Eisert},\ and\
  \citenamefont {Bloch}}]{Trotzky_2012}%
  \BibitemOpen
  \bibfield  {author} {\bibinfo {author} {\bibfnamefont {S.}~\bibnamefont
  {Trotzky}}, \bibinfo {author} {\bibfnamefont {Y.-A.}\ \bibnamefont {Chen}},
  \bibinfo {author} {\bibfnamefont {A.}~\bibnamefont {Flesch}}, \bibinfo
  {author} {\bibfnamefont {I.~P.}\ \bibnamefont {McCulloch}}, \bibinfo {author}
  {\bibfnamefont {U.}~\bibnamefont {Schollwöck}}, \bibinfo {author}
  {\bibfnamefont {J.}~\bibnamefont {Eisert}},\ and\ \bibinfo {author}
  {\bibfnamefont {I.}~\bibnamefont {Bloch}},\ }\bibfield  {title} {\bibinfo
  {title} {Probing the relaxation towards equilibrium in an isolated strongly
  correlated one-dimensional bose gas},\ }\href
  {https://doi.org/10.1038/nphys2232} {\bibfield  {journal} {\bibinfo
  {journal} {Nature Physics}\ }\textbf {\bibinfo {volume} {8}},\ \bibinfo
  {pages} {325–330} (\bibinfo {year} {2012})}\BibitemShut {NoStop}%
\bibitem [{\citenamefont {Zhang}\ \emph
  {et~al.}(2017{\natexlab{a}})\citenamefont {Zhang}, \citenamefont {Pagano},
  \citenamefont {Hess}, \citenamefont {Kyprianidis}, \citenamefont {Becker},
  \citenamefont {Kaplan}, \citenamefont {Gorshkov}, \citenamefont {Gong},\ and\
  \citenamefont {Monroe}}]{Zhang_2017}%
  \BibitemOpen
  \bibfield  {author} {\bibinfo {author} {\bibfnamefont {J.}~\bibnamefont
  {Zhang}}, \bibinfo {author} {\bibfnamefont {G.}~\bibnamefont {Pagano}},
  \bibinfo {author} {\bibfnamefont {P.~W.}\ \bibnamefont {Hess}}, \bibinfo
  {author} {\bibfnamefont {A.}~\bibnamefont {Kyprianidis}}, \bibinfo {author}
  {\bibfnamefont {P.}~\bibnamefont {Becker}}, \bibinfo {author} {\bibfnamefont
  {H.}~\bibnamefont {Kaplan}}, \bibinfo {author} {\bibfnamefont {A.~V.}\
  \bibnamefont {Gorshkov}}, \bibinfo {author} {\bibfnamefont {Z.-X.}\
  \bibnamefont {Gong}},\ and\ \bibinfo {author} {\bibfnamefont
  {C.}~\bibnamefont {Monroe}},\ }\bibfield  {title} {\bibinfo {title}
  {Observation of a many-body dynamical phase transition with a 53-qubit
  quantum simulator},\ }\href {https://doi.org/10.1038/nature24654} {\bibfield
  {journal} {\bibinfo  {journal} {Nature}\ }\textbf {\bibinfo {volume} {551}},\
  \bibinfo {pages} {601–604} (\bibinfo {year}
  {2017}{\natexlab{a}})}\BibitemShut {NoStop}%
\bibitem [{\citenamefont {Ebadi}\ \emph {et~al.}(2020)\citenamefont {Ebadi},
  \citenamefont {Wang}, \citenamefont {Levine}, \citenamefont {Keesling},
  \citenamefont {Semeghini}, \citenamefont {Omran}, \citenamefont {Bluvstein},
  \citenamefont {Samajdar}, \citenamefont {Pichler}, \citenamefont {Ho},
  \citenamefont {Choi}, \citenamefont {Sachdev}, \citenamefont {Greiner},
  \citenamefont {Vuletic},\ and\ \citenamefont {Lukin}}]{ebadi2020quantum}%
  \BibitemOpen
  \bibfield  {author} {\bibinfo {author} {\bibfnamefont {S.}~\bibnamefont
  {Ebadi}}, \bibinfo {author} {\bibfnamefont {T.~T.}\ \bibnamefont {Wang}},
  \bibinfo {author} {\bibfnamefont {H.}~\bibnamefont {Levine}}, \bibinfo
  {author} {\bibfnamefont {A.}~\bibnamefont {Keesling}}, \bibinfo {author}
  {\bibfnamefont {G.}~\bibnamefont {Semeghini}}, \bibinfo {author}
  {\bibfnamefont {A.}~\bibnamefont {Omran}}, \bibinfo {author} {\bibfnamefont
  {D.}~\bibnamefont {Bluvstein}}, \bibinfo {author} {\bibfnamefont
  {R.}~\bibnamefont {Samajdar}}, \bibinfo {author} {\bibfnamefont
  {H.}~\bibnamefont {Pichler}}, \bibinfo {author} {\bibfnamefont {W.~W.}\
  \bibnamefont {Ho}}, \bibinfo {author} {\bibfnamefont {S.}~\bibnamefont
  {Choi}}, \bibinfo {author} {\bibfnamefont {S.}~\bibnamefont {Sachdev}},
  \bibinfo {author} {\bibfnamefont {M.}~\bibnamefont {Greiner}}, \bibinfo
  {author} {\bibfnamefont {V.}~\bibnamefont {Vuletic}},\ and\ \bibinfo {author}
  {\bibfnamefont {M.~D.}\ \bibnamefont {Lukin}},\ }\bibfield  {title} {\bibinfo
  {title} {Quantum phases of matter on a 256-atom programmable quantum
  simulator},\ }\href {https://arxiv.org/pdf/2012.12281.pdf} {\bibfield
  {journal} {\bibinfo  {journal} {arXiv:2012.12281}\ } (\bibinfo {year}
  {2020})}\BibitemShut {NoStop}%
\bibitem [{\citenamefont {Scholl}\ \emph {et~al.}(2020)\citenamefont {Scholl},
  \citenamefont {Schuler}, \citenamefont {Williams}, \citenamefont
  {Eberharter}, \citenamefont {Barredo}, \citenamefont {Schymik}, \citenamefont
  {Lienhard}, \citenamefont {Henry}, \citenamefont {Lang}, \citenamefont
  {Lahaye}, \citenamefont {Läuchli},\ and\ \citenamefont
  {Browaeys}}]{scholl2020programmable}%
  \BibitemOpen
  \bibfield  {author} {\bibinfo {author} {\bibfnamefont {P.}~\bibnamefont
  {Scholl}}, \bibinfo {author} {\bibfnamefont {M.}~\bibnamefont {Schuler}},
  \bibinfo {author} {\bibfnamefont {H.~J.}\ \bibnamefont {Williams}}, \bibinfo
  {author} {\bibfnamefont {A.~A.}\ \bibnamefont {Eberharter}}, \bibinfo
  {author} {\bibfnamefont {D.}~\bibnamefont {Barredo}}, \bibinfo {author}
  {\bibfnamefont {K.-N.}\ \bibnamefont {Schymik}}, \bibinfo {author}
  {\bibfnamefont {V.}~\bibnamefont {Lienhard}}, \bibinfo {author}
  {\bibfnamefont {L.-P.}\ \bibnamefont {Henry}}, \bibinfo {author}
  {\bibfnamefont {T.~C.}\ \bibnamefont {Lang}}, \bibinfo {author}
  {\bibfnamefont {T.}~\bibnamefont {Lahaye}}, \bibinfo {author} {\bibfnamefont
  {A.~M.}\ \bibnamefont {Läuchli}},\ and\ \bibinfo {author} {\bibfnamefont
  {A.}~\bibnamefont {Browaeys}},\ }\bibfield  {title} {\bibinfo {title}
  {Programmable quantum simulation of 2d antiferromagnets with hundreds of
  rydberg atoms},\ }\href {https://arxiv.org/pdf/2012.12268.pdf} {\bibfield
  {journal} {\bibinfo  {journal} {arXiv:2012.12268}\ } (\bibinfo {year}
  {2020})}\BibitemShut {NoStop}%
\bibitem [{\citenamefont {Gogolin}\ and\ \citenamefont
  {Eisert}(2016)}]{gogolin_equilibration_2016}%
  \BibitemOpen
  \bibfield  {author} {\bibinfo {author} {\bibfnamefont {C.}~\bibnamefont
  {Gogolin}}\ and\ \bibinfo {author} {\bibfnamefont {J.}~\bibnamefont
  {Eisert}},\ }\bibfield  {title} {\bibinfo {title} {Equilibration,
  thermalisation, and the emergence of statistical mechanics in closed quantum
  systems},\ }\href {https://doi.org/10.1088/0034-4885/79/5/056001} {\bibfield
  {journal} {\bibinfo  {journal} {Rep. Prog. Phys.}\ }\textbf {\bibinfo
  {volume} {79}},\ \bibinfo {pages} {056001} (\bibinfo {year}
  {2016})}\BibitemShut {NoStop}%
\bibitem [{\citenamefont {Altman}\ and\ \citenamefont
  {Vosk}(2015)}]{altman_universal_2015}%
  \BibitemOpen
  \bibfield  {author} {\bibinfo {author} {\bibfnamefont {E.}~\bibnamefont
  {Altman}}\ and\ \bibinfo {author} {\bibfnamefont {R.}~\bibnamefont {Vosk}},\
  }\bibfield  {title} {\bibinfo {title} {Universal {Dynamics} and
  {Renormalization} in {Many}-{Body}-{Localized} {Systems}},\ }\href
  {https://doi.org/10.1146/annurev-conmatphys-031214-014701} {\bibfield
  {journal} {\bibinfo  {journal} {Annu. Rev. Condens. Matter Phys.}\ }\textbf
  {\bibinfo {volume} {6}},\ \bibinfo {pages} {383} (\bibinfo {year}
  {2015})}\BibitemShut {NoStop}%
\bibitem [{\citenamefont {Nandkishore}\ and\ \citenamefont
  {Huse}(2015)}]{nandkishore_many-body_2015}%
  \BibitemOpen
  \bibfield  {author} {\bibinfo {author} {\bibfnamefont {R.}~\bibnamefont
  {Nandkishore}}\ and\ \bibinfo {author} {\bibfnamefont {D.~A.}\ \bibnamefont
  {Huse}},\ }\bibfield  {title} {\bibinfo {title} {Many-{Body} {Localization}
  and {Thermalization} in {Quantum} {Statistical} {Mechanics}},\ }\href
  {https://doi.org/10.1146/annurev-conmatphys-031214-014726} {\bibfield
  {journal} {\bibinfo  {journal} {Annu. Rev. Condens. Matter Phys.}\ }\textbf
  {\bibinfo {volume} {6}},\ \bibinfo {pages} {15} (\bibinfo {year}
  {2015})}\BibitemShut {NoStop}%
\bibitem [{\citenamefont {Abanin}\ \emph {et~al.}(2019)\citenamefont {Abanin},
  \citenamefont {Altman}, \citenamefont {Bloch},\ and\ \citenamefont
  {Serbyn}}]{abanin_colloquium_2019}%
  \BibitemOpen
  \bibfield  {author} {\bibinfo {author} {\bibfnamefont {D.~A.}\ \bibnamefont
  {Abanin}}, \bibinfo {author} {\bibfnamefont {E.}~\bibnamefont {Altman}},
  \bibinfo {author} {\bibfnamefont {I.}~\bibnamefont {Bloch}},\ and\ \bibinfo
  {author} {\bibfnamefont {M.}~\bibnamefont {Serbyn}},\ }\bibfield  {title}
  {\bibinfo {title} {Colloquium : {Many}-body localization, thermalization, and
  entanglement},\ }\href {https://doi.org/10.1103/RevModPhys.91.021001}
  {\bibfield  {journal} {\bibinfo  {journal} {Rev. Mod. Phys.}\ }\textbf
  {\bibinfo {volume} {91}},\ \bibinfo {pages} {021001} (\bibinfo {year}
  {2019})}\BibitemShut {NoStop}%
\bibitem [{\citenamefont {Schreiber}\ \emph {et~al.}(2015)\citenamefont
  {Schreiber}, \citenamefont {Hodgman}, \citenamefont {Bordia}, \citenamefont
  {Luschen}, \citenamefont {Fischer}, \citenamefont {Vosk}, \citenamefont
  {Altman}, \citenamefont {Schneider},\ and\ \citenamefont
  {Bloch}}]{Schreiber_2015}%
  \BibitemOpen
  \bibfield  {author} {\bibinfo {author} {\bibfnamefont {M.}~\bibnamefont
  {Schreiber}}, \bibinfo {author} {\bibfnamefont {S.~S.}\ \bibnamefont
  {Hodgman}}, \bibinfo {author} {\bibfnamefont {P.}~\bibnamefont {Bordia}},
  \bibinfo {author} {\bibfnamefont {H.~P.}\ \bibnamefont {Luschen}}, \bibinfo
  {author} {\bibfnamefont {M.~H.}\ \bibnamefont {Fischer}}, \bibinfo {author}
  {\bibfnamefont {R.}~\bibnamefont {Vosk}}, \bibinfo {author} {\bibfnamefont
  {E.}~\bibnamefont {Altman}}, \bibinfo {author} {\bibfnamefont
  {U.}~\bibnamefont {Schneider}},\ and\ \bibinfo {author} {\bibfnamefont
  {I.}~\bibnamefont {Bloch}},\ }\bibfield  {title} {\bibinfo {title}
  {Observation of many-body localization of interacting fermions in a
  quasirandom optical lattice},\ }\href
  {https://doi.org/10.1126/science.aaa7432} {\bibfield  {journal} {\bibinfo
  {journal} {Science}\ }\textbf {\bibinfo {volume} {349}},\ \bibinfo {pages}
  {842–845} (\bibinfo {year} {2015})}\BibitemShut {NoStop}%
\bibitem [{\citenamefont {Smith}\ \emph {et~al.}(2016)\citenamefont {Smith},
  \citenamefont {Lee}, \citenamefont {Richerme}, \citenamefont {Neyenhuis},
  \citenamefont {Hess}, \citenamefont {Hauke}, \citenamefont {Heyl},
  \citenamefont {Huse},\ and\ \citenamefont {Monroe}}]{smith_many-body_2016}%
  \BibitemOpen
  \bibfield  {author} {\bibinfo {author} {\bibfnamefont {J.}~\bibnamefont
  {Smith}}, \bibinfo {author} {\bibfnamefont {A.}~\bibnamefont {Lee}}, \bibinfo
  {author} {\bibfnamefont {P.}~\bibnamefont {Richerme}}, \bibinfo {author}
  {\bibfnamefont {B.}~\bibnamefont {Neyenhuis}}, \bibinfo {author}
  {\bibfnamefont {P.~W.}\ \bibnamefont {Hess}}, \bibinfo {author}
  {\bibfnamefont {P.}~\bibnamefont {Hauke}}, \bibinfo {author} {\bibfnamefont
  {M.}~\bibnamefont {Heyl}}, \bibinfo {author} {\bibfnamefont {D.~A.}\
  \bibnamefont {Huse}},\ and\ \bibinfo {author} {\bibfnamefont
  {C.}~\bibnamefont {Monroe}},\ }\bibfield  {title} {\bibinfo {title}
  {Many-body localization in a quantum simulator with programmable random
  disorder},\ }\href {https://doi.org/10.1038/nphys3783} {\bibfield  {journal}
  {\bibinfo  {journal} {Nature Physics}\ }\textbf {\bibinfo {volume} {12}},\
  \bibinfo {pages} {907} (\bibinfo {year} {2016})}\BibitemShut {NoStop}%
\bibitem [{\citenamefont {Choi}\ \emph {et~al.}(2016)\citenamefont {Choi},
  \citenamefont {Hild}, \citenamefont {Zeiher}, \citenamefont {Schauss},
  \citenamefont {Rubio-Abadal}, \citenamefont {Yefsah}, \citenamefont
  {Khemani}, \citenamefont {Huse}, \citenamefont {Bloch},\ and\ \citenamefont
  {Gross}}]{Choi_2016}%
  \BibitemOpen
  \bibfield  {author} {\bibinfo {author} {\bibfnamefont {J.-y.}\ \bibnamefont
  {Choi}}, \bibinfo {author} {\bibfnamefont {S.}~\bibnamefont {Hild}}, \bibinfo
  {author} {\bibfnamefont {J.}~\bibnamefont {Zeiher}}, \bibinfo {author}
  {\bibfnamefont {P.}~\bibnamefont {Schauss}}, \bibinfo {author} {\bibfnamefont
  {A.}~\bibnamefont {Rubio-Abadal}}, \bibinfo {author} {\bibfnamefont
  {T.}~\bibnamefont {Yefsah}}, \bibinfo {author} {\bibfnamefont
  {V.}~\bibnamefont {Khemani}}, \bibinfo {author} {\bibfnamefont {D.~A.}\
  \bibnamefont {Huse}}, \bibinfo {author} {\bibfnamefont {I.}~\bibnamefont
  {Bloch}},\ and\ \bibinfo {author} {\bibfnamefont {C.}~\bibnamefont {Gross}},\
  }\bibfield  {title} {\bibinfo {title} {Exploring the many-body localization
  transition in two dimensions},\ }\href
  {https://doi.org/10.1126/science.aaf8834} {\bibfield  {journal} {\bibinfo
  {journal} {Science}\ }\textbf {\bibinfo {volume} {352}},\ \bibinfo {pages}
  {1547–1552} (\bibinfo {year} {2016})}\BibitemShut {NoStop}%
\bibitem [{\citenamefont {Roushan}\ \emph {et~al.}(2017)\citenamefont
  {Roushan}, \citenamefont {Neill}, \citenamefont {Tangpanitanon},
  \citenamefont {Bastidas}, \citenamefont {Megrant}, \citenamefont {Barends},
  \citenamefont {Chen}, \citenamefont {Chen}, \citenamefont {Chiaro},
  \citenamefont {Dunsworth}, \citenamefont {Fowler}, \citenamefont {Foxen},
  \citenamefont {Giustina}, \citenamefont {Jeffrey}, \citenamefont {Kelly},
  \citenamefont {Lucero}, \citenamefont {Mutus}, \citenamefont {Neeley},
  \citenamefont {Quintana}, \citenamefont {Sank}, \citenamefont {Vainsencher},
  \citenamefont {Wenner}, \citenamefont {White}, \citenamefont {Neven},
  \citenamefont {Angelakis},\ and\ \citenamefont
  {Martinis}}]{roushan_spectroscopic_2017}%
  \BibitemOpen
  \bibfield  {author} {\bibinfo {author} {\bibfnamefont {P.}~\bibnamefont
  {Roushan}}, \bibinfo {author} {\bibfnamefont {C.}~\bibnamefont {Neill}},
  \bibinfo {author} {\bibfnamefont {J.}~\bibnamefont {Tangpanitanon}}, \bibinfo
  {author} {\bibfnamefont {V.~M.}\ \bibnamefont {Bastidas}}, \bibinfo {author}
  {\bibfnamefont {A.}~\bibnamefont {Megrant}}, \bibinfo {author} {\bibfnamefont
  {R.}~\bibnamefont {Barends}}, \bibinfo {author} {\bibfnamefont
  {Y.}~\bibnamefont {Chen}}, \bibinfo {author} {\bibfnamefont {Z.}~\bibnamefont
  {Chen}}, \bibinfo {author} {\bibfnamefont {B.}~\bibnamefont {Chiaro}},
  \bibinfo {author} {\bibfnamefont {A.}~\bibnamefont {Dunsworth}}, \bibinfo
  {author} {\bibfnamefont {A.}~\bibnamefont {Fowler}}, \bibinfo {author}
  {\bibfnamefont {B.}~\bibnamefont {Foxen}}, \bibinfo {author} {\bibfnamefont
  {M.}~\bibnamefont {Giustina}}, \bibinfo {author} {\bibfnamefont
  {E.}~\bibnamefont {Jeffrey}}, \bibinfo {author} {\bibfnamefont
  {J.}~\bibnamefont {Kelly}}, \bibinfo {author} {\bibfnamefont
  {E.}~\bibnamefont {Lucero}}, \bibinfo {author} {\bibfnamefont
  {J.}~\bibnamefont {Mutus}}, \bibinfo {author} {\bibfnamefont
  {M.}~\bibnamefont {Neeley}}, \bibinfo {author} {\bibfnamefont
  {C.}~\bibnamefont {Quintana}}, \bibinfo {author} {\bibfnamefont
  {D.}~\bibnamefont {Sank}}, \bibinfo {author} {\bibfnamefont {A.}~\bibnamefont
  {Vainsencher}}, \bibinfo {author} {\bibfnamefont {J.}~\bibnamefont {Wenner}},
  \bibinfo {author} {\bibfnamefont {T.}~\bibnamefont {White}}, \bibinfo
  {author} {\bibfnamefont {H.}~\bibnamefont {Neven}}, \bibinfo {author}
  {\bibfnamefont {D.~G.}\ \bibnamefont {Angelakis}},\ and\ \bibinfo {author}
  {\bibfnamefont {J.}~\bibnamefont {Martinis}},\ }\bibfield  {title} {\bibinfo
  {title} {Spectroscopic signatures of localization with interacting photons in
  superconducting qubits},\ }\href {https://doi.org/10.1126/science.aao1401}
  {\bibfield  {journal} {\bibinfo  {journal} {Science}\ }\textbf {\bibinfo
  {volume} {358}},\ \bibinfo {pages} {1175} (\bibinfo {year}
  {2017})}\BibitemShut {NoStop}%
\bibitem [{\citenamefont {Yao}\ \emph {et~al.}(2016)\citenamefont {Yao},
  \citenamefont {Laumann}, \citenamefont {Cirac}, \citenamefont {Lukin},\ and\
  \citenamefont {Moore}}]{PhysRevLett.117.240601}%
  \BibitemOpen
  \bibfield  {author} {\bibinfo {author} {\bibfnamefont {N.~Y.}\ \bibnamefont
  {Yao}}, \bibinfo {author} {\bibfnamefont {C.~R.}\ \bibnamefont {Laumann}},
  \bibinfo {author} {\bibfnamefont {J.~I.}\ \bibnamefont {Cirac}}, \bibinfo
  {author} {\bibfnamefont {M.~D.}\ \bibnamefont {Lukin}},\ and\ \bibinfo
  {author} {\bibfnamefont {J.~E.}\ \bibnamefont {Moore}},\ }\bibfield  {title}
  {\bibinfo {title} {Quasi-many-body localization in translation-invariant
  systems},\ }\href {https://doi.org/10.1103/PhysRevLett.117.240601} {\bibfield
   {journal} {\bibinfo  {journal} {Phys. Rev. Lett.}\ }\textbf {\bibinfo
  {volume} {117}},\ \bibinfo {pages} {240601} (\bibinfo {year}
  {2016})}\BibitemShut {NoStop}%
\bibitem [{\citenamefont {Scherg}\ \emph {et~al.}(2021)\citenamefont {Scherg},
  \citenamefont {Kohlert}, \citenamefont {Sala}, \citenamefont {Pollmann},
  \citenamefont {Hebbe~Madhusudhana}, \citenamefont {Bloch},\ and\
  \citenamefont {Aidelsburger}}]{scherg_observing_2020}%
  \BibitemOpen
  \bibfield  {author} {\bibinfo {author} {\bibfnamefont {S.}~\bibnamefont
  {Scherg}}, \bibinfo {author} {\bibfnamefont {T.}~\bibnamefont {Kohlert}},
  \bibinfo {author} {\bibfnamefont {P.}~\bibnamefont {Sala}}, \bibinfo {author}
  {\bibfnamefont {F.}~\bibnamefont {Pollmann}}, \bibinfo {author}
  {\bibfnamefont {B.}~\bibnamefont {Hebbe~Madhusudhana}}, \bibinfo {author}
  {\bibfnamefont {I.}~\bibnamefont {Bloch}},\ and\ \bibinfo {author}
  {\bibfnamefont {M.}~\bibnamefont {Aidelsburger}},\ }\bibfield  {title}
  {\bibinfo {title} {Observing non-ergodicity due to kinetic constraints in
  tilted fermi-hubbard chains},\ }\bibfield  {journal} {\bibinfo  {journal}
  {Nature Communications}\ }\textbf {\bibinfo {volume} {12}},\ \href
  {https://doi.org/10.1038/s41467-021-24726-0} {10.1038/s41467-021-24726-0}
  (\bibinfo {year} {2021})\BibitemShut {NoStop}%
\bibitem [{\citenamefont {Guardado-Sanchez}\ \emph {et~al.}(2020)\citenamefont
  {Guardado-Sanchez}, \citenamefont {Morningstar}, \citenamefont {Spar},
  \citenamefont {Brown}, \citenamefont {Huse},\ and\ \citenamefont
  {Bakr}}]{PhysRevX.10.011042}%
  \BibitemOpen
  \bibfield  {author} {\bibinfo {author} {\bibfnamefont {E.}~\bibnamefont
  {Guardado-Sanchez}}, \bibinfo {author} {\bibfnamefont {A.}~\bibnamefont
  {Morningstar}}, \bibinfo {author} {\bibfnamefont {B.~M.}\ \bibnamefont
  {Spar}}, \bibinfo {author} {\bibfnamefont {P.~T.}\ \bibnamefont {Brown}},
  \bibinfo {author} {\bibfnamefont {D.~A.}\ \bibnamefont {Huse}},\ and\
  \bibinfo {author} {\bibfnamefont {W.~S.}\ \bibnamefont {Bakr}},\ }\bibfield
  {title} {\bibinfo {title} {Subdiffusion and heat transport in a tilted
  two-dimensional fermi-hubbard system},\ }\href
  {https://doi.org/10.1103/PhysRevX.10.011042} {\bibfield  {journal} {\bibinfo
  {journal} {Phys. Rev. X}\ }\textbf {\bibinfo {volume} {10}},\ \bibinfo
  {pages} {011042} (\bibinfo {year} {2020})}\BibitemShut {NoStop}%
\bibitem [{\citenamefont {Guo}\ \emph {et~al.}(2020)\citenamefont {Guo},
  \citenamefont {Cheng}, \citenamefont {Li}, \citenamefont {Xu}, \citenamefont
  {Zhang}, \citenamefont {Wang}, \citenamefont {Song}, \citenamefont {Liu},
  \citenamefont {Ren}, \citenamefont {Dong}, \citenamefont {Mondaini},\ and\
  \citenamefont {Wang}}]{guo2020stark}%
  \BibitemOpen
  \bibfield  {author} {\bibinfo {author} {\bibfnamefont {Q.}~\bibnamefont
  {Guo}}, \bibinfo {author} {\bibfnamefont {C.}~\bibnamefont {Cheng}}, \bibinfo
  {author} {\bibfnamefont {H.}~\bibnamefont {Li}}, \bibinfo {author}
  {\bibfnamefont {S.}~\bibnamefont {Xu}}, \bibinfo {author} {\bibfnamefont
  {P.}~\bibnamefont {Zhang}}, \bibinfo {author} {\bibfnamefont
  {Z.}~\bibnamefont {Wang}}, \bibinfo {author} {\bibfnamefont {C.}~\bibnamefont
  {Song}}, \bibinfo {author} {\bibfnamefont {W.}~\bibnamefont {Liu}}, \bibinfo
  {author} {\bibfnamefont {W.}~\bibnamefont {Ren}}, \bibinfo {author}
  {\bibfnamefont {H.}~\bibnamefont {Dong}}, \bibinfo {author} {\bibfnamefont
  {R.}~\bibnamefont {Mondaini}},\ and\ \bibinfo {author} {\bibfnamefont
  {H.}~\bibnamefont {Wang}},\ }\bibfield  {title} {\bibinfo {title} {Stark
  many-body localization on a superconducting quantum processor},\ }\href
  {https://arxiv.org/pdf/2011.13895.pdf} {\bibfield  {journal} {\bibinfo
  {journal} {arXiv:2011.13895}\ } (\bibinfo {year} {2020})}\BibitemShut
  {NoStop}%
\bibitem [{\citenamefont {Morong}\ \emph {et~al.}(2021)\citenamefont {Morong},
  \citenamefont {Liu}, \citenamefont {Becker}, \citenamefont {Collins},
  \citenamefont {Feng}, \citenamefont {Kyprianidis}, \citenamefont {Pagano},
  \citenamefont {You}, \citenamefont {Gorshkov},\ and\ \citenamefont
  {Monroe}}]{morong_observation_2021}%
  \BibitemOpen
  \bibfield  {author} {\bibinfo {author} {\bibfnamefont {W.}~\bibnamefont
  {Morong}}, \bibinfo {author} {\bibfnamefont {F.}~\bibnamefont {Liu}},
  \bibinfo {author} {\bibfnamefont {P.}~\bibnamefont {Becker}}, \bibinfo
  {author} {\bibfnamefont {K.~S.}\ \bibnamefont {Collins}}, \bibinfo {author}
  {\bibfnamefont {L.}~\bibnamefont {Feng}}, \bibinfo {author} {\bibfnamefont
  {A.}~\bibnamefont {Kyprianidis}}, \bibinfo {author} {\bibfnamefont
  {G.}~\bibnamefont {Pagano}}, \bibinfo {author} {\bibfnamefont
  {T.}~\bibnamefont {You}}, \bibinfo {author} {\bibfnamefont {A.~V.}\
  \bibnamefont {Gorshkov}},\ and\ \bibinfo {author} {\bibfnamefont
  {C.}~\bibnamefont {Monroe}},\ }\bibfield  {title} {\bibinfo {title}
  {Observation of {Stark} many-body localization without disorder},\ }\href
  {http://arxiv.org/abs/2102.07250} {\bibfield  {journal} {\bibinfo  {journal}
  {arXiv:2102.07250}\ } (\bibinfo {year} {2021})}\BibitemShut {NoStop}%
\bibitem [{\citenamefont {Moudgalya}\ \emph {et~al.}(2019)\citenamefont
  {Moudgalya}, \citenamefont {Prem}, \citenamefont {Nandkishore}, \citenamefont
  {Regnault},\ and\ \citenamefont {Bernevig}}]{moudgalya_thermalization_2019}%
  \BibitemOpen
  \bibfield  {author} {\bibinfo {author} {\bibfnamefont {S.}~\bibnamefont
  {Moudgalya}}, \bibinfo {author} {\bibfnamefont {A.}~\bibnamefont {Prem}},
  \bibinfo {author} {\bibfnamefont {R.}~\bibnamefont {Nandkishore}}, \bibinfo
  {author} {\bibfnamefont {N.}~\bibnamefont {Regnault}},\ and\ \bibinfo
  {author} {\bibfnamefont {B.~A.}\ \bibnamefont {Bernevig}},\ }\bibfield
  {title} {\bibinfo {title} {Thermalization and its absence within {Krylov}
  subspaces of a constrained {Hamiltonian}},\ }\href
  {http://arxiv.org/abs/1910.14048} {\bibfield  {journal} {\bibinfo  {journal}
  {arXiv:1910.14048}\ } (\bibinfo {year} {2019})}\BibitemShut {NoStop}%
\bibitem [{\citenamefont {Khemani}\ \emph {et~al.}(2020)\citenamefont
  {Khemani}, \citenamefont {Hermele},\ and\ \citenamefont
  {Nandkishore}}]{khemani_localization_2020}%
  \BibitemOpen
  \bibfield  {author} {\bibinfo {author} {\bibfnamefont {V.}~\bibnamefont
  {Khemani}}, \bibinfo {author} {\bibfnamefont {M.}~\bibnamefont {Hermele}},\
  and\ \bibinfo {author} {\bibfnamefont {R.}~\bibnamefont {Nandkishore}},\
  }\bibfield  {title} {\bibinfo {title} {Localization from {Hilbert} space
  shattering: {From} theory to physical realizations},\ }\href
  {https://doi.org/10.1103/PhysRevB.101.174204} {\bibfield  {journal} {\bibinfo
   {journal} {Phys. Rev. B}\ }\textbf {\bibinfo {volume} {101}},\ \bibinfo
  {pages} {174204} (\bibinfo {year} {2020})}\BibitemShut {NoStop}%
\bibitem [{\citenamefont {Doggen}\ \emph {et~al.}(2020)\citenamefont {Doggen},
  \citenamefont {Gornyi},\ and\ \citenamefont {Polyakov}}]{doggen2020stark}%
  \BibitemOpen
  \bibfield  {author} {\bibinfo {author} {\bibfnamefont {E.~V.~H.}\
  \bibnamefont {Doggen}}, \bibinfo {author} {\bibfnamefont {I.~V.}\
  \bibnamefont {Gornyi}},\ and\ \bibinfo {author} {\bibfnamefont {D.~G.}\
  \bibnamefont {Polyakov}},\ }\bibfield  {title} {\bibinfo {title} {Stark
  many-body localization: Evidence for hilbert-space shattering},\ }\href
  {https://arxiv.org/pdf/2012.13722.pdf} {\bibfield  {journal} {\bibinfo
  {journal} {arXiv:2012.13722}\ } (\bibinfo {year} {2020})}\BibitemShut
  {NoStop}%
\bibitem [{\citenamefont {Serbyn}\ \emph {et~al.}(2020)\citenamefont {Serbyn},
  \citenamefont {Abanin},\ and\ \citenamefont
  {Papi{\'c}}}]{serbyn_quantum_2020}%
  \BibitemOpen
  \bibfield  {author} {\bibinfo {author} {\bibfnamefont {M.}~\bibnamefont
  {Serbyn}}, \bibinfo {author} {\bibfnamefont {D.~A.}\ \bibnamefont {Abanin}},\
  and\ \bibinfo {author} {\bibfnamefont {Z.}~\bibnamefont {Papi{\'c}}},\
  }\bibfield  {title} {\bibinfo {title} {Quantum {Many}-{Body} {Scars} and
  {Weak} {Breaking} of {Ergodicity}},\ }\href {http://arxiv.org/abs/2011.09486}
  {\bibfield  {journal} {\bibinfo  {journal} {arXiv:2011.09486}\ } (\bibinfo
  {year} {2020})}\BibitemShut {NoStop}%
\bibitem [{\citenamefont {Khemani}\ and\ \citenamefont
  {Nandkishore}(2020)}]{khemani_local_2020}%
  \BibitemOpen
  \bibfield  {author} {\bibinfo {author} {\bibfnamefont {V.}~\bibnamefont
  {Khemani}}\ and\ \bibinfo {author} {\bibfnamefont {R.}~\bibnamefont
  {Nandkishore}},\ }\bibfield  {title} {\bibinfo {title} {Local constraints can
  globally shatter {Hilbert} space: a new route to quantum information
  protection},\ }\href {https://doi.org/10.1103/PhysRevB.101.174204} {\bibfield
   {journal} {\bibinfo  {journal} {Phys. Rev. B}\ }\textbf {\bibinfo {volume}
  {101}},\ \bibinfo {pages} {174204} (\bibinfo {year} {2020})}\BibitemShut
  {NoStop}%
\bibitem [{\citenamefont {Sala}\ \emph {et~al.}(2020)\citenamefont {Sala},
  \citenamefont {Rakovszky}, \citenamefont {Verresen}, \citenamefont {Knap},\
  and\ \citenamefont {Pollmann}}]{sala_ergodicity_2020}%
  \BibitemOpen
  \bibfield  {author} {\bibinfo {author} {\bibfnamefont {P.}~\bibnamefont
  {Sala}}, \bibinfo {author} {\bibfnamefont {T.}~\bibnamefont {Rakovszky}},
  \bibinfo {author} {\bibfnamefont {R.}~\bibnamefont {Verresen}}, \bibinfo
  {author} {\bibfnamefont {M.}~\bibnamefont {Knap}},\ and\ \bibinfo {author}
  {\bibfnamefont {F.}~\bibnamefont {Pollmann}},\ }\bibfield  {title} {\bibinfo
  {title} {Ergodicity {Breaking} {Arising} from {Hilbert} {Space}
  {Fragmentation} in {Dipole}-{Conserving} {Hamiltonians}},\ }\href
  {https://doi.org/10.1103/PhysRevX.10.011047} {\bibfield  {journal} {\bibinfo
  {journal} {Phys. Rev. X}\ }\textbf {\bibinfo {volume} {10}},\ \bibinfo
  {pages} {011047} (\bibinfo {year} {2020})}\BibitemShut {NoStop}%
\bibitem [{\citenamefont {Bernien}\ \emph {et~al.}(2017)\citenamefont
  {Bernien}, \citenamefont {Schwartz}, \citenamefont {Keesling}, \citenamefont
  {Levine}, \citenamefont {Omran}, \citenamefont {Pichler}, \citenamefont
  {Choi}, \citenamefont {Zibrov}, \citenamefont {Endres}, \citenamefont
  {Greiner}, \citenamefont {Vuleti{\'c}},\ and\ \citenamefont
  {Lukin}}]{bernien_probing_2017}%
  \BibitemOpen
  \bibfield  {author} {\bibinfo {author} {\bibfnamefont {H.}~\bibnamefont
  {Bernien}}, \bibinfo {author} {\bibfnamefont {S.}~\bibnamefont {Schwartz}},
  \bibinfo {author} {\bibfnamefont {A.}~\bibnamefont {Keesling}}, \bibinfo
  {author} {\bibfnamefont {H.}~\bibnamefont {Levine}}, \bibinfo {author}
  {\bibfnamefont {A.}~\bibnamefont {Omran}}, \bibinfo {author} {\bibfnamefont
  {H.}~\bibnamefont {Pichler}}, \bibinfo {author} {\bibfnamefont
  {S.}~\bibnamefont {Choi}}, \bibinfo {author} {\bibfnamefont {A.~S.}\
  \bibnamefont {Zibrov}}, \bibinfo {author} {\bibfnamefont {M.}~\bibnamefont
  {Endres}}, \bibinfo {author} {\bibfnamefont {M.}~\bibnamefont {Greiner}},
  \bibinfo {author} {\bibfnamefont {V.}~\bibnamefont {Vuleti{\'c}}},\ and\
  \bibinfo {author} {\bibfnamefont {M.~D.}\ \bibnamefont {Lukin}},\ }\bibfield
  {title} {\bibinfo {title} {Probing many-body dynamics on a 51-atom quantum
  simulator},\ }\href {https://doi.org/10.1038/nature24622} {\bibfield
  {journal} {\bibinfo  {journal} {Nature}\ }\textbf {\bibinfo {volume} {551}},\
  \bibinfo {pages} {579} (\bibinfo {year} {2017})}\BibitemShut {NoStop}%
\bibitem [{\citenamefont {Bluvstein}\ \emph {et~al.}(2021)\citenamefont
  {Bluvstein}, \citenamefont {Omran}, \citenamefont {Levine}, \citenamefont
  {Keesling}, \citenamefont {Semeghini}, \citenamefont {Ebadi}, \citenamefont
  {Wang}, \citenamefont {Michailidis}, \citenamefont {Maskara}, \citenamefont
  {Ho}, \citenamefont {Choi}, \citenamefont {Serbyn}, \citenamefont {Greiner},
  \citenamefont {Vuleti{\'c}},\ and\ \citenamefont
  {Lukin}}]{bluvstein_controlling_2021}%
  \BibitemOpen
  \bibfield  {author} {\bibinfo {author} {\bibfnamefont {D.}~\bibnamefont
  {Bluvstein}}, \bibinfo {author} {\bibfnamefont {A.}~\bibnamefont {Omran}},
  \bibinfo {author} {\bibfnamefont {H.}~\bibnamefont {Levine}}, \bibinfo
  {author} {\bibfnamefont {A.}~\bibnamefont {Keesling}}, \bibinfo {author}
  {\bibfnamefont {G.}~\bibnamefont {Semeghini}}, \bibinfo {author}
  {\bibfnamefont {S.}~\bibnamefont {Ebadi}}, \bibinfo {author} {\bibfnamefont
  {T.~T.}\ \bibnamefont {Wang}}, \bibinfo {author} {\bibfnamefont {A.~A.}\
  \bibnamefont {Michailidis}}, \bibinfo {author} {\bibfnamefont
  {N.}~\bibnamefont {Maskara}}, \bibinfo {author} {\bibfnamefont {W.~W.}\
  \bibnamefont {Ho}}, \bibinfo {author} {\bibfnamefont {S.}~\bibnamefont
  {Choi}}, \bibinfo {author} {\bibfnamefont {M.}~\bibnamefont {Serbyn}},
  \bibinfo {author} {\bibfnamefont {M.}~\bibnamefont {Greiner}}, \bibinfo
  {author} {\bibfnamefont {V.}~\bibnamefont {Vuleti{\'c}}},\ and\ \bibinfo
  {author} {\bibfnamefont {M.~D.}\ \bibnamefont {Lukin}},\ }\bibfield  {title}
  {\bibinfo {title} {Controlling quantum many-body dynamics in driven rydberg
  atom arrays},\ }\href {https://doi.org/10.1126/science.abg2530} {\bibfield
  {journal} {\bibinfo  {journal} {Science}\ }\textbf {\bibinfo {volume}
  {371}},\ \bibinfo {pages} {1355} (\bibinfo {year} {2021})}\BibitemShut
  {NoStop}%
\bibitem [{\citenamefont {De~Roeck}\ and\ \citenamefont
  {Verreet}(2019)}]{de_roeck_very_2019}%
  \BibitemOpen
  \bibfield  {author} {\bibinfo {author} {\bibfnamefont {W.}~\bibnamefont
  {De~Roeck}}\ and\ \bibinfo {author} {\bibfnamefont {V.}~\bibnamefont
  {Verreet}},\ }\bibfield  {title} {\bibinfo {title} {Very slow heating for
  weakly driven quantum many-body systems},\ }\href
  {http://arxiv.org/abs/1911.01998} {\bibfield  {journal} {\bibinfo  {journal}
  {arXiv:1911.01998}\ } (\bibinfo {year} {2019})}\BibitemShut {NoStop}%
\bibitem [{\citenamefont {Abanin}\ \emph {et~al.}(2017)\citenamefont {Abanin},
  \citenamefont {De~Roeck}, \citenamefont {Ho},\ and\ \citenamefont
  {Huveneers}}]{abanin_rigorous_2017}%
  \BibitemOpen
  \bibfield  {author} {\bibinfo {author} {\bibfnamefont {D.}~\bibnamefont
  {Abanin}}, \bibinfo {author} {\bibfnamefont {W.}~\bibnamefont {De~Roeck}},
  \bibinfo {author} {\bibfnamefont {W.~W.}\ \bibnamefont {Ho}},\ and\ \bibinfo
  {author} {\bibfnamefont {F.}~\bibnamefont {Huveneers}},\ }\bibfield  {title}
  {\bibinfo {title} {A {Rigorous} {Theory} of {Many}-{Body} {Prethermalization}
  for {Periodically} {Driven} and {Closed} {Quantum} {Systems}},\ }\href
  {https://doi.org/10.1007/s00220-017-2930-x} {\bibfield  {journal} {\bibinfo
  {journal} {Commun. Math. Phys.}\ }\textbf {\bibinfo {volume} {354}},\
  \bibinfo {pages} {809} (\bibinfo {year} {2017})}\BibitemShut {NoStop}%
\bibitem [{\citenamefont {Gromov}\ \emph {et~al.}(2020)\citenamefont {Gromov},
  \citenamefont {Lucas},\ and\ \citenamefont
  {Nandkishore}}]{gromov_fracton_2020}%
  \BibitemOpen
  \bibfield  {author} {\bibinfo {author} {\bibfnamefont {A.}~\bibnamefont
  {Gromov}}, \bibinfo {author} {\bibfnamefont {A.}~\bibnamefont {Lucas}},\ and\
  \bibinfo {author} {\bibfnamefont {R.~M.}\ \bibnamefont {Nandkishore}},\
  }\bibfield  {title} {\bibinfo {title} {Fracton hydrodynamics},\ }\href
  {http://arxiv.org/abs/2003.09429} {\bibfield  {journal} {\bibinfo  {journal}
  {arXiv:2003.09429}\ } (\bibinfo {year} {2020})}\BibitemShut {NoStop}%
\bibitem [{\citenamefont {Rubio-Abadal}\ \emph {et~al.}(2020)\citenamefont
  {Rubio-Abadal}, \citenamefont {Ippoliti}, \citenamefont {Hollerith},
  \citenamefont {Wei}, \citenamefont {Rui}, \citenamefont {Sondhi},
  \citenamefont {Khemani}, \citenamefont {Gross},\ and\ \citenamefont
  {Bloch}}]{PhysRevX.10.021044}%
  \BibitemOpen
  \bibfield  {author} {\bibinfo {author} {\bibfnamefont {A.}~\bibnamefont
  {Rubio-Abadal}}, \bibinfo {author} {\bibfnamefont {M.}~\bibnamefont
  {Ippoliti}}, \bibinfo {author} {\bibfnamefont {S.}~\bibnamefont {Hollerith}},
  \bibinfo {author} {\bibfnamefont {D.}~\bibnamefont {Wei}}, \bibinfo {author}
  {\bibfnamefont {J.}~\bibnamefont {Rui}}, \bibinfo {author} {\bibfnamefont
  {S.~L.}\ \bibnamefont {Sondhi}}, \bibinfo {author} {\bibfnamefont
  {V.}~\bibnamefont {Khemani}}, \bibinfo {author} {\bibfnamefont
  {C.}~\bibnamefont {Gross}},\ and\ \bibinfo {author} {\bibfnamefont
  {I.}~\bibnamefont {Bloch}},\ }\bibfield  {title} {\bibinfo {title} {Floquet
  prethermalization in a bose-hubbard system},\ }\href
  {https://doi.org/10.1103/PhysRevX.10.021044} {\bibfield  {journal} {\bibinfo
  {journal} {Phys. Rev. X}\ }\textbf {\bibinfo {volume} {10}},\ \bibinfo
  {pages} {021044} (\bibinfo {year} {2020})}\BibitemShut {NoStop}%
\bibitem [{\citenamefont {Yao}\ \emph {et~al.}(2017)\citenamefont {Yao},
  \citenamefont {Potter}, \citenamefont {Potirniche},\ and\ \citenamefont
  {Vishwanath}}]{yao_discrete_2017}%
  \BibitemOpen
  \bibfield  {author} {\bibinfo {author} {\bibfnamefont {N.~Y.}\ \bibnamefont
  {Yao}}, \bibinfo {author} {\bibfnamefont {A.~C.}\ \bibnamefont {Potter}},
  \bibinfo {author} {\bibfnamefont {I.-D.}\ \bibnamefont {Potirniche}},\ and\
  \bibinfo {author} {\bibfnamefont {A.}~\bibnamefont {Vishwanath}},\ }\bibfield
   {title} {\bibinfo {title} {Discrete {Time} {Crystals}: {Rigidity},
  {Criticality}, and {Realizations}},\ }\href
  {https://doi.org/10.1103/PhysRevLett.118.030401} {\bibfield  {journal}
  {\bibinfo  {journal} {Phys. Rev. Lett.}\ }\textbf {\bibinfo {volume} {118}},\
  \bibinfo {pages} {030401} (\bibinfo {year} {2017})}\BibitemShut {NoStop}%
\bibitem [{\citenamefont {Zhang}\ \emph
  {et~al.}(2017{\natexlab{b}})\citenamefont {Zhang}, \citenamefont {Hess},
  \citenamefont {Kyprianidis}, \citenamefont {Becker}, \citenamefont {Lee},
  \citenamefont {Smith}, \citenamefont {Pagano}, \citenamefont {Potirniche},
  \citenamefont {Potter}, \citenamefont {Vishwanath},\ and\ \citenamefont
  {et~al.}}]{zhang_observation_2017}%
  \BibitemOpen
  \bibfield  {author} {\bibinfo {author} {\bibfnamefont {J.}~\bibnamefont
  {Zhang}}, \bibinfo {author} {\bibfnamefont {P.~W.}\ \bibnamefont {Hess}},
  \bibinfo {author} {\bibfnamefont {A.}~\bibnamefont {Kyprianidis}}, \bibinfo
  {author} {\bibfnamefont {P.}~\bibnamefont {Becker}}, \bibinfo {author}
  {\bibfnamefont {A.}~\bibnamefont {Lee}}, \bibinfo {author} {\bibfnamefont
  {J.}~\bibnamefont {Smith}}, \bibinfo {author} {\bibfnamefont
  {G.}~\bibnamefont {Pagano}}, \bibinfo {author} {\bibfnamefont {I.-D.}\
  \bibnamefont {Potirniche}}, \bibinfo {author} {\bibfnamefont {A.~C.}\
  \bibnamefont {Potter}}, \bibinfo {author} {\bibfnamefont {A.}~\bibnamefont
  {Vishwanath}},\ and\ \bibinfo {author} {\bibnamefont {et~al.}},\ }\bibfield
  {title} {\bibinfo {title} {Observation of a discrete time crystal},\ }\href
  {https://doi.org/10.1038/nature21413} {\bibfield  {journal} {\bibinfo
  {journal} {Nature}\ }\textbf {\bibinfo {volume} {543}},\ \bibinfo {pages}
  {217–220} (\bibinfo {year} {2017}{\natexlab{b}})}\BibitemShut {NoStop}%
\bibitem [{\citenamefont {Choi}\ \emph {et~al.}(2017)\citenamefont {Choi},
  \citenamefont {Choi}, \citenamefont {Landig}, \citenamefont {Kucsko},
  \citenamefont {Zhou}, \citenamefont {Isoya}, \citenamefont {Jelezko},
  \citenamefont {Onoda}, \citenamefont {Sumiya}, \citenamefont {Khemani},
  \citenamefont {von Keyserlingk}, \citenamefont {Yao}, \citenamefont
  {Demler},\ and\ \citenamefont {Lukin}}]{choi_observation_2017}%
  \BibitemOpen
  \bibfield  {author} {\bibinfo {author} {\bibfnamefont {S.}~\bibnamefont
  {Choi}}, \bibinfo {author} {\bibfnamefont {J.}~\bibnamefont {Choi}}, \bibinfo
  {author} {\bibfnamefont {R.}~\bibnamefont {Landig}}, \bibinfo {author}
  {\bibfnamefont {G.}~\bibnamefont {Kucsko}}, \bibinfo {author} {\bibfnamefont
  {H.}~\bibnamefont {Zhou}}, \bibinfo {author} {\bibfnamefont {J.}~\bibnamefont
  {Isoya}}, \bibinfo {author} {\bibfnamefont {F.}~\bibnamefont {Jelezko}},
  \bibinfo {author} {\bibfnamefont {S.}~\bibnamefont {Onoda}}, \bibinfo
  {author} {\bibfnamefont {H.}~\bibnamefont {Sumiya}}, \bibinfo {author}
  {\bibfnamefont {V.}~\bibnamefont {Khemani}}, \bibinfo {author} {\bibfnamefont
  {C.}~\bibnamefont {von Keyserlingk}}, \bibinfo {author} {\bibfnamefont
  {N.~Y.}\ \bibnamefont {Yao}}, \bibinfo {author} {\bibfnamefont
  {E.}~\bibnamefont {Demler}},\ and\ \bibinfo {author} {\bibfnamefont {M.~D.}\
  \bibnamefont {Lukin}},\ }\bibfield  {title} {\bibinfo {title} {Observation of
  discrete time-crystalline order in a disordered dipolar many-body system},\
  }\href {https://doi.org/10.1038/nature21426} {\bibfield  {journal} {\bibinfo
  {journal} {Nature}\ }\textbf {\bibinfo {volume} {543}},\ \bibinfo {pages}
  {221} (\bibinfo {year} {2017})}\BibitemShut {NoStop}%
\bibitem [{\citenamefont {Else}\ \emph {et~al.}(2020)\citenamefont {Else},
  \citenamefont {Monroe}, \citenamefont {Nayak},\ and\ \citenamefont
  {Yao}}]{else_discrete_2020}%
  \BibitemOpen
  \bibfield  {author} {\bibinfo {author} {\bibfnamefont {D.~V.}\ \bibnamefont
  {Else}}, \bibinfo {author} {\bibfnamefont {C.}~\bibnamefont {Monroe}},
  \bibinfo {author} {\bibfnamefont {C.}~\bibnamefont {Nayak}},\ and\ \bibinfo
  {author} {\bibfnamefont {N.~Y.}\ \bibnamefont {Yao}},\ }\bibfield  {title}
  {\bibinfo {title} {Discrete {Time} {Crystals}},\ }\href
  {https://doi.org/10.1146/annurev-conmatphys-031119-050658} {\bibfield
  {journal} {\bibinfo  {journal} {Annual Review of Condensed Matter Physics}\
  }\textbf {\bibinfo {volume} {11}},\ \bibinfo {pages} {467} (\bibinfo {year}
  {2020})}\BibitemShut {NoStop}%
\bibitem [{\citenamefont {Barahona}(1982)}]{Barahona_1982}%
  \BibitemOpen
  \bibfield  {author} {\bibinfo {author} {\bibfnamefont {F.}~\bibnamefont
  {Barahona}},\ }\bibfield  {title} {\bibinfo {title} {On the computational
  complexity of ising spin glass models},\ }\href
  {https://doi.org/10.1088/0305-4470/15/10/028} {\bibfield  {journal} {\bibinfo
   {journal} {Journal of Physics A: Mathematical and General}\ }\textbf
  {\bibinfo {volume} {15}},\ \bibinfo {pages} {3241} (\bibinfo {year}
  {1982})}\BibitemShut {NoStop}%
\bibitem [{\citenamefont {Kalai}\ and\ \citenamefont
  {Kindler}(2014)}]{kalai2014gaussian}%
  \BibitemOpen
  \bibfield  {author} {\bibinfo {author} {\bibfnamefont {G.}~\bibnamefont
  {Kalai}}\ and\ \bibinfo {author} {\bibfnamefont {G.}~\bibnamefont
  {Kindler}},\ }\bibfield  {title} {\bibinfo {title} {Gaussian noise
  sensitivity and bosonsampling},\ }\href@noop {} {\bibfield  {journal}
  {\bibinfo  {journal} {arXiv:1409.3093}\ } (\bibinfo {year}
  {2014})}\BibitemShut {NoStop}%
\bibitem [{\citenamefont {Zhou}\ \emph {et~al.}(2020)\citenamefont {Zhou},
  \citenamefont {Stoudenmire},\ and\ \citenamefont
  {Waintal}}]{PhysRevX.10.041038}%
  \BibitemOpen
  \bibfield  {author} {\bibinfo {author} {\bibfnamefont {Y.}~\bibnamefont
  {Zhou}}, \bibinfo {author} {\bibfnamefont {E.~M.}\ \bibnamefont
  {Stoudenmire}},\ and\ \bibinfo {author} {\bibfnamefont {X.}~\bibnamefont
  {Waintal}},\ }\bibfield  {title} {\bibinfo {title} {What limits the
  simulation of quantum computers?},\ }\href
  {https://doi.org/10.1103/PhysRevX.10.041038} {\bibfield  {journal} {\bibinfo
  {journal} {Phys. Rev. X}\ }\textbf {\bibinfo {volume} {10}},\ \bibinfo
  {pages} {041038} (\bibinfo {year} {2020})}\BibitemShut {NoStop}%
\bibitem [{\citenamefont {Pan}\ and\ \citenamefont
  {Zhang}(2021)}]{pan2021simulating}%
  \BibitemOpen
  \bibfield  {author} {\bibinfo {author} {\bibfnamefont {F.}~\bibnamefont
  {Pan}}\ and\ \bibinfo {author} {\bibfnamefont {P.}~\bibnamefont {Zhang}},\
  }\bibfield  {title} {\bibinfo {title} {Simulating the sycamore quantum
  supremacy circuits},\ }\href {https://arxiv.org/pdf/2103.03074.pdf}
  {\bibfield  {journal} {\bibinfo  {journal} {arXiv:2103.03074}\ } (\bibinfo
  {year} {2021})}\BibitemShut {NoStop}%
\bibitem [{\citenamefont {Domb}\ and\ \citenamefont
  {Lebowitz}(1989)}]{PT_CM13}%
  \BibitemOpen
  \bibfield  {author} {\bibinfo {author} {\bibfnamefont {C.}~\bibnamefont
  {Domb}}\ and\ \bibinfo {author} {\bibfnamefont {J.~L.}\ \bibnamefont
  {Lebowitz}},\ }\href@noop {} {\emph {\bibinfo {title} {Phase Transitions and
  Critical Phenomena}}},\ Vol.~\bibinfo {volume} {13}\ (\bibinfo  {publisher}
  {London, Academic Press},\ \bibinfo {year} {1989})\BibitemShut {NoStop}%
\bibitem [{\citenamefont {Gelfand}\ and\ \citenamefont
  {Singh}(2000)}]{RRP_singh_high_order}%
  \BibitemOpen
  \bibfield  {author} {\bibinfo {author} {\bibfnamefont {M.~P.}\ \bibnamefont
  {Gelfand}}\ and\ \bibinfo {author} {\bibfnamefont {R.~R.~P.}\ \bibnamefont
  {Singh}},\ }\bibfield  {title} {\bibinfo {title} {High-order convergent
  expansions for quantum many particle systems},\ }\href
  {https://doi.org/10.1080/000187300243390} {\bibfield  {journal} {\bibinfo
  {journal} {Advances in Physics}\ }\textbf {\bibinfo {volume} {49}},\ \bibinfo
  {pages} {93} (\bibinfo {year} {2000})}\BibitemShut {NoStop}%
\bibitem [{\citenamefont {{Park}}\ and\ \citenamefont
  {{Khatami}}(2021)}]{2021arXiv210112721P}%
  \BibitemOpen
  \bibfield  {author} {\bibinfo {author} {\bibfnamefont {J.}~\bibnamefont
  {{Park}}}\ and\ \bibinfo {author} {\bibfnamefont {E.}~\bibnamefont
  {{Khatami}}},\ }\bibfield  {title} {\bibinfo {title} {{Thermodynamics of the
  disordered Hubbard model; studied via numerical linked-cluster expansions}},\
  }\href {https://arxiv.org/pdf/2101.12721.pdf} {\bibfield  {journal} {\bibinfo
   {journal} {arXiv:2101.12721}\ } (\bibinfo {year} {2021})}\BibitemShut
  {NoStop}%
\bibitem [{\citenamefont {Schollwöck}(2011)}]{Schollw_ck_2011}%
  \BibitemOpen
  \bibfield  {author} {\bibinfo {author} {\bibfnamefont {U.}~\bibnamefont
  {Schollwöck}},\ }\bibfield  {title} {\bibinfo {title} {The density-matrix
  renormalization group in the age of matrix product states},\ }\href
  {https://doi.org/10.1016/j.aop.2010.09.012} {\bibfield  {journal} {\bibinfo
  {journal} {Annals of Physics}\ }\textbf {\bibinfo {volume} {326}},\ \bibinfo
  {pages} {96–192} (\bibinfo {year} {2011})}\BibitemShut {NoStop}%
\bibitem [{\citenamefont {Verstraete}\ \emph {et~al.}(2008)\citenamefont
  {Verstraete}, \citenamefont {Murg},\ and\ \citenamefont
  {Cirac}}]{Verstraete_2008}%
  \BibitemOpen
  \bibfield  {author} {\bibinfo {author} {\bibfnamefont {F.}~\bibnamefont
  {Verstraete}}, \bibinfo {author} {\bibfnamefont {V.}~\bibnamefont {Murg}},\
  and\ \bibinfo {author} {\bibfnamefont {J.}~\bibnamefont {Cirac}},\ }\bibfield
   {title} {\bibinfo {title} {Matrix product states, projected entangled pair
  states, and variational renormalization group methods for quantum spin
  systems},\ }\href {https://doi.org/10.1080/14789940801912366} {\bibfield
  {journal} {\bibinfo  {journal} {Advances in Physics}\ }\textbf {\bibinfo
  {volume} {57}},\ \bibinfo {pages} {143–224} (\bibinfo {year}
  {2008})}\BibitemShut {NoStop}%
\bibitem [{\citenamefont {Paeckel}\ \emph {et~al.}(2019)\citenamefont
  {Paeckel}, \citenamefont {Köhler}, \citenamefont {Swoboda}, \citenamefont
  {Manmana}, \citenamefont {Schollwöck},\ and\ \citenamefont
  {Hubig}}]{PAECKEL2019167998}%
  \BibitemOpen
  \bibfield  {author} {\bibinfo {author} {\bibfnamefont {S.}~\bibnamefont
  {Paeckel}}, \bibinfo {author} {\bibfnamefont {T.}~\bibnamefont {Köhler}},
  \bibinfo {author} {\bibfnamefont {A.}~\bibnamefont {Swoboda}}, \bibinfo
  {author} {\bibfnamefont {S.~R.}\ \bibnamefont {Manmana}}, \bibinfo {author}
  {\bibfnamefont {U.}~\bibnamefont {Schollwöck}},\ and\ \bibinfo {author}
  {\bibfnamefont {C.}~\bibnamefont {Hubig}},\ }\bibfield  {title} {\bibinfo
  {title} {Time-evolution methods for matrix-product states},\ }\href
  {https://doi.org/https://doi.org/10.1016/j.aop.2019.167998} {\bibfield
  {journal} {\bibinfo  {journal} {Annals of Physics}\ }\textbf {\bibinfo
  {volume} {411}},\ \bibinfo {pages} {167998} (\bibinfo {year}
  {2019})}\BibitemShut {NoStop}%
\bibitem [{\citenamefont {White}\ \emph {et~al.}(2018)\citenamefont {White},
  \citenamefont {Zaletel}, \citenamefont {Mong},\ and\ \citenamefont
  {Refael}}]{white_quantum_2018}%
  \BibitemOpen
  \bibfield  {author} {\bibinfo {author} {\bibfnamefont {C.~D.}\ \bibnamefont
  {White}}, \bibinfo {author} {\bibfnamefont {M.}~\bibnamefont {Zaletel}},
  \bibinfo {author} {\bibfnamefont {R.~S.~K.}\ \bibnamefont {Mong}},\ and\
  \bibinfo {author} {\bibfnamefont {G.}~\bibnamefont {Refael}},\ }\bibfield
  {title} {\bibinfo {title} {Quantum dynamics of thermalizing systems},\ }\href
  {https://doi.org/10.1103/PhysRevB.97.035127} {\bibfield  {journal} {\bibinfo
  {journal} {Phys. Rev. B}\ }\textbf {\bibinfo {volume} {97}},\ \bibinfo
  {pages} {035127} (\bibinfo {year} {2018})}\BibitemShut {NoStop}%
\bibitem [{\citenamefont {Ye}\ \emph {et~al.}(2020)\citenamefont {Ye},
  \citenamefont {Machado}, \citenamefont {White}, \citenamefont {Mong},\ and\
  \citenamefont {Yao}}]{ye_emergent_2020}%
  \BibitemOpen
  \bibfield  {author} {\bibinfo {author} {\bibfnamefont {B.}~\bibnamefont
  {Ye}}, \bibinfo {author} {\bibfnamefont {F.}~\bibnamefont {Machado}},
  \bibinfo {author} {\bibfnamefont {C.~D.}\ \bibnamefont {White}}, \bibinfo
  {author} {\bibfnamefont {R.~S.~K.}\ \bibnamefont {Mong}},\ and\ \bibinfo
  {author} {\bibfnamefont {N.~Y.}\ \bibnamefont {Yao}},\ }\bibfield  {title}
  {\bibinfo {title} {Emergent {Hydrodynamics} in {Nonequilibrium} {Quantum}
  {Systems}},\ }\href {https://doi.org/10.1103/PhysRevLett.125.030601}
  {\bibfield  {journal} {\bibinfo  {journal} {Phys. Rev. Lett.}\ }\textbf
  {\bibinfo {volume} {125}},\ \bibinfo {pages} {030601} (\bibinfo {year}
  {2020})}\BibitemShut {NoStop}%
\bibitem [{\citenamefont {Scherg}\ \emph {et~al.}(2018)\citenamefont {Scherg},
  \citenamefont {Kohlert}, \citenamefont {Herbrych}, \citenamefont {Stolpp},
  \citenamefont {Bordia}, \citenamefont {Schneider}, \citenamefont
  {Heidrich-Meisner}, \citenamefont {Bloch},\ and\ \citenamefont
  {Aidelsburger}}]{scherg_nonequilibrium_2018}%
  \BibitemOpen
  \bibfield  {author} {\bibinfo {author} {\bibfnamefont {S.}~\bibnamefont
  {Scherg}}, \bibinfo {author} {\bibfnamefont {T.}~\bibnamefont {Kohlert}},
  \bibinfo {author} {\bibfnamefont {J.}~\bibnamefont {Herbrych}}, \bibinfo
  {author} {\bibfnamefont {J.}~\bibnamefont {Stolpp}}, \bibinfo {author}
  {\bibfnamefont {P.}~\bibnamefont {Bordia}}, \bibinfo {author} {\bibfnamefont
  {U.}~\bibnamefont {Schneider}}, \bibinfo {author} {\bibfnamefont
  {F.}~\bibnamefont {Heidrich-Meisner}}, \bibinfo {author} {\bibfnamefont
  {I.}~\bibnamefont {Bloch}},\ and\ \bibinfo {author} {\bibfnamefont
  {M.}~\bibnamefont {Aidelsburger}},\ }\bibfield  {title} {\bibinfo {title}
  {Nonequilibrium {Mass} {Transport} in the {1D} {Fermi}-{Hubbard} {Model}},\
  }\href {https://doi.org/10.1103/PhysRevLett.121.130402} {\bibfield  {journal}
  {\bibinfo  {journal} {Phys. Rev. Lett.}\ }\textbf {\bibinfo {volume} {121}},\
  \bibinfo {pages} {130402} (\bibinfo {year} {2018})}\BibitemShut {NoStop}%
\bibitem [{\citenamefont {Chanda}\ \emph {et~al.}(2020)\citenamefont {Chanda},
  \citenamefont {Sierant},\ and\ \citenamefont
  {Zakrzewski}}]{PhysRevB.101.035148}%
  \BibitemOpen
  \bibfield  {author} {\bibinfo {author} {\bibfnamefont {T.}~\bibnamefont
  {Chanda}}, \bibinfo {author} {\bibfnamefont {P.}~\bibnamefont {Sierant}},\
  and\ \bibinfo {author} {\bibfnamefont {J.}~\bibnamefont {Zakrzewski}},\
  }\bibfield  {title} {\bibinfo {title} {Time dynamics with matrix product
  states: Many-body localization transition of large systems revisited},\
  }\href {https://doi.org/10.1103/PhysRevB.101.035148} {\bibfield  {journal}
  {\bibinfo  {journal} {Phys. Rev. B}\ }\textbf {\bibinfo {volume} {101}},\
  \bibinfo {pages} {035148} (\bibinfo {year} {2020})}\BibitemShut {NoStop}%
\bibitem [{\citenamefont {Uhlmann}(1976)}]{Uhlmann_1976}%
  \BibitemOpen
  \bibfield  {author} {\bibinfo {author} {\bibfnamefont {A.}~\bibnamefont
  {Uhlmann}},\ }\bibfield  {title} {\bibinfo {title} {The "transition
  probability" in the state space of a *-algebra},\ }\href
  {https://doi.org/10.1016/0034-4877(76)90060-4} {\bibfield  {journal}
  {\bibinfo  {journal} {Reports on mathematical physics}\ }\textbf {\bibinfo
  {volume} {9}},\ \bibinfo {pages} {273} (\bibinfo {year} {1976})}\BibitemShut
  {NoStop}%
\bibitem [{\citenamefont {Hauru}\ and\ \citenamefont
  {Vidal}(2018)}]{PhysRevA.98.042316}%
  \BibitemOpen
  \bibfield  {author} {\bibinfo {author} {\bibfnamefont {M.}~\bibnamefont
  {Hauru}}\ and\ \bibinfo {author} {\bibfnamefont {G.}~\bibnamefont {Vidal}},\
  }\bibfield  {title} {\bibinfo {title} {Uhlmann fidelities from tensor
  networks},\ }\href {https://doi.org/10.1103/PhysRevA.98.042316} {\bibfield
  {journal} {\bibinfo  {journal} {Phys. Rev. A}\ }\textbf {\bibinfo {volume}
  {98}},\ \bibinfo {pages} {042316} (\bibinfo {year} {2018})}\BibitemShut
  {NoStop}%
\bibitem [{\citenamefont {L\"uschen}\ \emph
  {et~al.}(2017{\natexlab{a}})\citenamefont {L\"uschen}, \citenamefont
  {Bordia}, \citenamefont {Hodgman}, \citenamefont {Schreiber}, \citenamefont
  {Sarkar}, \citenamefont {Daley}, \citenamefont {Fischer}, \citenamefont
  {Altman}, \citenamefont {Bloch},\ and\ \citenamefont
  {Schneider}}]{PhysRevX.7.011034}%
  \BibitemOpen
  \bibfield  {author} {\bibinfo {author} {\bibfnamefont {H.~P.}\ \bibnamefont
  {L\"uschen}}, \bibinfo {author} {\bibfnamefont {P.}~\bibnamefont {Bordia}},
  \bibinfo {author} {\bibfnamefont {S.~S.}\ \bibnamefont {Hodgman}}, \bibinfo
  {author} {\bibfnamefont {M.}~\bibnamefont {Schreiber}}, \bibinfo {author}
  {\bibfnamefont {S.}~\bibnamefont {Sarkar}}, \bibinfo {author} {\bibfnamefont
  {A.~J.}\ \bibnamefont {Daley}}, \bibinfo {author} {\bibfnamefont {M.~H.}\
  \bibnamefont {Fischer}}, \bibinfo {author} {\bibfnamefont {E.}~\bibnamefont
  {Altman}}, \bibinfo {author} {\bibfnamefont {I.}~\bibnamefont {Bloch}},\ and\
  \bibinfo {author} {\bibfnamefont {U.}~\bibnamefont {Schneider}},\ }\bibfield
  {title} {\bibinfo {title} {Signatures of many-body localization in a
  controlled open quantum system},\ }\href
  {https://doi.org/10.1103/PhysRevX.7.011034} {\bibfield  {journal} {\bibinfo
  {journal} {Phys. Rev. X}\ }\textbf {\bibinfo {volume} {7}},\ \bibinfo {pages}
  {011034} (\bibinfo {year} {2017}{\natexlab{a}})}\BibitemShut {NoStop}%
\bibitem [{\citenamefont {Bordia}\ \emph {et~al.}(2016)\citenamefont {Bordia},
  \citenamefont {L\"uschen}, \citenamefont {Hodgman}, \citenamefont
  {Schreiber}, \citenamefont {Bloch},\ and\ \citenamefont
  {Schneider}}]{PhysRevLett.116.140401}%
  \BibitemOpen
  \bibfield  {author} {\bibinfo {author} {\bibfnamefont {P.}~\bibnamefont
  {Bordia}}, \bibinfo {author} {\bibfnamefont {H.~P.}\ \bibnamefont
  {L\"uschen}}, \bibinfo {author} {\bibfnamefont {S.~S.}\ \bibnamefont
  {Hodgman}}, \bibinfo {author} {\bibfnamefont {M.}~\bibnamefont {Schreiber}},
  \bibinfo {author} {\bibfnamefont {I.}~\bibnamefont {Bloch}},\ and\ \bibinfo
  {author} {\bibfnamefont {U.}~\bibnamefont {Schneider}},\ }\bibfield  {title}
  {\bibinfo {title} {Coupling identical one-dimensional many-body localized
  systems},\ }\href {https://doi.org/10.1103/PhysRevLett.116.140401} {\bibfield
   {journal} {\bibinfo  {journal} {Phys. Rev. Lett.}\ }\textbf {\bibinfo
  {volume} {116}},\ \bibinfo {pages} {140401} (\bibinfo {year}
  {2016})}\BibitemShut {NoStop}%
\bibitem [{\citenamefont {Bordia}\ \emph {et~al.}(2017)\citenamefont {Bordia},
  \citenamefont {L\"uschen}, \citenamefont {Scherg}, \citenamefont
  {Gopalakrishnan}, \citenamefont {Knap}, \citenamefont {Schneider},\ and\
  \citenamefont {Bloch}}]{PhysRevX.7.041047}%
  \BibitemOpen
  \bibfield  {author} {\bibinfo {author} {\bibfnamefont {P.}~\bibnamefont
  {Bordia}}, \bibinfo {author} {\bibfnamefont {H.}~\bibnamefont {L\"uschen}},
  \bibinfo {author} {\bibfnamefont {S.}~\bibnamefont {Scherg}}, \bibinfo
  {author} {\bibfnamefont {S.}~\bibnamefont {Gopalakrishnan}}, \bibinfo
  {author} {\bibfnamefont {M.}~\bibnamefont {Knap}}, \bibinfo {author}
  {\bibfnamefont {U.}~\bibnamefont {Schneider}},\ and\ \bibinfo {author}
  {\bibfnamefont {I.}~\bibnamefont {Bloch}},\ }\bibfield  {title} {\bibinfo
  {title} {Probing slow relaxation and many-body localization in
  two-dimensional quasiperiodic systems},\ }\href
  {https://doi.org/10.1103/PhysRevX.7.041047} {\bibfield  {journal} {\bibinfo
  {journal} {Phys. Rev. X}\ }\textbf {\bibinfo {volume} {7}},\ \bibinfo {pages}
  {041047} (\bibinfo {year} {2017})}\BibitemShut {NoStop}%
\bibitem [{\citenamefont {Bera}\ \emph {et~al.}(2015)\citenamefont {Bera},
  \citenamefont {Schomerus}, \citenamefont {Heidrich-Meisner},\ and\
  \citenamefont {Bardarson}}]{PhysRevLett.115.046603}%
  \BibitemOpen
  \bibfield  {author} {\bibinfo {author} {\bibfnamefont {S.}~\bibnamefont
  {Bera}}, \bibinfo {author} {\bibfnamefont {H.}~\bibnamefont {Schomerus}},
  \bibinfo {author} {\bibfnamefont {F.}~\bibnamefont {Heidrich-Meisner}},\ and\
  \bibinfo {author} {\bibfnamefont {J.~H.}\ \bibnamefont {Bardarson}},\
  }\bibfield  {title} {\bibinfo {title} {Many-body localization characterized
  from a one-particle perspective},\ }\href
  {https://doi.org/10.1103/PhysRevLett.115.046603} {\bibfield  {journal}
  {\bibinfo  {journal} {Phys. Rev. Lett.}\ }\textbf {\bibinfo {volume} {115}},\
  \bibinfo {pages} {046603} (\bibinfo {year} {2015})}\BibitemShut {NoStop}%
\bibitem [{\citenamefont {Anderson}(1958)}]{PhysRev.109.1492}%
  \BibitemOpen
  \bibfield  {author} {\bibinfo {author} {\bibfnamefont {P.~W.}\ \bibnamefont
  {Anderson}},\ }\bibfield  {title} {\bibinfo {title} {Absence of diffusion in
  certain random lattices},\ }\href {https://doi.org/10.1103/PhysRev.109.1492}
  {\bibfield  {journal} {\bibinfo  {journal} {Phys. Rev.}\ }\textbf {\bibinfo
  {volume} {109}},\ \bibinfo {pages} {1492} (\bibinfo {year}
  {1958})}\BibitemShut {NoStop}%
\bibitem [{\citenamefont {Reichl}\ and\ \citenamefont
  {Mueller}(2016)}]{PhysRevA.93.031601}%
  \BibitemOpen
  \bibfield  {author} {\bibinfo {author} {\bibfnamefont {M.~D.}\ \bibnamefont
  {Reichl}}\ and\ \bibinfo {author} {\bibfnamefont {E.~J.}\ \bibnamefont
  {Mueller}},\ }\bibfield  {title} {\bibinfo {title} {Dynamics of
  pattern-loaded fermions in bichromatic optical lattices},\ }\href
  {https://doi.org/10.1103/PhysRevA.93.031601} {\bibfield  {journal} {\bibinfo
  {journal} {Phys. Rev. A}\ }\textbf {\bibinfo {volume} {93}},\ \bibinfo
  {pages} {031601} (\bibinfo {year} {2016})}\BibitemShut {NoStop}%
\bibitem [{k_s()}]{k_sigma_shell}%
  \BibitemOpen
  \href@noop {} {}\bibinfo {note} {This set will be symmetric with respect to
  $i_r$ if $\kappa_{\sigma}$ is even. If $\kappa_{\sigma}$ is odd, there are
  two choices for the shell; we may pick any one of them.}\BibitemShut {Stop}%
\bibitem [{\citenamefont {Rigol}\ \emph {et~al.}(2006)\citenamefont {Rigol},
  \citenamefont {Bryant},\ and\ \citenamefont {Singh}}]{PhysRevLett.97.187202}%
  \BibitemOpen
  \bibfield  {author} {\bibinfo {author} {\bibfnamefont {M.}~\bibnamefont
  {Rigol}}, \bibinfo {author} {\bibfnamefont {T.}~\bibnamefont {Bryant}},\ and\
  \bibinfo {author} {\bibfnamefont {R.~R.~P.}\ \bibnamefont {Singh}},\
  }\bibfield  {title} {\bibinfo {title} {Numerical linked-cluster approach to
  quantum lattice models},\ }\href
  {https://doi.org/10.1103/PhysRevLett.97.187202} {\bibfield  {journal}
  {\bibinfo  {journal} {Phys. Rev. Lett.}\ }\textbf {\bibinfo {volume} {97}},\
  \bibinfo {pages} {187202} (\bibinfo {year} {2006})}\BibitemShut {NoStop}%
\bibitem [{\citenamefont {Mallayya}\ and\ \citenamefont
  {Rigol}(2017)}]{PhysRevE.95.033302}%
  \BibitemOpen
  \bibfield  {author} {\bibinfo {author} {\bibfnamefont {K.}~\bibnamefont
  {Mallayya}}\ and\ \bibinfo {author} {\bibfnamefont {M.}~\bibnamefont
  {Rigol}},\ }\bibfield  {title} {\bibinfo {title} {Numerical linked cluster
  expansions for quantum quenches in one-dimensional lattices},\ }\href
  {https://doi.org/10.1103/PhysRevE.95.033302} {\bibfield  {journal} {\bibinfo
  {journal} {Phys. Rev. E}\ }\textbf {\bibinfo {volume} {95}},\ \bibinfo
  {pages} {033302} (\bibinfo {year} {2017})}\BibitemShut {NoStop}%
\bibitem [{\citenamefont {Hazzard}\ \emph {et~al.}(2014)\citenamefont
  {Hazzard}, \citenamefont {Gadway}, \citenamefont {Foss-Feig}, \citenamefont
  {Yan}, \citenamefont {Moses}, \citenamefont {Covey}, \citenamefont {Yao},
  \citenamefont {Lukin}, \citenamefont {Ye}, \citenamefont {Jin},\ and\
  \citenamefont {Rey}}]{PhysRevLett.113.195302}%
  \BibitemOpen
  \bibfield  {author} {\bibinfo {author} {\bibfnamefont {K.~R.~A.}\
  \bibnamefont {Hazzard}}, \bibinfo {author} {\bibfnamefont {B.}~\bibnamefont
  {Gadway}}, \bibinfo {author} {\bibfnamefont {M.}~\bibnamefont {Foss-Feig}},
  \bibinfo {author} {\bibfnamefont {B.}~\bibnamefont {Yan}}, \bibinfo {author}
  {\bibfnamefont {S.~A.}\ \bibnamefont {Moses}}, \bibinfo {author}
  {\bibfnamefont {J.~P.}\ \bibnamefont {Covey}}, \bibinfo {author}
  {\bibfnamefont {N.~Y.}\ \bibnamefont {Yao}}, \bibinfo {author} {\bibfnamefont
  {M.~D.}\ \bibnamefont {Lukin}}, \bibinfo {author} {\bibfnamefont
  {J.}~\bibnamefont {Ye}}, \bibinfo {author} {\bibfnamefont {D.~S.}\
  \bibnamefont {Jin}},\ and\ \bibinfo {author} {\bibfnamefont {A.~M.}\
  \bibnamefont {Rey}},\ }\bibfield  {title} {\bibinfo {title} {Many-body
  dynamics of dipolar molecules in an optical lattice},\ }\href
  {https://doi.org/10.1103/PhysRevLett.113.195302} {\bibfield  {journal}
  {\bibinfo  {journal} {Phys. Rev. Lett.}\ }\textbf {\bibinfo {volume} {113}},\
  \bibinfo {pages} {195302} (\bibinfo {year} {2014})}\BibitemShut {NoStop}%
\bibitem [{\citenamefont {Aubry}\ and\ \citenamefont
  {Andr{\'e}}(1980)}]{Aubry80}%
  \BibitemOpen
  \bibfield  {author} {\bibinfo {author} {\bibfnamefont {S.}~\bibnamefont
  {Aubry}}\ and\ \bibinfo {author} {\bibfnamefont {G.}~\bibnamefont
  {Andr{\'e}}},\ }\bibfield  {title} {\bibinfo {title} {Analyticity breaking
  and {A}nderson localization in incommensurate lattices},\ }\href@noop {}
  {\bibfield  {journal} {\bibinfo  {journal} {Ann. Israel Phys. Soc.}\ }\textbf
  {\bibinfo {volume} {3}} (\bibinfo {year} {1980})}\BibitemShut {NoStop}%
\bibitem [{\citenamefont {Kohlert}\ \emph {et~al.}(2021)\citenamefont
  {Kohlert}, \citenamefont {Scherg}, \citenamefont {Sala}, \citenamefont
  {Pollmann}, \citenamefont {Madhusudhana}, \citenamefont {Bloch},\ and\
  \citenamefont {Aidelsburger}}]{HS_fragmentation}%
  \BibitemOpen
  \bibfield  {author} {\bibinfo {author} {\bibfnamefont {T.}~\bibnamefont
  {Kohlert}}, \bibinfo {author} {\bibfnamefont {S.}~\bibnamefont {Scherg}},
  \bibinfo {author} {\bibfnamefont {P.}~\bibnamefont {Sala}}, \bibinfo {author}
  {\bibfnamefont {F.}~\bibnamefont {Pollmann}}, \bibinfo {author}
  {\bibfnamefont {B.~H.}\ \bibnamefont {Madhusudhana}}, \bibinfo {author}
  {\bibfnamefont {I.}~\bibnamefont {Bloch}},\ and\ \bibinfo {author}
  {\bibfnamefont {M.}~\bibnamefont {Aidelsburger}},\ }\bibfield  {title}
  {\bibinfo {title} {Experimental realization of fragmented models in tilted
  fermi-hubbard chains},\ }\href {https://arxiv.org/abs/2106.15586} {\bibfield
  {journal} {\bibinfo  {journal} {arXiv:2106.15586}\ } (\bibinfo {year}
  {2021})}\BibitemShut {NoStop}%
\bibitem [{\citenamefont {Bardarson}\ \emph {et~al.}(2012)\citenamefont
  {Bardarson}, \citenamefont {Pollmann},\ and\ \citenamefont
  {Moore}}]{PhysRevLett.109.017202}%
  \BibitemOpen
  \bibfield  {author} {\bibinfo {author} {\bibfnamefont {J.~H.}\ \bibnamefont
  {Bardarson}}, \bibinfo {author} {\bibfnamefont {F.}~\bibnamefont
  {Pollmann}},\ and\ \bibinfo {author} {\bibfnamefont {J.~E.}\ \bibnamefont
  {Moore}},\ }\bibfield  {title} {\bibinfo {title} {Unbounded growth of
  entanglement in models of many-body localization},\ }\href
  {https://doi.org/10.1103/PhysRevLett.109.017202} {\bibfield  {journal}
  {\bibinfo  {journal} {Phys. Rev. Lett.}\ }\textbf {\bibinfo {volume} {109}},\
  \bibinfo {pages} {017202} (\bibinfo {year} {2012})}\BibitemShut {NoStop}%
\bibitem [{\citenamefont {Bauer}\ and\ \citenamefont
  {Nayak}(2013)}]{Bauer_2013}%
  \BibitemOpen
  \bibfield  {author} {\bibinfo {author} {\bibfnamefont {B.}~\bibnamefont
  {Bauer}}\ and\ \bibinfo {author} {\bibfnamefont {C.}~\bibnamefont {Nayak}},\
  }\bibfield  {title} {\bibinfo {title} {Area laws in a many-body localized
  state and its implications for topological order},\ }\href
  {https://doi.org/10.1088/1742-5468/2013/09/p09005} {\bibfield  {journal}
  {\bibinfo  {journal} {Journal of Statistical Mechanics: Theory and
  Experiment}\ }\textbf {\bibinfo {volume} {2013}},\ \bibinfo {pages} {P09005}
  (\bibinfo {year} {2013})}\BibitemShut {NoStop}%
\bibitem [{\citenamefont {Kiefer-Emmanouilidis}\ \emph
  {et~al.}(2020)\citenamefont {Kiefer-Emmanouilidis}, \citenamefont {Unanyan},
  \citenamefont {Fleischhauer},\ and\ \citenamefont
  {Sirker}}]{PhysRevLett.124.243601}%
  \BibitemOpen
  \bibfield  {author} {\bibinfo {author} {\bibfnamefont {M.}~\bibnamefont
  {Kiefer-Emmanouilidis}}, \bibinfo {author} {\bibfnamefont {R.}~\bibnamefont
  {Unanyan}}, \bibinfo {author} {\bibfnamefont {M.}~\bibnamefont
  {Fleischhauer}},\ and\ \bibinfo {author} {\bibfnamefont {J.}~\bibnamefont
  {Sirker}},\ }\bibfield  {title} {\bibinfo {title} {Evidence for unbounded
  growth of the number entropy in many-body localized phases},\ }\href
  {https://doi.org/10.1103/PhysRevLett.124.243601} {\bibfield  {journal}
  {\bibinfo  {journal} {Phys. Rev. Lett.}\ }\textbf {\bibinfo {volume} {124}},\
  \bibinfo {pages} {243601} (\bibinfo {year} {2020})}\BibitemShut {NoStop}%
\bibitem [{\citenamefont {Luitz}\ and\ \citenamefont
  {Lev}(2020)}]{luitz2020slow}%
  \BibitemOpen
  \bibfield  {author} {\bibinfo {author} {\bibfnamefont {D.~J.}\ \bibnamefont
  {Luitz}}\ and\ \bibinfo {author} {\bibfnamefont {Y.~B.}\ \bibnamefont
  {Lev}},\ }\bibfield  {title} {\bibinfo {title} {Is there slow particle
  transport in the mbl phase?},\ }\href {https://arxiv.org/pdf/2007.13767.pdf}
  {\bibfield  {journal} {\bibinfo  {journal} {arXiv:2007.13767}\ } (\bibinfo
  {year} {2020})}\BibitemShut {NoStop}%
\bibitem [{\citenamefont {Duan}(2011)}]{PhysRevLett.107.180502}%
  \BibitemOpen
  \bibfield  {author} {\bibinfo {author} {\bibfnamefont {L.-M.}\ \bibnamefont
  {Duan}},\ }\bibfield  {title} {\bibinfo {title} {Entanglement detection in
  the vicinity of arbitrary dicke states},\ }\href
  {https://doi.org/10.1103/PhysRevLett.107.180502} {\bibfield  {journal}
  {\bibinfo  {journal} {Phys. Rev. Lett.}\ }\textbf {\bibinfo {volume} {107}},\
  \bibinfo {pages} {180502} (\bibinfo {year} {2011})}\BibitemShut {NoStop}%
\bibitem [{\citenamefont {Schuch}\ \emph {et~al.}(2008)\citenamefont {Schuch},
  \citenamefont {Wolf}, \citenamefont {Verstraete},\ and\ \citenamefont
  {Cirac}}]{PhysRevLett.100.030504}%
  \BibitemOpen
  \bibfield  {author} {\bibinfo {author} {\bibfnamefont {N.}~\bibnamefont
  {Schuch}}, \bibinfo {author} {\bibfnamefont {M.~M.}\ \bibnamefont {Wolf}},
  \bibinfo {author} {\bibfnamefont {F.}~\bibnamefont {Verstraete}},\ and\
  \bibinfo {author} {\bibfnamefont {J.~I.}\ \bibnamefont {Cirac}},\ }\bibfield
  {title} {\bibinfo {title} {Entropy scaling and simulability by matrix product
  states},\ }\href {https://doi.org/10.1103/PhysRevLett.100.030504} {\bibfield
  {journal} {\bibinfo  {journal} {Phys. Rev. Lett.}\ }\textbf {\bibinfo
  {volume} {100}},\ \bibinfo {pages} {030504} (\bibinfo {year}
  {2008})}\BibitemShut {NoStop}%
\bibitem [{\citenamefont {Bharath}\ and\ \citenamefont
  {Ravishankar}(2014)}]{PhysRevA.89.062110}%
  \BibitemOpen
  \bibfield  {author} {\bibinfo {author} {\bibfnamefont {H.~M.}\ \bibnamefont
  {Bharath}}\ and\ \bibinfo {author} {\bibfnamefont {V.}~\bibnamefont
  {Ravishankar}},\ }\bibfield  {title} {\bibinfo {title} {Classical simulation
  of entangled states},\ }\href {https://doi.org/10.1103/PhysRevA.89.062110}
  {\bibfield  {journal} {\bibinfo  {journal} {Phys. Rev. A}\ }\textbf {\bibinfo
  {volume} {89}},\ \bibinfo {pages} {062110} (\bibinfo {year}
  {2014})}\BibitemShut {NoStop}%
\bibitem [{\citenamefont {Omran}\ \emph {et~al.}(2019)\citenamefont {Omran},
  \citenamefont {Levine}, \citenamefont {Keesling}, \citenamefont {Semeghini},
  \citenamefont {Wang}, \citenamefont {Ebadi}, \citenamefont {Bernien},
  \citenamefont {Zibrov}, \citenamefont {Pichler}, \citenamefont {Choi},\ and\
  \citenamefont {et~al.}}]{Omran_2019}%
  \BibitemOpen
  \bibfield  {author} {\bibinfo {author} {\bibfnamefont {A.}~\bibnamefont
  {Omran}}, \bibinfo {author} {\bibfnamefont {H.}~\bibnamefont {Levine}},
  \bibinfo {author} {\bibfnamefont {A.}~\bibnamefont {Keesling}}, \bibinfo
  {author} {\bibfnamefont {G.}~\bibnamefont {Semeghini}}, \bibinfo {author}
  {\bibfnamefont {T.~T.}\ \bibnamefont {Wang}}, \bibinfo {author}
  {\bibfnamefont {S.}~\bibnamefont {Ebadi}}, \bibinfo {author} {\bibfnamefont
  {H.}~\bibnamefont {Bernien}}, \bibinfo {author} {\bibfnamefont {A.~S.}\
  \bibnamefont {Zibrov}}, \bibinfo {author} {\bibfnamefont {H.}~\bibnamefont
  {Pichler}}, \bibinfo {author} {\bibfnamefont {S.}~\bibnamefont {Choi}},\ and\
  \bibinfo {author} {\bibnamefont {et~al.}},\ }\bibfield  {title} {\bibinfo
  {title} {Generation and manipulation of schrödinger cat states in rydberg
  atom arrays},\ }\href {https://doi.org/10.1126/science.aax9743} {\bibfield
  {journal} {\bibinfo  {journal} {Science}\ }\textbf {\bibinfo {volume}
  {365}},\ \bibinfo {pages} {570–574} (\bibinfo {year} {2019})}\BibitemShut
  {NoStop}%
\bibitem [{\citenamefont {L\"uschen}\ \emph
  {et~al.}(2017{\natexlab{b}})\citenamefont {L\"uschen}, \citenamefont
  {Bordia}, \citenamefont {Scherg}, \citenamefont {Alet}, \citenamefont
  {Altman}, \citenamefont {Schneider},\ and\ \citenamefont
  {Bloch}}]{PhysRevLett.119.260401}%
  \BibitemOpen
  \bibfield  {author} {\bibinfo {author} {\bibfnamefont {H.~P.}\ \bibnamefont
  {L\"uschen}}, \bibinfo {author} {\bibfnamefont {P.}~\bibnamefont {Bordia}},
  \bibinfo {author} {\bibfnamefont {S.}~\bibnamefont {Scherg}}, \bibinfo
  {author} {\bibfnamefont {F.}~\bibnamefont {Alet}}, \bibinfo {author}
  {\bibfnamefont {E.}~\bibnamefont {Altman}}, \bibinfo {author} {\bibfnamefont
  {U.}~\bibnamefont {Schneider}},\ and\ \bibinfo {author} {\bibfnamefont
  {I.}~\bibnamefont {Bloch}},\ }\bibfield  {title} {\bibinfo {title}
  {Observation of slow dynamics near the many-body localization transition in
  one-dimensional quasiperiodic systems},\ }\href
  {https://doi.org/10.1103/PhysRevLett.119.260401} {\bibfield  {journal}
  {\bibinfo  {journal} {Phys. Rev. Lett.}\ }\textbf {\bibinfo {volume} {119}},\
  \bibinfo {pages} {260401} (\bibinfo {year} {2017}{\natexlab{b}})}\BibitemShut
  {NoStop}%
\bibitem [{\citenamefont {Torres-Herrera}\ and\ \citenamefont
  {Santos}(2015)}]{PhysRevB.92.014208}%
  \BibitemOpen
  \bibfield  {author} {\bibinfo {author} {\bibfnamefont {E.~J.}\ \bibnamefont
  {Torres-Herrera}}\ and\ \bibinfo {author} {\bibfnamefont {L.~F.}\
  \bibnamefont {Santos}},\ }\bibfield  {title} {\bibinfo {title} {Dynamics at
  the many-body localization transition},\ }\href
  {https://doi.org/10.1103/PhysRevB.92.014208} {\bibfield  {journal} {\bibinfo
  {journal} {Phys. Rev. B}\ }\textbf {\bibinfo {volume} {92}},\ \bibinfo
  {pages} {014208} (\bibinfo {year} {2015})}\BibitemShut {NoStop}%
\bibitem [{\citenamefont {Luitz}\ \emph {et~al.}(2016)\citenamefont {Luitz},
  \citenamefont {Laflorencie},\ and\ \citenamefont
  {Alet}}]{PhysRevB.93.060201}%
  \BibitemOpen
  \bibfield  {author} {\bibinfo {author} {\bibfnamefont {D.~J.}\ \bibnamefont
  {Luitz}}, \bibinfo {author} {\bibfnamefont {N.}~\bibnamefont {Laflorencie}},\
  and\ \bibinfo {author} {\bibfnamefont {F.}~\bibnamefont {Alet}},\ }\bibfield
  {title} {\bibinfo {title} {Extended slow dynamical regime close to the
  many-body localization transition},\ }\href
  {https://doi.org/10.1103/PhysRevB.93.060201} {\bibfield  {journal} {\bibinfo
  {journal} {Phys. Rev. B}\ }\textbf {\bibinfo {volume} {93}},\ \bibinfo
  {pages} {060201} (\bibinfo {year} {2016})}\BibitemShut {NoStop}%
\bibitem [{\citenamefont {Gopalakrishnan}\ \emph {et~al.}(2016)\citenamefont
  {Gopalakrishnan}, \citenamefont {Agarwal}, \citenamefont {Demler},
  \citenamefont {Huse},\ and\ \citenamefont {Knap}}]{PhysRevB.93.134206}%
  \BibitemOpen
  \bibfield  {author} {\bibinfo {author} {\bibfnamefont {S.}~\bibnamefont
  {Gopalakrishnan}}, \bibinfo {author} {\bibfnamefont {K.}~\bibnamefont
  {Agarwal}}, \bibinfo {author} {\bibfnamefont {E.~A.}\ \bibnamefont {Demler}},
  \bibinfo {author} {\bibfnamefont {D.~A.}\ \bibnamefont {Huse}},\ and\
  \bibinfo {author} {\bibfnamefont {M.}~\bibnamefont {Knap}},\ }\bibfield
  {title} {\bibinfo {title} {Griffiths effects and slow dynamics in nearly
  many-body localized systems},\ }\href
  {https://doi.org/10.1103/PhysRevB.93.134206} {\bibfield  {journal} {\bibinfo
  {journal} {Phys. Rev. B}\ }\textbf {\bibinfo {volume} {93}},\ \bibinfo
  {pages} {134206} (\bibinfo {year} {2016})}\BibitemShut {NoStop}%
\bibitem [{\citenamefont {Agarwal}\ \emph {et~al.}(2015)\citenamefont
  {Agarwal}, \citenamefont {Gopalakrishnan}, \citenamefont {Knap},
  \citenamefont {M\"uller},\ and\ \citenamefont
  {Demler}}]{PhysRevLett.114.160401}%
  \BibitemOpen
  \bibfield  {author} {\bibinfo {author} {\bibfnamefont {K.}~\bibnamefont
  {Agarwal}}, \bibinfo {author} {\bibfnamefont {S.}~\bibnamefont
  {Gopalakrishnan}}, \bibinfo {author} {\bibfnamefont {M.}~\bibnamefont
  {Knap}}, \bibinfo {author} {\bibfnamefont {M.}~\bibnamefont {M\"uller}},\
  and\ \bibinfo {author} {\bibfnamefont {E.}~\bibnamefont {Demler}},\
  }\bibfield  {title} {\bibinfo {title} {Anomalous diffusion and griffiths
  effects near the many-body localization transition},\ }\href
  {https://doi.org/10.1103/PhysRevLett.114.160401} {\bibfield  {journal}
  {\bibinfo  {journal} {Phys. Rev. Lett.}\ }\textbf {\bibinfo {volume} {114}},\
  \bibinfo {pages} {160401} (\bibinfo {year} {2015})}\BibitemShut {NoStop}%
\bibitem [{\citenamefont {\ifmmode \check{Z}\else
  \v{Z}\fi{}nidari\ifmmode~\check{c}\else \v{c}\fi{}}\ \emph
  {et~al.}(2016)\citenamefont {\ifmmode \check{Z}\else
  \v{Z}\fi{}nidari\ifmmode~\check{c}\else \v{c}\fi{}}, \citenamefont
  {Scardicchio},\ and\ \citenamefont {Varma}}]{PhysRevLett.117.040601}%
  \BibitemOpen
  \bibfield  {author} {\bibinfo {author} {\bibfnamefont {M.}~\bibnamefont
  {\ifmmode \check{Z}\else \v{Z}\fi{}nidari\ifmmode~\check{c}\else
  \v{c}\fi{}}}, \bibinfo {author} {\bibfnamefont {A.}~\bibnamefont
  {Scardicchio}},\ and\ \bibinfo {author} {\bibfnamefont {V.~K.}\ \bibnamefont
  {Varma}},\ }\bibfield  {title} {\bibinfo {title} {Diffusive and subdiffusive
  spin transport in the ergodic phase of a many-body localizable system},\
  }\href {https://doi.org/10.1103/PhysRevLett.117.040601} {\bibfield  {journal}
  {\bibinfo  {journal} {Phys. Rev. Lett.}\ }\textbf {\bibinfo {volume} {117}},\
  \bibinfo {pages} {040601} (\bibinfo {year} {2016})}\BibitemShut {NoStop}%
\bibitem [{\citenamefont {Molnar}\ \emph {et~al.}(2015)\citenamefont {Molnar},
  \citenamefont {Schuch}, \citenamefont {Verstraete},\ and\ \citenamefont
  {Cirac}}]{PhysRevB.91.045138}%
  \BibitemOpen
  \bibfield  {author} {\bibinfo {author} {\bibfnamefont {A.}~\bibnamefont
  {Molnar}}, \bibinfo {author} {\bibfnamefont {N.}~\bibnamefont {Schuch}},
  \bibinfo {author} {\bibfnamefont {F.}~\bibnamefont {Verstraete}},\ and\
  \bibinfo {author} {\bibfnamefont {J.~I.}\ \bibnamefont {Cirac}},\ }\bibfield
  {title} {\bibinfo {title} {Approximating gibbs states of local hamiltonians
  efficiently with projected entangled pair states},\ }\href
  {https://doi.org/10.1103/PhysRevB.91.045138} {\bibfield  {journal} {\bibinfo
  {journal} {Phys. Rev. B}\ }\textbf {\bibinfo {volume} {91}},\ \bibinfo
  {pages} {045138} (\bibinfo {year} {2015})}\BibitemShut {NoStop}%
\bibitem [{\citenamefont {Wu}\ \emph {et~al.}(2019)\citenamefont {Wu},
  \citenamefont {Schnell}, \citenamefont {Tomasi}, \citenamefont {Heyl},\ and\
  \citenamefont {Eckardt}}]{Wu_2019}%
  \BibitemOpen
  \bibfield  {author} {\bibinfo {author} {\bibfnamefont {L.-N.}\ \bibnamefont
  {Wu}}, \bibinfo {author} {\bibfnamefont {A.}~\bibnamefont {Schnell}},
  \bibinfo {author} {\bibfnamefont {G.~D.}\ \bibnamefont {Tomasi}}, \bibinfo
  {author} {\bibfnamefont {M.}~\bibnamefont {Heyl}},\ and\ \bibinfo {author}
  {\bibfnamefont {A.}~\bibnamefont {Eckardt}},\ }\bibfield  {title} {\bibinfo
  {title} {Describing many-body localized systems in thermal environments},\
  }\href {https://doi.org/10.1088/1367-2630/ab25a4} {\bibfield  {journal}
  {\bibinfo  {journal} {New Journal of Physics}\ }\textbf {\bibinfo {volume}
  {21}},\ \bibinfo {pages} {063026} (\bibinfo {year} {2019})}\BibitemShut
  {NoStop}%
\bibitem [{\citenamefont {Srednicki}(2007)}]{srednicki_2007}%
  \BibitemOpen
  \bibfield  {author} {\bibinfo {author} {\bibfnamefont {M.}~\bibnamefont
  {Srednicki}},\ }\href@noop {} {\emph {\bibinfo {title} {Quantum Field
  Theory}}}\ (\bibinfo  {publisher} {Cambridge University Press},\ \bibinfo
  {year} {2007})\BibitemShut {NoStop}%
\bibitem [{\citenamefont {Brezinski}\ and\ \citenamefont
  {Zaglia}(2013)}]{brezinski2013extrapolation}%
  \BibitemOpen
  \bibfield  {author} {\bibinfo {author} {\bibfnamefont {C.}~\bibnamefont
  {Brezinski}}\ and\ \bibinfo {author} {\bibfnamefont {M.}~\bibnamefont
  {Zaglia}},\ }\href {https://books.google.de/books?id=WGviBQAAQBAJ} {\emph
  {\bibinfo {title} {Extrapolation Methods: Theory and Practice}}},\ ISSN\
  (\bibinfo  {publisher} {Elsevier Science},\ \bibinfo {year}
  {2013})\BibitemShut {NoStop}%
\bibitem [{\citenamefont {Golub}\ and\ \citenamefont
  {Van~Loan}(2013)}]{golub2013matrix}%
  \BibitemOpen
  \bibfield  {author} {\bibinfo {author} {\bibfnamefont {G.}~\bibnamefont
  {Golub}}\ and\ \bibinfo {author} {\bibfnamefont {C.}~\bibnamefont
  {Van~Loan}},\ }\href {https://books.google.de/books?id=X5YfsuCWpxMC} {\emph
  {\bibinfo {title} {Matrix Computations}}},\ Johns Hopkins Studies in the
  Mathematical Sciences\ (\bibinfo  {publisher} {Johns Hopkins University
  Press},\ \bibinfo {year} {2013})\BibitemShut {NoStop}%
\bibitem [{\citenamefont {Hauschild}\ and\ \citenamefont
  {Pollmann}(2018)}]{Hauschild_2018}%
  \BibitemOpen
  \bibfield  {author} {\bibinfo {author} {\bibfnamefont {J.}~\bibnamefont
  {Hauschild}}\ and\ \bibinfo {author} {\bibfnamefont {F.}~\bibnamefont
  {Pollmann}},\ }\bibfield  {title} {\bibinfo {title} {Efficient numerical
  simulations with tensor networks: Tensor network python (tenpy)},\ }\bibfield
   {journal} {\bibinfo  {journal} {SciPost Physics Lecture Notes}\ }\href
  {https://doi.org/10.21468/scipostphyslectnotes.5}
  {10.21468/scipostphyslectnotes.5} (\bibinfo {year} {2018})\BibitemShut
  {NoStop}%
\bibitem [{\citenamefont {Uhlmann}(2011)}]{Uhlmann_2011}%
  \BibitemOpen
  \bibfield  {author} {\bibinfo {author} {\bibfnamefont {A.}~\bibnamefont
  {Uhlmann}},\ }\bibfield  {title} {\bibinfo {title} {Transition probability
  (fidelity) and its relatives},\ }\href
  {https://link.springer.com/article/10.1007/s10701-009-9381-y} {\bibfield
  {journal} {\bibinfo  {journal} {Foundations of physics}\ }\textbf {\bibinfo
  {volume} {41}},\ \bibinfo {pages} {288} (\bibinfo {year} {2011})}\BibitemShut
  {NoStop}%
\bibitem [{\citenamefont {Zhou}\ \emph {et~al.}(2008)\citenamefont {Zhou},
  \citenamefont {Or\'us},\ and\ \citenamefont
  {Vidal}}]{PhysRevLett.100.080601}%
  \BibitemOpen
  \bibfield  {author} {\bibinfo {author} {\bibfnamefont {H.-Q.}\ \bibnamefont
  {Zhou}}, \bibinfo {author} {\bibfnamefont {R.}~\bibnamefont {Or\'us}},\ and\
  \bibinfo {author} {\bibfnamefont {G.}~\bibnamefont {Vidal}},\ }\bibfield
  {title} {\bibinfo {title} {Ground state fidelity from tensor network
  representations},\ }\href {https://doi.org/10.1103/PhysRevLett.100.080601}
  {\bibfield  {journal} {\bibinfo  {journal} {Phys. Rev. Lett.}\ }\textbf
  {\bibinfo {volume} {100}},\ \bibinfo {pages} {080601} (\bibinfo {year}
  {2008})}\BibitemShut {NoStop}%
\bibitem [{\citenamefont {Zhou}(2007)}]{zhou2007renormalization}%
  \BibitemOpen
  \bibfield  {author} {\bibinfo {author} {\bibfnamefont {H.-Q.}\ \bibnamefont
  {Zhou}},\ }\href@noop {} {\bibinfo {title} {Renormalization group flows and
  quantum phase transitions: fidelity versus entanglement}} (\bibinfo {year}
  {2007}),\ \Eprint {https://arxiv.org/abs/0704.2945} {arXiv:0704.2945
  [cond-mat.stat-mech]} \BibitemShut {NoStop}%
\bibitem [{\citenamefont {Hartmann}\ \emph {et~al.}(2004)\citenamefont
  {Hartmann}, \citenamefont {Keck}, \citenamefont {Korsch},\ and\ \citenamefont
  {Mossmann}}]{hartmann_dynamics_2004}%
  \BibitemOpen
  \bibfield  {author} {\bibinfo {author} {\bibfnamefont {T.}~\bibnamefont
  {Hartmann}}, \bibinfo {author} {\bibfnamefont {F.}~\bibnamefont {Keck}},
  \bibinfo {author} {\bibfnamefont {H.~J.}\ \bibnamefont {Korsch}},\ and\
  \bibinfo {author} {\bibfnamefont {S.}~\bibnamefont {Mossmann}},\ }\bibfield
  {title} {\bibinfo {title} {Dynamics of {Bloch} oscillations},\ }\href
  {https://doi.org/10.1088/1367-2630/6/1/002} {\bibfield  {journal} {\bibinfo
  {journal} {New J. Phys.}\ }\textbf {\bibinfo {volume} {6}},\ \bibinfo {pages}
  {2} (\bibinfo {year} {2004})}\BibitemShut {NoStop}%
\bibitem [{\citenamefont {Sebby-Strabley}\ \emph {et~al.}(2006)\citenamefont
  {Sebby-Strabley}, \citenamefont {Anderlini}, \citenamefont {Jessen},\ and\
  \citenamefont {Porto}}]{sebby-strabley_lattice_2006}%
  \BibitemOpen
  \bibfield  {author} {\bibinfo {author} {\bibfnamefont {J.}~\bibnamefont
  {Sebby-Strabley}}, \bibinfo {author} {\bibfnamefont {M.}~\bibnamefont
  {Anderlini}}, \bibinfo {author} {\bibfnamefont {P.~S.}\ \bibnamefont
  {Jessen}},\ and\ \bibinfo {author} {\bibfnamefont {J.~V.}\ \bibnamefont
  {Porto}},\ }\bibfield  {title} {\bibinfo {title} {Lattice of double wells for
  manipulating pairs of cold atoms},\ }\href
  {https://doi.org/10.1103/PhysRevA.73.033605} {\bibfield  {journal} {\bibinfo
  {journal} {Phys. Rev. A}\ }\textbf {\bibinfo {volume} {73}},\ \bibinfo
  {pages} {033605} (\bibinfo {year} {2006})}\BibitemShut {NoStop}%
\bibitem [{\citenamefont {F{\"o}lling}\ \emph {et~al.}(2007)\citenamefont
  {F{\"o}lling}, \citenamefont {Trotzky}, \citenamefont {Cheinet},
  \citenamefont {Feld}, \citenamefont {Saers}, \citenamefont {Widera},
  \citenamefont {M{\"u}ller},\ and\ \citenamefont
  {Bloch}}]{foelling_direct_2007}%
  \BibitemOpen
  \bibfield  {author} {\bibinfo {author} {\bibfnamefont {S.}~\bibnamefont
  {F{\"o}lling}}, \bibinfo {author} {\bibfnamefont {S.}~\bibnamefont
  {Trotzky}}, \bibinfo {author} {\bibfnamefont {P.}~\bibnamefont {Cheinet}},
  \bibinfo {author} {\bibfnamefont {M.}~\bibnamefont {Feld}}, \bibinfo {author}
  {\bibfnamefont {R.}~\bibnamefont {Saers}}, \bibinfo {author} {\bibfnamefont
  {A.}~\bibnamefont {Widera}}, \bibinfo {author} {\bibfnamefont
  {T.}~\bibnamefont {M{\"u}ller}},\ and\ \bibinfo {author} {\bibfnamefont
  {I.}~\bibnamefont {Bloch}},\ }\bibfield  {title} {\bibinfo {title} {Direct
  observation of second-order atom tunnelling},\ }\href
  {https://doi.org/10.1038/nature06112} {\bibfield  {journal} {\bibinfo
  {journal} {Nature}\ }\textbf {\bibinfo {volume} {448}} (\bibinfo {year}
  {2007})}\BibitemShut {NoStop}%
\bibitem [{\citenamefont {Kohlert}\ \emph {et~al.}(2019)\citenamefont
  {Kohlert}, \citenamefont {Scherg}, \citenamefont {Li}, \citenamefont
  {L\"uschen}, \citenamefont {Das~Sarma}, \citenamefont {Bloch},\ and\
  \citenamefont {Aidelsburger}}]{PhysRevLett.122.170403}%
  \BibitemOpen
  \bibfield  {author} {\bibinfo {author} {\bibfnamefont {T.}~\bibnamefont
  {Kohlert}}, \bibinfo {author} {\bibfnamefont {S.}~\bibnamefont {Scherg}},
  \bibinfo {author} {\bibfnamefont {X.}~\bibnamefont {Li}}, \bibinfo {author}
  {\bibfnamefont {H.~P.}\ \bibnamefont {L\"uschen}}, \bibinfo {author}
  {\bibfnamefont {S.}~\bibnamefont {Das~Sarma}}, \bibinfo {author}
  {\bibfnamefont {I.}~\bibnamefont {Bloch}},\ and\ \bibinfo {author}
  {\bibfnamefont {M.}~\bibnamefont {Aidelsburger}},\ }\bibfield  {title}
  {\bibinfo {title} {Observation of many-body localization in a one-dimensional
  system with a single-particle mobility edge},\ }\href
  {https://doi.org/10.1103/PhysRevLett.122.170403} {\bibfield  {journal}
  {\bibinfo  {journal} {Phys. Rev. Lett.}\ }\textbf {\bibinfo {volume} {122}},\
  \bibinfo {pages} {170403} (\bibinfo {year} {2019})}\BibitemShut {NoStop}%
\end{thebibliography}%

\end{document}